\documentclass[12pt,a4paper]{article}
\usepackage{jheplike,tocloft,ifpdf,tocloft,rotating,amsmath,mathrsfs,multirow,mathtools,xcolor,cancel,tikz}
\usetikzlibrary{calc,decorations.markings,shapes.geometric}
\usepackage[backgroundcolor=green!5, linecolor=blue!60]{todonotes}

\def\twoLoopCorners{1}
\def\colorTableQ{1}

\newcommand{\resetGraphDefaults}{
\definecolor{emphA}{rgb}{0.75,0.0,0.325}
\definecolor{emphB}{rgb}{0.2,0.605,0}
\definecolor{emphC}{rgb}{0,0,0.975}
\definecolor{ndotColor}{rgb}{0.65,0.25,0.25}

\definecolor{optLegColour}{rgb}{0.5,0.5,0.5}
\definecolor{legColour}{rgb}{0.35,0.35,0.35}
\definecolor{legLabelColour}{rgb}{0,0,0}
\definecolor{mhvblue}{rgb}{0.6,0.6,0.7765}
\definecolor{ampgrey}{rgb}{0.9,0.9,0.9}
\def\ampSize{(1*\figScale*7pt)}

\def\figScale{1}
\def\legSpread{4}
\def\lineThickness{(1*\figScale)}
\def\dotSize{10*\figScale}
\def\legLen{0.75*0.32*\figScale}
\def\extLegLen{0.75*0.32*\figScale}
\def\labelDist{0.4*\figScale}
\tikzset{intA/.style={black,line width=\lineThickness,line cap=round,rounded corners=0.5pt}}
\tikzset{intB/.style={black,line width=\lineThickness,line cap=round,rounded corners=0.5pt}}
\tikzset{intC/.style={black,line width=\lineThickness,line cap=round,rounded corners=0.5pt}}

\tikzset{int/.style={black,line width=\lineThickness,line cap=round,rounded corners=0.5pt}}
\tikzset{intH/.style={emphA,line width=\lineThickness,line cap=round,rounded corners=0.5pt}}
\tikzset{ddot/.style={fill=black,circle,minimum size=0.35*\dotSize,inner sep=0}}
}
\tikzset{fullamp/.style={coordinate,minimum size=1.5*\ampSize,ball color=black!20,circle,text=white,inner sep=0}}
\tikzset{fullmhv/.style={coordinate,minimum size=0.9*\ampSize,ball color=mhvblue,circle,text=white,inner sep=0}}
\tikzset{fullmhvBar/.style={coordinate,minimum size=0.9*\ampSize,ball color=white,circle,text=white,inner sep=0}}
\tikzset{mhv/.style={fill=mhvblue,circle,draw=black,line width=\lineThickness,minimum size=0.8*\ampSize,text=white,inner sep=0}}
\tikzset{mhvBar/.style={fill=white,circle,draw=black,line width=\lineThickness,minimum size=0.8*\ampSize,text=white,inner sep=0}}

\resetGraphDefaults

\newcommand{\decorateA}{\definecolor{ndotColor}{rgb}{0.75,0.0,0.325}}
\newcommand{\decorateB}{\definecolor{ndotColor}{rgb}{0.2,0.605,0}}
\newcommand{\decorateC}{\definecolor{ndotColor}{rgb}{0,0,0.975}}
\newcommand{\decorateNo}{\definecolor{ndotColor}{rgb}{0.65,0.25,0.25}}

\newcommand{\emphEdges}{
\tikzset{intA/.style={emphA,line width=\lineThickness,line cap=round,rounded corners=0.5pt}}
\tikzset{intB/.style={emphB,line width=\lineThickness,line cap=round,rounded corners=0.5pt}}
\tikzset{intC/.style={emphC,line width=\lineThickness,line cap=round,rounded corners=0.5pt}}
}

\resetGraphDefaults

\tikzset{ndot/.style={transform shape,scale=0.25*\figScale,aspect=0.55,shape=diamond,fill=ndotColor}}

\tikzset{directedLoopArrow/.style={transform shape, scale=0.125*\figScale,shape=dart,aspect=0.5,fill,draw}}
\tikzset{markedEdge/.style={draw=none,decoration={markings,mark connection node=connode,mark=at position 0.5 with {\node[ndot] (connode) {};}},postaction={decorate}}}
\tikzset{doublemarkedEdge/.style={draw=none,decoration={markings,mark connection node=connode,mark=at position 0.33 with {\node[ndotR] (connode) {};},mark=at position 0.66 with {\node[ndotR] (connode) {};}},postaction={decorate}}}
\tikzset{directedLoopEdge/.style={black,line width=\lineThickness,line cap=round,decoration={markings,mark connection node=connode,mark=at position 0.5 with {\node[directedLoopArrow] (connode) {};}},postaction={decorate}}}

\tikzset{directedLoopEdgeReversed/.style={black,line width=\lineThickness,line cap=round,decoration={markings,mark connection node=connode,mark=at position 0.5 with {\node[rotate=180,directedLoopArrow] (connode) {};}},postaction={decorate}}}

\def\markStroke{0.65}

\tikzset{ndot/.style={transform shape,scale=0.35*\figScale,aspect=0.65,draw=ndotColor,line width=\markStroke*\figScale,shape=circle,fill=none}}
\tikzset{markedEdgeR/.style={draw=none,decoration={markings,mark connection node=connode,mark=at position 0.5 with {\node[ndot] (connode) {};}},postaction={decorate}}}
\tikzset{ndotR/.style={transform shape,scale=0.35*\figScale,aspect=0.65,draw=ndotColor,line width=\markStroke*\figScale,shape=circle,fill=white}}
\tikzset{ndotLayerA/.style={transform shape,scale=0.3*\figScale,aspect=0.65,shape=circle,fill=none,draw=none,line width=1pt,rounded corners=2pt}}
\tikzset{ndotLayerB/.style={transform shape,scale=0.35*\figScale,aspect=0.65,shape=circle,fill=white,draw=ndotColor,line width=\markStroke*\figScale,rounded corners=2pt}}
\tikzset{markedEdge/.style={draw=none,decoration={markings,mark connection node=mark,mark=at position 0.5 with {\node[ndotLayerA] (mark) {};},mark=at position 0.5 with {%\draw[int,line width=2.*\markStroke*\figScale,white,rounded corners=0.5pt](mark.north west)--(mark.south east);\draw[int,line width=2.*\markStroke*\figScale,white,rounded corners=0.5pt](mark.north east)--(mark.south west);\draw[int,line width=\markStroke*\figScale,ndotColor,rounded corners=0.5pt](mark.north west)--(mark.south east);\draw[int,line width=\markStroke*\figScale,ndotColor,rounded corners=0.5pt](mark.north east)--(mark.south west);
\node[ndotLayerB] (mark) {};
}},postaction={decorate}}}

\newcommand{\singleLegLabelled}[3]{\node at #1 [ddot]{};\draw[int] #1--($#1+(#2:\extLegLen*1.25)$);
\node at ($#1+(#2:\labelDist*1.2)$)[]{{\footnotesize #3}};}

\newcommand{\leg}[2]{
\fill[legColour] #1--($#1+(#2+\legSpread*3:\extLegLen)$)--($#1+(#2-\legSpread*3:\extLegLen)$);
\node at #1 [ddot]{};
}
\newcommand{\emptyLeg}[2]{
\node at #1 [ddot]{};
}
\newcommand{\legLabelled}[3]{\node at #1 [ddot]{};\fill[legColour] #1--($#1+(#2+\legSpread*3:\extLegLen)$)--($#1+(#2-\legSpread*3:\extLegLen)$);\node at ($#1+(#2:\labelDist)$)[]{{\footnotesize #3}};}

\newcommand{\arrow}[4]{\coordinate (aa0) at ($#1!0.5!#2$);\node[#4,transform shape, scale=0.125*\figScale,shape=dart,fill,draw,line width=\lineThickness,line cap=round,rotate=#3] at (aa0) {};}

\newcommand{\twoLoopVacGraph}[4]{\def\ephScale{#4*\figScale}\def\edgeLen{0.75*\ephScale}\def\angleA{180/#1}\def\angleC{180/#3}
\ifthenelse{#2=0}{
\coordinate (v0) at (0,0);
\ifthenelse{#1=0}{}{\draw[draw=none] (v0) arc (0:360:\edgeLen/2 and \edgeLen/2)\foreach\a in {0,...,#1}{coordinate[pos=1/#1*\a-1/#1] (a\a)}};
\ifthenelse{#3=0}{}{\draw[draw=none] (v0) arc (180:540:\edgeLen/2 and \edgeLen/2)\foreach\c in {0,...,#3}{coordinate[pos=1/#3*\c-1/#3] (c\c)}};
\ifthenelse{#1=1}{\draw[intA](v0) arc (0:360:\edgeLen/1.125*0.3 and \edgeLen/2*0.3)}{
\ifthenelse{#1=2}{\draw[intA](v0) arc (0:360:\edgeLen/1.125*0.45 and \edgeLen/2*0.45)coordinate[pos=0.5](a1);\emptyLeg{(a1)}{180};}{
\foreach\a[remember=\a as \la] in {0,...,#1}{\ifthenelse{\a=0}{}{\draw[intA](a\la)--(a\a);\ifthenelse{\a=1}{}{}}}
\foreach\a[remember=\a as \la] in {0,...,#1}{\ifthenelse{\a=0}{}{\ifthenelse{\a=1}{}{\emptyLeg{(a\a)}{2*\angleA*\a-2*\angleA+0}}}}
}};
\ifthenelse{#3=1}{\draw[intC](v0) arc (180:540:\edgeLen/1.125*0.3 and \edgeLen/2*0.3)}{\ifthenelse{#3=2}{\draw[intC](v0) arc (180:540:\edgeLen/1.125*0.45 and \edgeLen/2*0.45)coordinate[pos=0.5](c1);\emptyLeg{(c1)}{0};}{
\foreach\c[remember=\c as \lc] in {0,...,#3}{\ifthenelse{\c=0}{}{\draw[intC](c\lc)--(c\c);\ifthenelse{\c=1}{}{}}}
\foreach\c[remember=\c as \lc] in {0,...,#3}{\ifthenelse{\c=0}{}{\ifthenelse{\c=1}{}{\emptyLeg{(c\c)}{2*\angleC*\c-2*\angleC+180}}}}
}};
\ifthenelse{#3=0}{\emptyLeg{(v0)}{0}}{\emptyLeg{(v0)}{90}};
}{\ifthenelse{\lengthtest{#3pt>#1pt}}{\def\angleB{0}}{\def\angleB{180}};
\coordinate (v0) at (90:0.5*\edgeLen);
\coordinate (v1) at (-90:0.5*\edgeLen);

\ifthenelse{#1=1}{\draw[intA] (v0) arc (90:270:\edgeLen/2.5 and \edgeLen/2)}{
\ifthenelse{\twoLoopCorners=0}{
\draw[intA] (v0) arc (90:270:\edgeLen/1.45 and \edgeLen/2)\foreach\a in {2,...,#1}{coordinate[pos=1/#1*\a-1/#1] (a\a)};
\foreach\a in {2,...,#1}{\emptyLeg{(a\a)}{\angleA*\a-\angleA+90}};
}{
\draw[draw=none] (v0) arc (90:270:\edgeLen/1.45 and \edgeLen/2)\foreach\a in {0,...,#1}{coordinate[pos=1/#1*\a] (a\a)};
\foreach\a[remember=\a as \la] in {0,...,#1}{\ifthenelse{\a=0}{}{\draw[intA](a\la)--(a\a);}};
\foreach\a in {1,...,#1}{\ifthenelse{\a=#1}{}{\emptyLeg{(a\a)}{\angleA*\a+90}}};
}};

\ifthenelse{#3=1}{\draw[intC] (v0) arc (90:-90:\edgeLen/2.5 and \edgeLen/2)}{

\ifthenelse{\twoLoopCorners=0}{\draw[intC] (v0) arc (90:-90:\edgeLen/1.35 and \edgeLen/2) \foreach\c in {2,...,#3}{coordinate[pos=1/#3*\c-1/#3] (c\c)};
\foreach\c in {2,...,#3}{\emptyLeg{(c\c)}{-\angleC*\c+\angleC+90}}}{
\draw[draw=none] (v0) arc (90:-90:\edgeLen/1.45 and \edgeLen/2) \foreach\c in {0,...,#3}{coordinate[pos=1/#3*\c] (c\c)};
\foreach\c[remember=\c as \lc] in {0,...,#3}{\ifthenelse{\c=0}{}{\draw[intC](c\lc)--(c\c);}};
\foreach\c in {1,...,#3}{\ifthenelse{\c=#3}{}{\emptyLeg{(c\c)}{-\angleC*\c+90}}};
}};
\ifthenelse{#2=1}{\draw[intB] (v0) -- (v1);\emptyLeg{(v1)}{-90};\ifthenelse{#3=1\OR #1=1}{\emptyLeg{(v0)}{90}}{\emptyLeg{(v0)}{90}}}{\draw[intB] (v0) -- (v1) \foreach\b in {2,...,#2}{coordinate[pos=1/#2*\b-1/#2] (b\b)};
\foreach\b in {2,...,#2}{\emptyLeg{(b\b)}{\angleB}};\emptyLeg{(v1)}{-90};\emptyLeg{(v0)}{90};};
}}

\newcommand{\wheelVacGraphWithLabels}[7]{%
\ifthenelse{\colorTableQ=1}{\tikzset{intA/.style={emphA,line width=\lineThickness,line cap=round}}%
\tikzset{intB/.style={emphB,line width=\lineThickness,line cap=round}}%
\tikzset{intC/.style={emphC,line width=\lineThickness,line cap=round}}}{}%
$\fwbox{55pt}{\begin{tikzpicture}[scale=#7*\figScale,baseline=-3.05]\useasboundingbox($\figScale*(-0.85,-0.85)$) rectangle ($\figScale*(1,0.85)$);
\ifthenelse{\showTikzQ=1}{\draw[draw=none]($\figScale*(-0.85,-0.85)$) rectangle ($\figScale*(1,0.85)$);
\def\rotn{0}\def\edgeLen{0.75*\figScale}\def\angleA{120/#1}\def\angleB{120/#2}\def\angleC{120/#3}\def\innerRad{\edgeLen/1.3}
\coordinate(v0)at(0,0);
\coordinate(v1)at(0+\rotn:\innerRad);
\coordinate(v2)at(120+\rotn:\innerRad);
\coordinate(v3)at(240+\rotn:\innerRad);
\draw[directedLoopEdge,emphA](v0)--(v1); \draw[directedLoopEdge,emphC](v0)--(v2);\draw[directedLoopEdge,emphB](v0)--(v3);
\foreach\a in {0,...,#1}{\ifthenelse{\a=0}{\coordinate (a\a) at (v3);}{\ifthenelse{\a=#1}{\coordinate (a\a) at (v2);}{\coordinate (a\a) at ($(v0)+(240+\rotn-\a*\angleA:\edgeLen/1.0125)$);}}};
\foreach\b in {0,...,#2}{\ifthenelse{\b=0}{\coordinate (b\b) at (v2);}{\ifthenelse{\b=#2}{\coordinate (b\b) at (v1);}{\coordinate (b\b) at ($(v0)+(120+\rotn-\b*\angleB:\edgeLen/1.0125)$);}}};
\foreach\c in {0,...,#3}{\ifthenelse{\c=0}{\coordinate (c\c) at (v1);}{\ifthenelse{\c=#3}{\coordinate (c\c) at (v3);}{\coordinate (c\c) at ($(v0)+(0+\rotn-\c*\angleC:\edgeLen/1.0125)$);}}};
\foreach\d in {0,...,#4}{\coordinate (d\d) at ($(v0)!\d/#4!(v1)$);
\ifthenelse{\NOT\d=0 \AND \NOT\d=#4}{\leg{(d\d)}{90+\rotn+180};}{};};
\foreach\e in {0,...,#5}{\coordinate (e\e) at ($(v0)!\e/#5!(v3)$);
\ifthenelse{\NOT\e=0 \AND \NOT\e=#5}{\leg{(e\e)}{150-0*180+\rotn};}{};};
\foreach\f in {0,...,#6}{\coordinate (f\f) at ($(v0)!\f/#6!(v2)$);
\ifthenelse{\NOT\f=0 \AND \NOT\f=#6}{\leg{(f\f)}{30-0*180+\rotn};}{};};
\ifthenelse{#1=1}{\draw[directedLoopEdge,emphA] (v3) arc (240+\rotn:120+\rotn:\innerRad) node[pos=0.5,below=1pt,left=0.25pt] {$\ell_1$};; 
}{\foreach\a[remember=\a as \la] in {0,...,#1}{\ifthenelse{\a=0}{}{\draw[intA](a\la)--(a\a);}}};
\ifthenelse{#2=1}{\draw[directedLoopEdge,emphB] (v2) arc (120+\rotn:0+\rotn:\innerRad) node[pos=0.5,above=1pt] {$\ell_2$};}
{\foreach\b[remember=\b as \lb] in {0,...,#2}{\ifthenelse{\b=0}{}{\draw[intB](b\lb)--(b\b);}};};
\ifthenelse{#3=1}{\draw[directedLoopEdge,emphC] (v1) arc (0+\rotn:-120+\rotn:\innerRad) node[pos=0.5,below=3pt,right=1pt] {$\ell_3$};;
}{\foreach\c[remember=\c as \lc] in {0,...,#3}{\ifthenelse{\c=0}{}{\draw[intC](c\lc)--(c\c);}};};
\ifthenelse{#1=1}{}{\foreach\a in {1,...,#1}{\ifthenelse{\a=#1}{}{\leg{(a\a)}{240+\rotn-\a*\angleA};}}};
\ifthenelse{#2=1}{}{\foreach\b in {1,...,#2}{\ifthenelse{\b=#2}{}{\leg{(b\b)}{120+\rotn-\b*\angleB};}}};
\ifthenelse{#3=1}{}{\foreach\c in {1,...,#3}{\ifthenelse{\c=#3}{}{\leg{(c\c)}{0+\rotn-\c*\angleC};}}};
\emptyLeg{(v1)}{0+\rotn};\emptyLeg{(v2)}{120+\rotn};\emptyLeg{(v3)}{240+\rotn};\emptyLeg{(v0)}{180+\rotn};
}{\draw[line width=0.125]($\figScale*(-0.85,-0.85)$) rectangle ($\figScale*(1,0.85)$);
\node at (0,0) {$\threeLoopWheelName{#1}{#2}{#3}{#4}{#5}{#6}$};}
\end{tikzpicture}%
\tikzset{intA/.style={black,line width=\lineThickness,line cap=round}}\tikzset{intB/.style={black,line width=\lineThickness,line cap=round}}\tikzset{intC/.style={black,line width=\lineThickness,line cap=round}}}$}

\newcommand{\ladderThreeVacGraph}[7]{
\def\ephScale{#7*\figScale}\def\edgeLen{0.75*\ephScale}\def\angleAa{180/#1}\def\angleAb{180/#2}\def\angleBa{180/#3}\def\angleBb{180/#4}\def\angleCc{180/#6}
\ifthenelse{#6=0}{
\coordinate (v1) at (0,0);
\coordinate (v2) at (v1);
\coordinate (v3) at ($(v1)+(-90:\edgeLen)$);
\coordinate (v4) at (v3);
\ifthenelse{#1=1}{\draw[intA] (v1) arc (90:270:\edgeLen/1.15 and \edgeLen/2);
}{\draw[draw=none] (v1) arc (90:270:\edgeLen/1.15 and \edgeLen/2)\foreach\a in {0,...,#1}{coordinate[pos=1/#1*\a] (aa\a)};
\foreach\a[remember=\a as \la] in {0,...,#1}{\ifthenelse{\a=0}{}{\draw[intA](aa\la)--(aa\a);}};
\foreach\a in {1,...,#1}{\ifthenelse{\a=#1}{}{\leg{(aa\a)}{\angleAa*\a+90}}};
}

\ifthenelse{#2=1}{\draw[intA] (v1) arc (90:270:\edgeLen/12 and \edgeLen/2);
}{
\draw[draw=none] (v1) arc (90:270:\edgeLen/4.5 and \edgeLen/2)\foreach\a in {0,...,#2}{coordinate[pos=(((1/#2*\a)*(3+(1/#2*\a)*(2*(1/#2*\a)-3)))*1/2)] (ab\a)};
\foreach\a[remember=\a as \la] in {0,...,#2}{\ifthenelse{\a=0}{}{\draw[intA](ab\la)--(ab\a);}};
\foreach\a in {1,...,#2}{\ifthenelse{\a=#2}{}{\leg{(ab\a)}{180}}};
}

\ifthenelse{#3=1}{\draw[intC] (v3) arc (-90:90:\edgeLen/4.5 and \edgeLen/2);
}{%\draw[draw=none] (v3) arc (-90:90:\edgeLen/4.5 and \edgeLen/2)\foreach\b in {0,...,#3}{coordinate[pos=1/#3*\b] (ba\b)};
\draw[draw=none] (v3) arc (-90:90:\edgeLen/4.5 and \edgeLen/2)\foreach\b in {0,...,#3}{coordinate[pos=(((1/#3*\b)*(3+(1/#3*\b)*(2*(1/#3*\b)-3)))*1/2)] (ba\b)};

\foreach\b[remember=\b as \lb] in {0,...,#3}{\ifthenelse{\b=0}{}{\draw[intC](ba\lb)--(ba\b);}};
\foreach\b in {1,...,#3}{\ifthenelse{\b=#3}{}{\leg{(ba\b)}{0}}};
}

\ifthenelse{#4=1}{\draw[intC] (v3) arc (-90:90:\edgeLen/1.15 and \edgeLen/2);
}{\draw[draw=none] (v3) arc (-90:90:\edgeLen/1.15 and \edgeLen/2)\foreach\b in {0,...,#4}{coordinate[pos=1/#4*\b] (bb\b)};
\foreach\b[remember=\b as \lb] in {0,...,#4}{\ifthenelse{\b=0}{}{\draw[intC](bb\lb)--(bb\b);}};
\foreach\b in {1,...,#4}{\ifthenelse{\b=#4}{}{\leg{(bb\b)}{\angleBb*\b-90}}};
}
\optLeg{(v1)}{90};\optLeg{(v3)}{-90};

}{

\ifthenelse{#5=0}{%
\coordinate (v1) at (0,0);
\coordinate (v2) at (v1);
\coordinate (v3) at ($(v2)+(-50:\edgeLen)$);
\coordinate (v4) at (230:\edgeLen);

\ifthenelse{#1=1}{\draw[intA,rotate=-40] (v1) arc (90:270:\edgeLen/1.65 and \edgeLen/2);
}{\draw[draw=none,rotate=-40] (v1) arc (90:270:\edgeLen/1.65 and \edgeLen/2)\foreach\a in {0,...,#1}{coordinate[pos=1/#1*\a] (aa\a)};
\foreach\a[remember=\a as \la] in {0,...,#1}{\ifthenelse{\a=0}{}{\draw[intA](aa\la)--(aa\a);}};
\foreach\a in {1,...,#1}{\ifthenelse{\a=#1}{}{\leg{(aa\a)}{\angleAa*\a+90-40}}};
}

\ifthenelse{#4=1}{\draw[intC,rotate=40] (v3) arc (-90:90:\edgeLen/1.65 and \edgeLen/2);
}{\draw[draw=none,rotate=40] (v3) arc (-90:90:\edgeLen/1.65 and \edgeLen/2)\foreach\b in {0,...,#4}{coordinate[pos=1/#4*\b] (bb\b)};
\foreach\b[remember=\b as \lb] in {0,...,#4}{\ifthenelse{\b=0}{}{\draw[intC](bb\lb)--(bb\b);}};
\foreach\b in {1,...,#4}{\ifthenelse{\b=#4}{}{\leg{(bb\b)}{\angleBb*\b-90+40}}};
}

\ifthenelse{#6=1}{\draw[intB] (v3)--(v4);%\draw[intB,rotate=-30] (v3) arc (0:-120:0.746*\edgeLen/1 and 0.746*\edgeLen/1);
}{\draw[draw=none,rotate=-30] (v3) arc (0:-120:0.746*\edgeLen/1 and 0.746*\edgeLen/1)\foreach\c in {0,...,#6}{coordinate[pos=1/#6*\c] (cc\c)};
\foreach\c[remember=\c as \lc] in {0,...,#6}{\ifthenelse{\c=0}{}{\draw[intB](cc\lc)--(cc\c);}};
\foreach\c in {1,...,#6}{\ifthenelse{\c=#6}{}{\leg{(cc\c)}{-\angleCc*\c+0}}};
}

\draw[intA] (v4)--(v1)\foreach\a in {0,...,#2}{coordinate[pos=1/#2*\a] (ab\a)};
\foreach\a in {1,...,#2}{\ifthenelse{\a=#2}{}{\leg{(ab\a)}{-40}}};
\draw[intC] (v2)--(v3)\foreach\b in {0,...,#3}{coordinate[pos=1/#3*\b] (ba\b)};
\foreach\b in {1,...,#3}{\ifthenelse{\b=#3}{}{\leg{(ba\b)}{40+180}}};

\optLeg{(v1)}{90};
\optLeg{(v4)}{180+35};
\optLeg{(v3)}{-35};

}{

\coordinate (v1) at (-0.75*\edgeLen/2,0);
\coordinate (v2) at (0:0.75*\edgeLen/2);
\coordinate (v3) at ($(v2)+(-90:\edgeLen)$);
\coordinate (v4) at ($(v1)+(-90:\edgeLen)$);

\draw[directedLoopEdgeReversed,emphB] (v1)--(v2)\foreach\c in {0,...,#5}{coordinate[pos=1/#5*\c] (ca\c)};
\foreach\c in {1,...,#5}{\ifthenelse{\c=#5}{}{\leg{(ca\c)}{90}}};
\draw[directedLoopEdgeReversed,emphB] (v3)--(v4)\foreach\c in {0,...,#6}{coordinate[pos=1/#6*\c] (cc\c)};
\foreach\c in {1,...,#6}{\ifthenelse{\c=#6}{}{\leg{(cc\c)}{-90}}};

\draw[directedLoopEdge,emphA] (v4)--(v1)\foreach\a in {0,...,#2}{coordinate[pos=1/#2*\a] (ab\a)};
\foreach\a in {1,...,#2}{\ifthenelse{\a=#2}{}{\leg{(ab\a)}{0}}};

\ifthenelse{#1=1}{\draw[directedLoopEdge,emphA] (v1) arc (90:270:\edgeLen/1.75 and \edgeLen/2);}
{
\draw[draw=none] (v1) arc (90:270:\edgeLen/1.75 and \edgeLen/2)\foreach\a in {0,...,#1}{coordinate[pos=1/#1*\a] (aa\a)};
\foreach\a[remember=\a as \la] in {0,...,#1}{\ifthenelse{\a=0}{}{\draw[directedLoopEdge,emphA](aa\la)--(aa\a);}};
\foreach\a in {1,...,#1}{\ifthenelse{\a=#1}{}{\leg{(aa\a)}{\angleAa*\a+90}}};
}

\draw[directedLoopEdge,emphC] (v2)--(v3)\foreach\b in {0,...,#3}{coordinate[pos=1/#3*\b] (ba\b)};
\foreach\b in {1,...,#3}{\ifthenelse{\b=#3}{}{\leg{(ba\b)}{180}}};
\ifthenelse{#4=1}{\draw[directedLoopEdge,emphC] (v3) arc (-90:90:\edgeLen/1.75 and \edgeLen/2);}
{
\draw[draw=none] (v3) arc (-90:90:\edgeLen/1.75 and \edgeLen/2)\foreach\b in {0,...,#4}{coordinate[pos=1/#4*\b] (bb\b)};
\foreach\b[remember=\b as \lb] in {0,...,#4}{\ifthenelse{\b=0}{}{\draw[directedLoopEdge,emphC](bb\lb)--(bb\b);}};
\foreach\b in {1,...,#4}{\ifthenelse{\b=#4}{}{\leg{(bb\b)}{\angleBb*\b-90}}};
}
\emptyLeg{(v1)}{120};
\emptyLeg{(v2)}{60};
\emptyLeg{(v3)}{-60};
\emptyLeg{(v4)}{-120};
}
}
}

\newcommand{\optLeg}[2]{\coordinate (aa0) at #1;
\coordinate (aa5) at ($#1+(#2+\legSpread*3:\extLegLen)$);
\coordinate (bb5) at ($#1+(#2-\legSpread*3:\extLegLen)$);
\coordinate (aa1) at ($(aa0)!0.4!(aa5)$);\coordinate (aa2) at ($(aa0)!0.55!(aa5)$);\coordinate (aa3) at ($(aa0)!0.7!(aa5)$);\coordinate (aa4) at ($(aa0)!0.85!(aa5)$);
\coordinate (bb1) at ($(aa0)!0.4!(bb5)$);\coordinate (bb2) at ($(aa0)!0.55!(bb5)$);\coordinate (bb3) at ($(aa0)!0.7!(bb5)$);\coordinate (bb4) at ($(aa0)!0.85!(bb5)$);
\fill[optLegColour] (aa0)--(aa1)--(bb1)--(aa0);
\fill[optLegColour] (aa2)--(aa3)--(bb3)--(bb2)--(aa2);
\fill[optLegColour] (aa4)--(aa5)--(bb5)--(bb4)--(aa4);
\node at #1 [ddot]{};
}

\newcommand{\tikzBox}[2][0.5]{\begin{tikzpicture}[scale=1,baseline=-3.05,rotate=0]\useasboundingbox ($\figScale*(-#1,-#1)$) rectangle ($\figScale*(#1,#1)$);#2\end{tikzpicture}}

\newcommand{\oneLoopGraphElement}[2][0]{\def\rotn{-360/#2}\def\edgeLen{0.75*\figScale}
\ifthenelse{#2=3}{\coordinate (v0) at (0,-\edgeLen/9)}{\coordinate (v0) at (0,0)};
\ifthenelse{#2=1}{\draw[int]($(v0)+(\edgeLen/2*0.3,0)$) arc (0:360:\edgeLen/2*0.3 and \edgeLen/1.125*0.3)coordinate[pos=0.25](e1);\leg{($(v0)-(0,\edgeLen/1.125*0.3)$)}{-90}}{\ifthenelse{#2=2}{
\draw[int]($(v0)+(\edgeLen/1.125*0.45,0)$) arc (0:360:\edgeLen/1.125*0.45 and \edgeLen/2*0.45)coordinate[pos=0.5](a1)coordinate[pos=0](a0)coordinate[pos=0.25](e1)coordinate[pos=0.75](e2);\leg{(a1)}{180};\leg{(a0)}{0};
}{
\foreach\a in {0,...,#2}{\coordinate (a\a) at ($(v0)+(\a*\rotn-90-\rotn/2:\edgeLen/2)$);};
\foreach\a[remember=\a as \la] in {0,...,#2}{\ifthenelse{\a=0}{}{\draw[int](a\la)--(a\a)coordinate (e\a) at ($(a\la)!0.5!(a\a)$);}};
\foreach\a in {1,...,#2}{\leg{(a\a)}{\a*\rotn-90-\rotn/2};};
}};
}

\newcommand{\oneLoopGraphSkeleton}[2][0]{\begin{scope}[rotate=#1]\def\rotn{-360/#2}\def\edgeLen{0.75*\figScale}
\ifthenelse{#2=3}{\coordinate (v0) at (0,-\edgeLen/9)}{\coordinate (v0) at (0,0)};
\ifthenelse{#2=1}{\draw[draw=none]($(v0)+(\edgeLen/2*0.3,0)$) arc (0:360:\edgeLen/2*0.3 and \edgeLen/1.125*0.3);\leg{($(v0)-(0,\edgeLen/1.125*0.3)$)}{-90}}{\ifthenelse{#2=2}{
\draw[draw=none]($(v0)+(\edgeLen/1.125*0.45,0)$) arc (0:360:\edgeLen/1.125*0.45 and \edgeLen/2*0.45)coordinate[pos=0.5](a1)coordinate[pos=0](a0);%\leg{(a1)}{180};\leg{(a0)}{0};
}{
\foreach\a in {0,...,#2}{\coordinate (a\a) at ($(v0)+(\a*\rotn-90-\rotn/2:\edgeLen/2)$);};
\foreach\a[remember=\a as \la] in {0,...,#2}{\ifthenelse{\a=0}{}{\draw[draw=none](a\la)--(a\a);}};
}};\end{scope}
}

\newcommand{\oneLoopGraph}[2][0]{
\begin{tikzpicture}[scale=\figScale,baseline=-3.05,rotate=#1]
\ifthenelse{\showTikzQ=1}{\useasboundingbox ($\figScale*(-0.6,-0.6)$) rectangle ($\figScale*(0.6,0.6)$);
\oneLoopGraphElement[#1]{#2}}{\draw [line width=0.15] ($\figScale*(-0.6,-0.6)$) rectangle ($\figScale*(0.6,0.6)$);}
\end{tikzpicture}}

\newcommand{\twoLoopTableFigure}[3]{\begin{tikzpicture}[scale=\figScale,baseline=-3.05]
\ifthenelse{\showTikzQ=1}{\useasboundingbox ($\figScale*(-0.95,-0.625)$) rectangle ($\figScale*(0.95,0.625)$);
\draw [draw=none] ($\figScale*(-0.95,-0.625)$) rectangle ($\figScale*(0.95,0.625)$);
\ifthenelse{\colorTableQ=1}{\tikzset{intA/.style={emphA,line width=\lineThickness,line cap=round}}
\tikzset{intB/.style={emphB,line width=\lineThickness,line cap=round}}
\tikzset{intC/.style={emphC,line width=\lineThickness,line cap=round}}}{};
\twoLoopGraph{#1}{#2}{#3}{1}
\tikzset{intA/.style={black,line width=\lineThickness,line cap=round}}
\tikzset{intB/.style={black,line width=\lineThickness,line cap=round}}
\tikzset{intC/.style={black,line width=\lineThickness,line cap=round}}
}{\draw [line width=0.15] ($\figScale*(-0.95,-0.625)$) rectangle ($\figScale*(0.95,0.625)$);
\node at (0,0) {$\twoLoopLadderName{#1}{#2}{#3}$};}
\end{tikzpicture}}

\newcommand{\twoLoopGraph}[4]{\def\ephScale{#4*\figScale}\def\edgeLen{0.75*\ephScale}\def\angleA{180/#1}\def\angleC{180/#3}
\ifthenelse{#2=0}{
\coordinate (v0) at (0,0);
\ifthenelse{#1=0}{}{\draw[draw=none] (v0) arc (0:360:\edgeLen/2 and \edgeLen/2)\foreach\a in {0,...,#1}{coordinate[pos=1/#1*\a-1/#1] (a\a)}};
\ifthenelse{#3=0}{}{\draw[draw=none] (v0) arc (180:540:\edgeLen/2 and \edgeLen/2)\foreach\c in {0,...,#3}{coordinate[pos=1/#3*\c-1/#3] (c\c)}};
\ifthenelse{#1=1}{\draw[intA](v0) arc (0:360:\edgeLen/1.125*0.3 and \edgeLen/2*0.3)}{
\ifthenelse{#1=2}{\draw[intA](v0) arc (0:360:\edgeLen/1.125*0.45 and \edgeLen/2*0.45)coordinate[pos=0.5](a1);\leg{(a1)}{180};}{
\foreach\a[remember=\a as \la] in {0,...,#1}{\ifthenelse{\a=0}{}{\draw[intA](a\la)--(a\a);\ifthenelse{\a=1}{}{}}}
\foreach\a[remember=\a as \la] in {0,...,#1}{\ifthenelse{\a=0}{}{\ifthenelse{\a=1}{}{\leg{(a\a)}{2*\angleA*\a-2*\angleA+0}}}}
}};
\ifthenelse{#3=1}{\draw[intC](v0) arc (180:540:\edgeLen/1.125*0.3 and \edgeLen/2*0.3)}{\ifthenelse{#3=2}{\draw[intC](v0) arc (180:540:\edgeLen/1.125*0.45 and \edgeLen/2*0.45)coordinate[pos=0.5](c1);\leg{(c1)}{0};}{
\foreach\c[remember=\c as \lc] in {0,...,#3}{\ifthenelse{\c=0}{}{\draw[intC](c\lc)--(c\c);\ifthenelse{\c=1}{}{}}}
\foreach\c[remember=\c as \lc] in {0,...,#3}{\ifthenelse{\c=0}{}{\ifthenelse{\c=1}{}{\leg{(c\c)}{2*\angleC*\c-2*\angleC+180}}}}
}};
\ifthenelse{#3=0}{\leg{(v0)}{0}}{\optLeg{(v0)}{90}};
}{\ifthenelse{\lengthtest{#3pt>#1pt}}{\def\angleB{0}}{\def\angleB{180}};
\coordinate (v0) at (90:0.5*\edgeLen);
\coordinate (v1) at (-90:0.5*\edgeLen);

\ifthenelse{#1=1}{\draw[intA] (v0) arc (90:270:\edgeLen/2.5 and \edgeLen/2)}{
\ifthenelse{\twoLoopCorners=0}{
\draw[intA] (v0) arc (90:270:\edgeLen/1.45 and \edgeLen/2)\foreach\a in {2,...,#1}{coordinate[pos=1/#1*\a-1/#1] (a\a)};
\foreach\a in {2,...,#1}{\leg{(a\a)}{\angleA*\a-\angleA+90}};
}{
\draw[draw=none] (v0) arc (90:270:\edgeLen/1.45 and \edgeLen/2)\foreach\a in {0,...,#1}{coordinate[pos=1/#1*\a] (a\a)};
\foreach\a[remember=\a as \la] in {0,...,#1}{\ifthenelse{\a=0}{}{\draw[intA](a\la)--(a\a);}};
\foreach\a in {1,...,#1}{\ifthenelse{\a=#1}{}{\leg{(a\a)}{\angleA*\a+90}}};
}};

\ifthenelse{#3=1}{\draw[intC] (v0) arc (90:-90:\edgeLen/2.5 and \edgeLen/2)}{

\ifthenelse{\twoLoopCorners=0}{\draw[intC] (v0) arc (90:-90:\edgeLen/1.35 and \edgeLen/2) \foreach\c in {2,...,#3}{coordinate[pos=1/#3*\c-1/#3] (c\c)};
\foreach\c in {2,...,#3}{\leg{(c\c)}{-\angleC*\c+\angleC+90}}}{
\draw[draw=none] (v0) arc (90:-90:\edgeLen/1.45 and \edgeLen/2) \foreach\c in {0,...,#3}{coordinate[pos=1/#3*\c] (c\c)};
\foreach\c[remember=\c as \lc] in {0,...,#3}{\ifthenelse{\c=0}{}{\draw[intC](c\lc)--(c\c);}};
\foreach\c in {1,...,#3}{\ifthenelse{\c=#3}{}{\leg{(c\c)}{-\angleC*\c+90}}};
}};
\ifthenelse{#2=1}{\draw[intB] (v0) -- (v1);\optLeg{(v1)}{-90};\ifthenelse{#3=1\OR #1=1}{\leg{(v0)}{90}}{\optLeg{(v0)}{90}}}{\draw[intB] (v0) -- (v1) \foreach\b in {2,...,#2}{coordinate[pos=1/#2*\b-1/#2] (b\b)};
\foreach\b in {2,...,#2}{\leg{(b\b)}{\angleB}};\optLeg{(v1)}{-90};\optLeg{(v0)}{90};};
}}

\newcommand{\wheelGraph}[6]{%
\ifthenelse{\colorTableQ=1}{\tikzset{intA/.style={emphA,line width=\lineThickness,line cap=round}}%
\tikzset{intB/.style={emphB,line width=\lineThickness,line cap=round}}%
\tikzset{intC/.style={emphC,line width=\lineThickness,line cap=round}}}{}%
$\fwbox{55pt}{\begin{tikzpicture}[scale=\figScale,baseline=-3.05]\useasboundingbox($\figScale*(-0.85,-0.85)$) rectangle ($\figScale*(1,0.85)$);
\ifthenelse{\showTikzQ=1}{\draw[draw=none]($\figScale*(-0.85,-0.85)$) rectangle ($\figScale*(1,0.85)$);
\def\rotn{0}\def\edgeLen{0.75*\figScale}\def\angleA{120/#1}\def\angleB{120/#2}\def\angleC{120/#3}\def\innerRad{\edgeLen/1.3}
\coordinate(v0)at(0,0);
\coordinate(v1)at(0+\rotn:\innerRad);
\coordinate(v2)at(120+\rotn:\innerRad);
\coordinate(v3)at(240+\rotn:\innerRad);
\draw[intA](v0)--(v1);\draw[intC](v0)--(v2);\draw[intB](v0)--(v3);
\foreach\a in {0,...,#1}{\ifthenelse{\a=0}{\coordinate (a\a) at (v3);}{\ifthenelse{\a=#1}{\coordinate (a\a) at (v2);}{\coordinate (a\a) at ($(v0)+(240+\rotn-\a*\angleA:\edgeLen/1.0125)$);}}};
\foreach\b in {0,...,#2}{\ifthenelse{\b=0}{\coordinate (b\b) at (v2);}{\ifthenelse{\b=#2}{\coordinate (b\b) at (v1);}{\coordinate (b\b) at ($(v0)+(120+\rotn-\b*\angleB:\edgeLen/1.0125)$);}}};
\foreach\c in {0,...,#3}{\ifthenelse{\c=0}{\coordinate (c\c) at (v1);}{\ifthenelse{\c=#3}{\coordinate (c\c) at (v3);}{\coordinate (c\c) at ($(v0)+(0+\rotn-\c*\angleC:\edgeLen/1.0125)$);}}};
\foreach\d in {0,...,#4}{\coordinate (d\d) at ($(v0)!\d/#4!(v1)$);
\ifthenelse{\NOT\d=0 \AND \NOT\d=#4}{\leg{(d\d)}{90+\rotn+180};}{};};
\foreach\e in {0,...,#5}{\coordinate (e\e) at ($(v0)!\e/#5!(v3)$);
\ifthenelse{\NOT\e=0 \AND \NOT\e=#5}{\leg{(e\e)}{150-0*180+\rotn};}{};};
\foreach\f in {0,...,#6}{\coordinate (f\f) at ($(v0)!\f/#6!(v2)$);
\ifthenelse{\NOT\f=0 \AND \NOT\f=#6}{\leg{(f\f)}{30-0*180+\rotn};}{};};
\ifthenelse{#1=1}{\draw[intA] (v3) arc (240+\rotn:120+\rotn:\innerRad);
}{\foreach\a[remember=\a as \la] in {0,...,#1}{\ifthenelse{\a=0}{}{\draw[intA](a\la)--(a\a);}}};
\ifthenelse{#2=1}{\draw[intB] (v2) arc (120+\rotn:0+\rotn:\innerRad);}
{\foreach\b[remember=\b as \lb] in {0,...,#2}{\ifthenelse{\b=0}{}{\draw[intB](b\lb)--(b\b);}};};
\ifthenelse{#3=1}{\draw[intC] (v1) arc (0+\rotn:-120+\rotn:\innerRad);
}{\foreach\c[remember=\c as \lc] in {0,...,#3}{\ifthenelse{\c=0}{}{\draw[intC](c\lc)--(c\c);}};};
\ifthenelse{#1=1}{}{\foreach\a in {1,...,#1}{\ifthenelse{\a=#1}{}{\leg{(a\a)}{240+\rotn-\a*\angleA};}}};
\ifthenelse{#2=1}{}{\foreach\b in {1,...,#2}{\ifthenelse{\b=#2}{}{\leg{(b\b)}{120+\rotn-\b*\angleB};}}};
\ifthenelse{#3=1}{}{\foreach\c in {1,...,#3}{\ifthenelse{\c=#3}{}{\leg{(c\c)}{0+\rotn-\c*\angleC};}}};
\optLeg{(v1)}{0+\rotn};\optLeg{(v2)}{120+\rotn};\optLeg{(v3)}{240+\rotn};\optLeg{(v0)}{180+\rotn};
}{\draw[line width=0.125]($\figScale*(-0.85,-0.85)$) rectangle ($\figScale*(1,0.85)$);
\node at (0,0) {$\threeLoopWheelName{#1}{#2}{#3}{#4}{#5}{#6}$};}
\end{tikzpicture}%
\tikzset{intA/.style={black,line width=\lineThickness,line cap=round}}\tikzset{intB/.style={black,line width=\lineThickness,line cap=round}}\tikzset{intC/.style={black,line width=\lineThickness,line cap=round}}}$}

\newcommand{\ladderThreeFigure}[6]{\fwbox{55pt}{\begin{tikzpicture}[scale=1,baseline=-3.05,rotate=0]\useasboundingbox ($\figScale*(-0.95,-1.125)$) rectangle ($\figScale*(0.95,0.375)$);
\draw[draw=none] ($\figScale*(-0.95,-1.125)$) rectangle ($\figScale*(0.95,0.375)$);
\ifthenelse{\showTikzQ=1}{
\ladderThreeGraph{#1}{#2}{#3}{#4}{#5}{#6}}{
\draw[line width=0.125] ($\figScale*(-0.95,-1.125)$) rectangle ($\figScale*(0.95,0.375)$);
\node at ($(0,-0.375*\figScale)$) {$\threeLoopLadderName{#1}{#2}{#3}{#4}{#5}{#6}$};}\end{tikzpicture}}}

\newcommand{\ladderThreeGraph}[6]{
\def\ephScale{\figScale}\def\edgeLen{0.75*\ephScale}\def\angleAa{180/#1}\def\angleAb{180/#2}\def\angleBa{180/#3}\def\angleBb{180/#4}\def\angleCc{180/#6}
\ifthenelse{#6=0}{
\coordinate (v1) at (0,0);
\coordinate (v2) at (v1);
\coordinate (v3) at ($(v1)+(-90:\edgeLen)$);
\coordinate (v4) at (v3);
\ifthenelse{#1=1}{\draw[intA] (v1) arc (90:270:\edgeLen/1.15 and \edgeLen/2);
}{\draw[draw=none] (v1) arc (90:270:\edgeLen/1.15 and \edgeLen/2)\foreach\a in {0,...,#1}{coordinate[pos=1/#1*\a] (aa\a)};
\foreach\a[remember=\a as \la] in {0,...,#1}{\ifthenelse{\a=0}{}{\draw[intA](aa\la)--(aa\a);}};
\foreach\a in {1,...,#1}{\ifthenelse{\a=#1}{}{\leg{(aa\a)}{\angleAa*\a+90}}};
}

\ifthenelse{#2=1}{\draw[intA] (v1) arc (90:270:\edgeLen/12 and \edgeLen/2);
}{
\draw[draw=none] (v1) arc (90:270:\edgeLen/4.5 and \edgeLen/2)\foreach\a in {0,...,#2}{coordinate[pos=(((1/#2*\a)*(3+(1/#2*\a)*(2*(1/#2*\a)-3)))*1/2)] (ab\a)};
\foreach\a[remember=\a as \la] in {0,...,#2}{\ifthenelse{\a=0}{}{\draw[intA](ab\la)--(ab\a);}};
\foreach\a in {1,...,#2}{\ifthenelse{\a=#2}{}{\leg{(ab\a)}{180}}};
}

\ifthenelse{#3=1}{\draw[intC] (v3) arc (-90:90:\edgeLen/4.5 and \edgeLen/2);
}{%\draw[draw=none] (v3) arc (-90:90:\edgeLen/4.5 and \edgeLen/2)\foreach\b in {0,...,#3}{coordinate[pos=1/#3*\b] (ba\b)};
\draw[draw=none] (v3) arc (-90:90:\edgeLen/4.5 and \edgeLen/2)\foreach\b in {0,...,#3}{coordinate[pos=(((1/#3*\b)*(3+(1/#3*\b)*(2*(1/#3*\b)-3)))*1/2)] (ba\b)};

\foreach\b[remember=\b as \lb] in {0,...,#3}{\ifthenelse{\b=0}{}{\draw[intC](ba\lb)--(ba\b);}};
\foreach\b in {1,...,#3}{\ifthenelse{\b=#3}{}{\leg{(ba\b)}{0}}};
}

\ifthenelse{#4=1}{\draw[intC] (v3) arc (-90:90:\edgeLen/1.15 and \edgeLen/2);
}{\draw[draw=none] (v3) arc (-90:90:\edgeLen/1.15 and \edgeLen/2)\foreach\b in {0,...,#4}{coordinate[pos=1/#4*\b] (bb\b)};
\foreach\b[remember=\b as \lb] in {0,...,#4}{\ifthenelse{\b=0}{}{\draw[intC](bb\lb)--(bb\b);}};
\foreach\b in {1,...,#4}{\ifthenelse{\b=#4}{}{\leg{(bb\b)}{\angleBb*\b-90}}};
}
\optLeg{(v1)}{90};\optLeg{(v3)}{-90};

}{

\ifthenelse{#5=0}{%
\coordinate (v1) at (0,0);
\coordinate (v2) at (v1);
\coordinate (v3) at ($(v2)+(-50:\edgeLen)$);
\coordinate (v4) at (230:\edgeLen);

\ifthenelse{#1=1}{\draw[intA,rotate=-40] (v1) arc (90:270:\edgeLen/1.65 and \edgeLen/2);
}{\draw[draw=none,rotate=-40] (v1) arc (90:270:\edgeLen/1.65 and \edgeLen/2)\foreach\a in {0,...,#1}{coordinate[pos=1/#1*\a] (aa\a)};
\foreach\a[remember=\a as \la] in {0,...,#1}{\ifthenelse{\a=0}{}{\draw[intA](aa\la)--(aa\a);}};
\foreach\a in {1,...,#1}{\ifthenelse{\a=#1}{}{\leg{(aa\a)}{\angleAa*\a+90-40}}};
}

\ifthenelse{#4=1}{\draw[intC,rotate=40] (v3) arc (-90:90:\edgeLen/1.65 and \edgeLen/2);
}{\draw[draw=none,rotate=40] (v3) arc (-90:90:\edgeLen/1.65 and \edgeLen/2)\foreach\b in {0,...,#4}{coordinate[pos=1/#4*\b] (bb\b)};
\foreach\b[remember=\b as \lb] in {0,...,#4}{\ifthenelse{\b=0}{}{\draw[intC](bb\lb)--(bb\b);}};
\foreach\b in {1,...,#4}{\ifthenelse{\b=#4}{}{\leg{(bb\b)}{\angleBb*\b-90+40}}};
}

\ifthenelse{#6=1}{\draw[intB] (v3)--(v4);%\draw[intB,rotate=-30] (v3) arc (0:-120:0.746*\edgeLen/1 and 0.746*\edgeLen/1);
}{\draw[draw=none,rotate=-30] (v3) arc (0:-120:0.746*\edgeLen/1 and 0.746*\edgeLen/1)\foreach\c in {0,...,#6}{coordinate[pos=1/#6*\c] (cc\c)};
\foreach\c[remember=\c as \lc] in {0,...,#6}{\ifthenelse{\c=0}{}{\draw[intB](cc\lc)--(cc\c);}};
\foreach\c in {1,...,#6}{\ifthenelse{\c=#6}{}{\leg{(cc\c)}{-\angleCc*\c+0}}};
}

\draw[intA] (v4)--(v1)\foreach\a in {0,...,#2}{coordinate[pos=1/#2*\a] (ab\a)};
\foreach\a in {1,...,#2}{\ifthenelse{\a=#2}{}{\leg{(ab\a)}{-40}}};
\draw[intC] (v2)--(v3)\foreach\b in {0,...,#3}{coordinate[pos=1/#3*\b] (ba\b)};
\foreach\b in {1,...,#3}{\ifthenelse{\b=#3}{}{\leg{(ba\b)}{40+180}}};

\optLeg{(v1)}{90};
\optLeg{(v4)}{180+35};
\optLeg{(v3)}{-35};

}{

\coordinate (v1) at (-0.75*\edgeLen/2,0);
\coordinate (v2) at (0:0.75*\edgeLen/2);
\coordinate (v3) at ($(v2)+(-90:\edgeLen)$);
\coordinate (v4) at ($(v1)+(-90:\edgeLen)$);

\draw[intB] (v1)--(v2)\foreach\c in {0,...,#5}{coordinate[pos=1/#5*\c] (ca\c)};
\foreach\c in {1,...,#5}{\ifthenelse{\c=#5}{}{\leg{(ca\c)}{90}}};
\draw[intB] (v3)--(v4)\foreach\c in {0,...,#6}{coordinate[pos=1/#6*\c] (cc\c)};
\foreach\c in {1,...,#6}{\ifthenelse{\c=#6}{}{\leg{(cc\c)}{-90}}};

\draw[intA] (v4)--(v1)\foreach\a in {0,...,#2}{coordinate[pos=1/#2*\a] (ab\a)};
\foreach\a in {1,...,#2}{\ifthenelse{\a=#2}{}{\leg{(ab\a)}{0}}};

\ifthenelse{#1=1}{\draw[intA] (v1) arc (90:270:\edgeLen/1.75 and \edgeLen/2);}
{
\draw[draw=none] (v1) arc (90:270:\edgeLen/1.75 and \edgeLen/2)\foreach\a in {0,...,#1}{coordinate[pos=1/#1*\a] (aa\a)};
\foreach\a[remember=\a as \la] in {0,...,#1}{\ifthenelse{\a=0}{}{\draw[intA](aa\la)--(aa\a);}};
\foreach\a in {1,...,#1}{\ifthenelse{\a=#1}{}{\leg{(aa\a)}{\angleAa*\a+90}}};
}

\draw[intC] (v2)--(v3)\foreach\b in {0,...,#3}{coordinate[pos=1/#3*\b] (ba\b)};
\foreach\b in {1,...,#3}{\ifthenelse{\b=#3}{}{\leg{(ba\b)}{180}}};
\ifthenelse{#4=1}{\draw[intC] (v3) arc (-90:90:\edgeLen/1.75 and \edgeLen/2);}
{
\draw[draw=none] (v3) arc (-90:90:\edgeLen/1.75 and \edgeLen/2)\foreach\b in {0,...,#4}{coordinate[pos=1/#4*\b] (bb\b)};
\foreach\b[remember=\b as \lb] in {0,...,#4}{\ifthenelse{\b=0}{}{\draw[intC](bb\lb)--(bb\b);}};
\foreach\b in {1,...,#4}{\ifthenelse{\b=#4}{}{\leg{(bb\b)}{\angleBb*\b-90}}};
}
\optLeg{(v1)}{120};
\optLeg{(v2)}{60};
\optLeg{(v3)}{-60};
\optLeg{(v4)}{-120};
}
}
}

\let\olditemize\itemize\renewcommand{\itemize}{\vspace{-2pt}\olditemize\setlength{\itemsep}{1pt}\setlength{\parskip}{0pt}\setlength{\parsep}{-0pt}}
\let\oldenumerate\enumerate\renewcommand{\enumerate}{\vspace{-4pt}\oldenumerate\setlength{\itemsep}{1pt}\setlength{\parskip}{0pt}\setlength{\parsep}{0pt}}
\makeatletter\renewcommand\section{\addtocontents{toc}{\protect\addvspace{-2.25\p@}}\@startsection {section}{1}{\z@}{-0.0ex \@plus .2ex \@minus 0.2ex}{1ex \@plus.1ex\@minus .5ex}{\normalfont\large\bfseries}}
\renewcommand\subsection{\addtocontents{toc}{\protect\addvspace{-2.5\p@}}\@startsection {subsection}{1}{\z@}{0.5ex \@plus .2ex \@minus 0.2ex}{0.75ex \@plus.1ex\@minus .5ex}{\normalfont\bfseries}}

\def\twoLoopCorners{1}
\def\colorTableQ{1}
\def\showTikzQ{0}

\definecolor{mhvBlue}{rgb}{0.3,0.2,0.75}
\definecolor{fRed}{rgb}{0.48,0.02824,0.18824}
\definecolor{cut2}{rgb}{0.18824,0.18824,0.48}
\definecolor{cut1}{rgb}{0.48,0.02824,0.18824}
\definecolor{hblue}{rgb}{0,0,0.575}
\definecolor{hred}{rgb}{0.475,0.0,0.15}
\definecolor{dred}{rgb}{0.575,0.4,0.45}

\newcommand{\ca}{{\color{emphA}a}}
\newcommand{\cb}{{\color{emphB}b}}
\newcommand{\cc}{{\color{emphC}c}}

\newcommand{\cca}[1]{{\color{emphA}#1}}
\newcommand{\ccb}[1]{{\color{emphB}#1}}
\newcommand{\ccc}[1]{{\color{emphC}#1}}

\newcommand{\tred}[1]{{\color{hred}#1}}

\newcommand{\lam}[1]{\lambda_{#1}}
\newcommand{\lamt}[1]{\widetilde{\lambda}_{#1}}

\newcommand{\basis}{\mathfrak{B}}

\DeclareMathOperator*{\mathspan}{\mathrm{span}}
\newcommand{\eq}[1]{\vspace{-0.5pt}\begin{equation}#1\vspace{-0.5pt}\end{equation}}

\newcommand{\fwbox}[2]{\text{\makebox[#1][c]{$\hspace{-150pt}\displaystyle#2\hspace{-150pt}$}}}
\newcommand{\fwboxL}[2]{\text{\makebox[#1][l]{$#2$}}}
\newcommand{\fwboxR}[2]{\text{\makebox[#1][r]{$#2$}}}
\newcommand{\equivR}{\fwbox{14.5pt}{\hspace{-0pt}\fwboxR{0pt}{\raisebox{0.47pt}{\hspace{1.25pt}:\hspace{-4pt}}}=\fwboxL{0pt}{}}}
\newcommand{\equivL}{\fwbox{14.5pt}{\fwboxR{0pt}{}=\fwboxL{0pt}{\raisebox{0.47pt}{\hspace{-4pt}:\hspace{1.25pt}}}}}
\newcommand{\fig}[3]{\raisebox{#1}{\includegraphics[scale=#2]{#3}}}
\newcommand{\Bigger}[1]{\raisebox{-2.25pt}{\scalebox{1.75}{$#1$}}}
\newcommand{\bigger}[1]{\raisebox{-0.95pt}{\scalebox{1.25}{$#1$}}}
\newcommand{\mi}{\raisebox{0.75pt}{\scalebox{0.75}{$\hspace{-0.5pt}\,-\,\hspace{-0.5pt}$}}}
\newcommand{\pl}{\raisebox{0.75pt}{\scalebox{0.75}{$\hspace{-0.5pt}\,+\,\hspace{-0.5pt}$}}}
\renewcommand{\phi}{\varphi}

\renewcommand{\hat}{\widehat}

\newcommand{\ab}[1]{\langle #1\rangle}

\newcommand{\x}[2]{{\color{black}(}\hspace{-0.85pt}{\color{black}#1}\hspace{-0.25pt}{\color{black}|}\hspace{-0.25pt}{\color{black}#2}\hspace{-0.85pt}{\color{black})}}

\newcommand{\Span}[1]{\hspace{-2pt}\mathspan_{\fwboxR{20pt}{\substack{~\\[-6pt]~\hspace{2pt}\text{{\scriptsize $#1$}\hspace{-4pt}}}}}}

\definecolor{hblue}{rgb}{0,0,0.575}
\definecolor{hred}{rgb}{0.575,0.0,0.225}
\definecolor{totalCount}{rgb}{0,0,0.575}
\definecolor{topCount}{rgb}{0.575,0.0,0.225}
\definecolor{dim}{rgb}{0.55,0.55,0.55}
\definecolor{deemph}{rgb}{0.25,0.25,0.25}

\newcommand{\squeezeB}{\hspace{-0.5pt}}
\newcommand{\squeezeC}{\hspace{-0.5pt}}
\newcommand{\squeezeD}{\hspace{-0.5pt}}
\newcommand{\ellK}[1]{[\squeezeD\ell\squeezeD]{}^{#1}}
\newcommand{\ellKd}[2]{[\squeezeD\ell\squeezeD]{}^{#1}_{#2}}
\newcommand{\ellKhat}[1]{\widehat{\hspace{-0.75pt}[\squeezeD\ell\squeezeD]\hspace{2.5pt}}{\hspace{-2.0pt}}^{#1}}

\newcommand{\numBasic}[1]{{\color{black}[}\squeezeD #1\squeezeD{\color{black}]}}
\newcommand{\numDiff}[2]{{\color{black}[}\squeezeD#1\squeezeD\mi\squeezeD#2\squeezeD{\color{black}]}}

\newcommand{\defn}[3]{\hypertarget{#1}{ ~\\\textbullet\,\textbf{Definition:}\,{\color{hred}\textbf{\emph{#2}}}---}#3}
\newcommand{\subdefn}[3]{\hypertarget{#1}{{\color{hred}\textbf{\emph{#2}}}---}#3}

\newcommand{\threeLoopTensorNameW}[6]{
\text{{\scriptsize$
{\color{emphA}\ifthenelse{#1=0}{}{\ifthenelse{#1=1}{\numBasic{1}}{\numBasic{1}^{\squeezeB #1}\squeezeB}\squeezeC}}
{\color{emphB}\ifthenelse{#2=0}{}{\ifthenelse{#2=1}{\numBasic{2}}{\numBasic{2}^{\squeezeB #2}\squeezeB}\squeezeC}}
{\color{emphC}\ifthenelse{#3=0}{}{\ifthenelse{#3=1}{\numBasic{3}}{\numBasic{3}^{\squeezeB #3}\squeezeB}\squeezeC}}
{\color{emphC}\ifthenelse{#4=0}{}{\ifthenelse{#4=1}{\numDiff{1}{2}}{\numDiff{1}{2}^{\squeezeB #4}\squeezeB}\squeezeC}}
{\color{emphA}\ifthenelse{#6=0}{}{\ifthenelse{#6=1}{\numDiff{2}{3}}{\numDiff{2}{3}^{\squeezeB #6}\squeezeB}\squeezeC}}
{\color{emphB}\ifthenelse{#5=0}{}{\ifthenelse{#5=1}{\numDiff{1}{3}}{\numDiff{1}{3}^{\squeezeB #5}\squeezeB}\squeezeC}}
$}}}

\newcommand{\threeLoopTensorNameWb}[6]{
\text{{\normalsize$
{\color{black}\ifthenelse{#1=0}{}{\ifthenelse{#1=1}{\numBasic{\cca1}}{\numBasic{\cca1}^{\squeezeB \cca{#1}}\squeezeB}\squeezeC}}
{\color{black}\ifthenelse{#2=0}{}{\ifthenelse{#2=1}{\numBasic{\ccb2}}{\numBasic{\ccb2}^{\squeezeB \ccb{#2}}\squeezeB}\squeezeC}}
{\color{black}\ifthenelse{#3=0}{}{\ifthenelse{#3=1}{\numBasic{\ccc3}}{\numBasic{\ccc3}^{\squeezeB \ccc{#3}}\squeezeB}\squeezeC}}
{\color{black}\ifthenelse{#4=0}{}{\ifthenelse{#4=1}{\ccc{\numDiff{1}{2}}}{\ccc{\numDiff{1}{2}}^{\squeezeB \ccc{#4}}\squeezeB}\squeezeC}}
{\color{black}\ifthenelse{#6=0}{}{\ifthenelse{#6=1}{\cca{\numDiff{2}{3}}}{\cca{\numDiff{2}{3}}^{\squeezeB \cca{#6}}\squeezeB}\squeezeC}}
{\color{black}\ifthenelse{#5=0}{}{\ifthenelse{#5=1}{\ccb{\numDiff{1}{3}}}{\ccb{\numDiff{1}{3}}^{\squeezeB \ccb{#5}}\squeezeB}\squeezeC}}
$}}}

\newcommand{\threeLoopTensorNameL}[6]{
\text{{\scriptsize$
{\color{emphA}\ifthenelse{#1=0}{}{\ifthenelse{#1=1}{\numBasic{1}}{\numBasic{1}^{\squeezeB #1}\squeezeB}\squeezeC}}
{\color{emphA}\ifthenelse{#4=0}{}{\ifthenelse{#4=1}{\numDiff{1}{2}}{\numDiff{1}{2}^{\squeezeB #4}\squeezeB}\squeezeC}}
{\color{emphB}\ifthenelse{#2=0}{}{\ifthenelse{#2=1}{\numBasic{2}}{\numBasic{2}^{\squeezeB #2}\squeezeB}\squeezeC}}
{\color{emphC}\ifthenelse{#6=0}{}{\ifthenelse{#6=1}{\numDiff{2}{3}}{\numDiff{2}{3}^{\squeezeB #6}\squeezeB}\squeezeC}}
{\color{emphC}\ifthenelse{#3=0}{}{\ifthenelse{#3=1}{\numBasic{3}}{\numBasic{3}^{\squeezeB #3}\squeezeB}\squeezeC}}
{\color{black}\ifthenelse{#5=0}{}{\ifthenelse{#5=1}{\numDiff{1}{3}}{\numDiff{1}{3}^{\squeezeB #5}\squeezeB}\squeezeC}}
$}}}

\newcommand{\threeLoopTensorNameLb}[6]{
\text{{\normalsize$
{\color{black}\ifthenelse{#1=0}{}{\ifthenelse{#1=1}{\numBasic{\cca{1}}}{\numBasic{1}^{\squeezeB \cca{#1}}\squeezeB}\squeezeC}}
{\color{black}\ifthenelse{#4=0}{}{\ifthenelse{#4=1}{\cca{\numDiff{1}{2}}}{\numDiff{1}{2}^{\squeezeB \cca{#4}}\squeezeB}\squeezeC}}
{\color{black}\ifthenelse{#2=0}{}{\ifthenelse{#2=1}{\ccb{\numBasic{2}}}{\numBasic{\ccb{2}}^{\squeezeB \ccb{#2}}\squeezeB}\squeezeC}}
{\color{black}\ifthenelse{#6=0}{}{\ifthenelse{#6=1}{\ccc{\numDiff{2}{3}}}{\ccc{\numDiff{2}{3}}^{\squeezeB \ccc{#6}}\squeezeB}\squeezeC}}
{\color{black}\ifthenelse{#3=0}{}{\ifthenelse{#3=1}{\numBasic{\ccc3}}{\numBasic{\ccc{3}}^{\squeezeB \ccc{#3}}\squeezeB}\squeezeC}}
{\color{black}\ifthenelse{#5=0}{}{\ifthenelse{#5=1}{\numDiff{1}{3}}{\numDiff{1}{3}^{\squeezeB \ccc{#5}}\squeezeB}\squeezeC}}
$}}}

\newcommand{\tensorDecomp}[3]{ {
\color{totalCount}#1} = {\color{topCount}#2} {\color{dim}+#3}}

\def\colorLabelsQ{1}

\newcommand{\twoLoopLadderName}[3]{\ifthenelse{\colorLabelsQ=1}{\text{\normalsize$\Gamma$}\fwboxL{22pt}{\hspace{-3.5pt}\raisebox{-2pt}{\scalebox{0.8}{\text{{\small$\raisebox{1.5pt}{\text{{\scriptsize$[$}}}{\color{emphA}#1},\!{\color{emphB}#2},\!{\color{emphC}#3}\raisebox{1.5pt}{\text{{\scriptsize$]$}}}$}}}}}}{
\text{\normalsize$\Gamma$}\fwboxL{22pt}{\hspace{-3.5pt}\raisebox{-2pt}{\scalebox{0.8}{\text{{\small$\raisebox{1.5pt}{\text{{\scriptsize$[$}}}{\color{black}#1},\!{\color{black}#2},\!{\color{black}#3}\raisebox{1.5pt}{\text{{\scriptsize$]$}}}$}}}}}
}}

\newcommand{\twoLoopLadderNameText}[3]{\ifthenelse{\colorLabelsQ=1}{\text{{\color{black}\normalsize$\Gamma$}}\fwboxL{17.5pt}{\hspace{-3.5pt}\raisebox{-0.5pt}{\scalebox{0.8}{\text{{\small$\raisebox{0.5pt}{\text{{\scriptsize$[$}}}{\color{emphA}#1},\!{\color{emphB}#2},\!{\color{emphC}#3}\raisebox{0.5pt}{\text{{\scriptsize$]$}}}$}}}}}}{
\text{{\color{black}\normalsize$\Gamma$}}\fwboxL{17.5pt}{\hspace{-3.5pt}\raisebox{-0.5pt}{\scalebox{0.8}{\text{{\small$\raisebox{0.5pt}{\text{{\scriptsize$[$}}}{\color{black}#1},\!{\color{black}#2},\!{\color{black}#3}\raisebox{0.5pt}{\text{{\scriptsize$]$}}}$}}}}}
}}

\newcommand{\threeLoopLadderName}[6]{
\ifthenelse{\colorLabelsQ=1}{\fwbox{40pt}{\raisebox{-2.5pt}{\scalebox{1.5}{$\mathcal{L}$}}{\scriptsize\begin{array}{@{}c@{}}\fwbox{28pt}{(\hspace{-1pt}{\color{emphA}#1},\!{\color{emphA}#2}\hspace{-1pt})(\hspace{-1pt}{\color{emphC}#3},\!{\color{emphC}#4}\hspace{-1pt})}\\[-1.5pt]\fwbox{28pt}{(\hspace{-1pt}{\color{emphB}#5},\!{\color{emphB}#6}\hspace{-1pt})}\end{array}}}}{
\fwbox{40pt}{\raisebox{-2.5pt}{\scalebox{1.5}{$\mathcal{L}$}}{\scriptsize\begin{array}{@{}c@{}}\fwbox{28pt}{(\hspace{-1pt}#1,\!#2\hspace{-1pt})(\hspace{-1pt}#3,\!#4\hspace{-1pt})}\\[-1.5pt]\fwbox{28pt}{(\hspace{-1pt}#5,\!#6\hspace{-1pt})}\end{array}}}}}

\newcommand{\threeLoopWheelName}[6]{\ifthenelse{\colorLabelsQ=1}{\fwbox{40pt}{\raisebox{-2.5pt}{\scalebox{1.5}{$\mathcal{W}$}}{\scriptsize\begin{array}{@{}c@{}}\fwbox{21pt}{(\hspace{-1pt}{\color{emphA}#1},\!{\color{emphB}#2},\!{\color{emphC}#3}\hspace{-1pt})}\\[-1.5pt]\fwbox{21pt}{(\hspace{-1pt}{\color{emphA}#4},\!{\color{emphB}#5},\!{\color{emphC}#6}\hspace{-1pt})}\end{array}}}}{
\fwbox{40pt}{\raisebox{-2.5pt}{\scalebox{1.5}{$\mathcal{W}$}}{\scriptsize\begin{array}{@{}c@{}}\fwbox{21pt}{(\hspace{-1pt}#1,\!#2,\!#3\hspace{-1pt})}\\[-1.5pt]\fwbox{21pt}{(\hspace{-1pt}#4,\!#5,\!#6\hspace{-1pt})}\end{array}}}}
}

\newcommand{\threeLoopLadderNamePrime}[6]{\fwbox{50pt}{\raisebox{-2.5pt}{\scalebox{1.5}{$\mathcal{L}$}}{\scriptsize\begin{array}{@{}c@{}}\fwbox{44pt}{(\hspace{-1pt}#1,\!#2\hspace{-1pt})(\hspace{-1pt}#3,\!#4\hspace{-1pt})}\\[-1.5pt]\fwbox{34pt}{(\hspace{-1pt}#5,\!#6\hspace{-1pt})}\end{array}}}}
\newcommand{\threeLoopWheelNamePrime}[6]{\fwbox{50pt}{\raisebox{-2.5pt}{\scalebox{1.5}{$\mathcal{W}$}}{\scriptsize\begin{array}{@{}c@{}}(\fwbox{28pt}{\hspace{-1pt}#1,\!#2,\!#3\hspace{-1pt}})\\[-1.5pt]\fwbox{21pt}{(\fwbox{28pt}{\hspace{-1pt}#4,\!#5,\!#6\hspace{-1pt}})}\end{array}}}}

\newcommand{\threeLoopWheelDataTablePrecompiled}{\fwbox{0pt}{\fig{-291pt}{1}{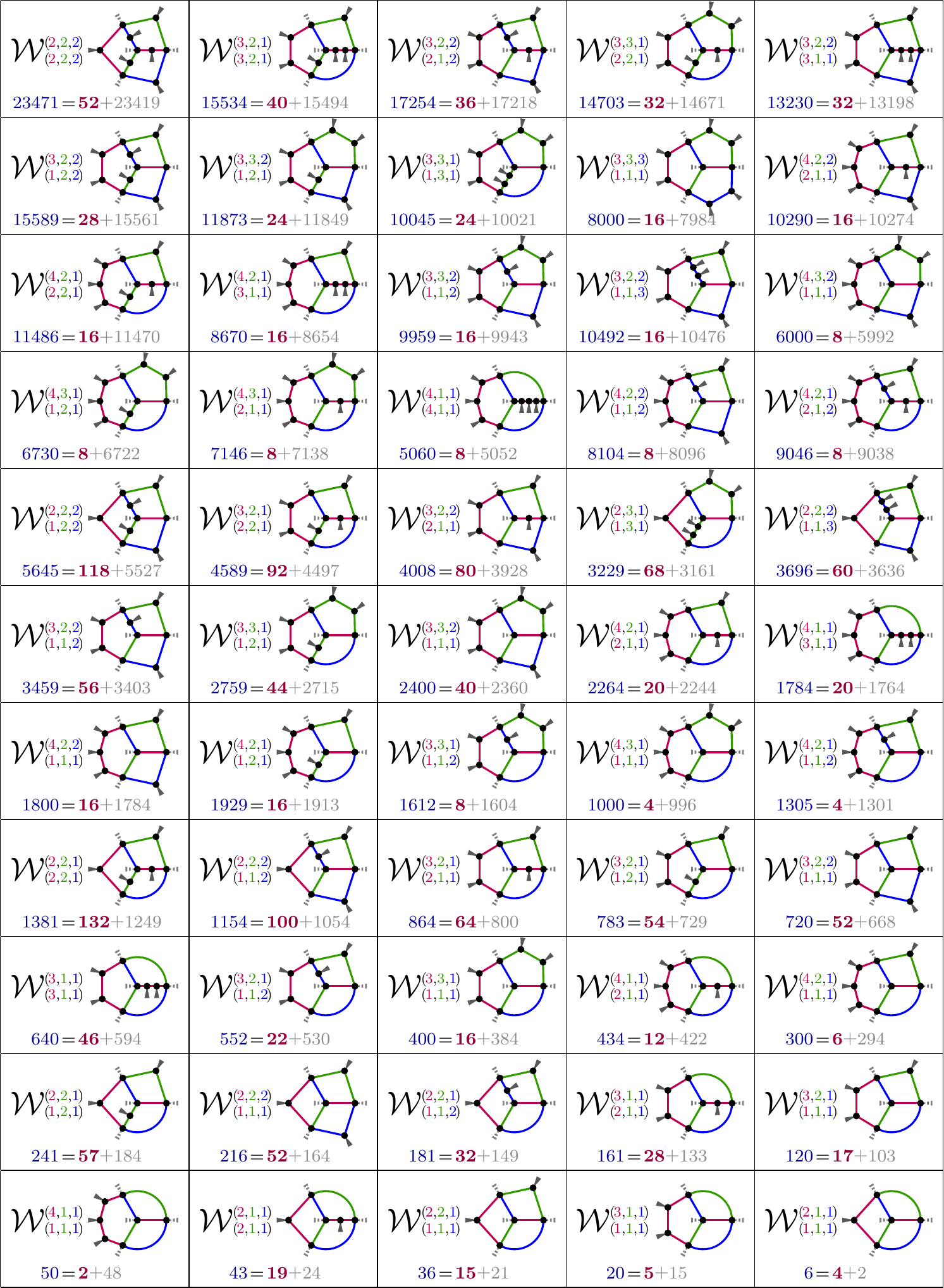}}}
\newcommand{\threeLoopLadderDataTableOnePrecompiled}{\fwbox{0pt}{\fig{-291pt}{1}{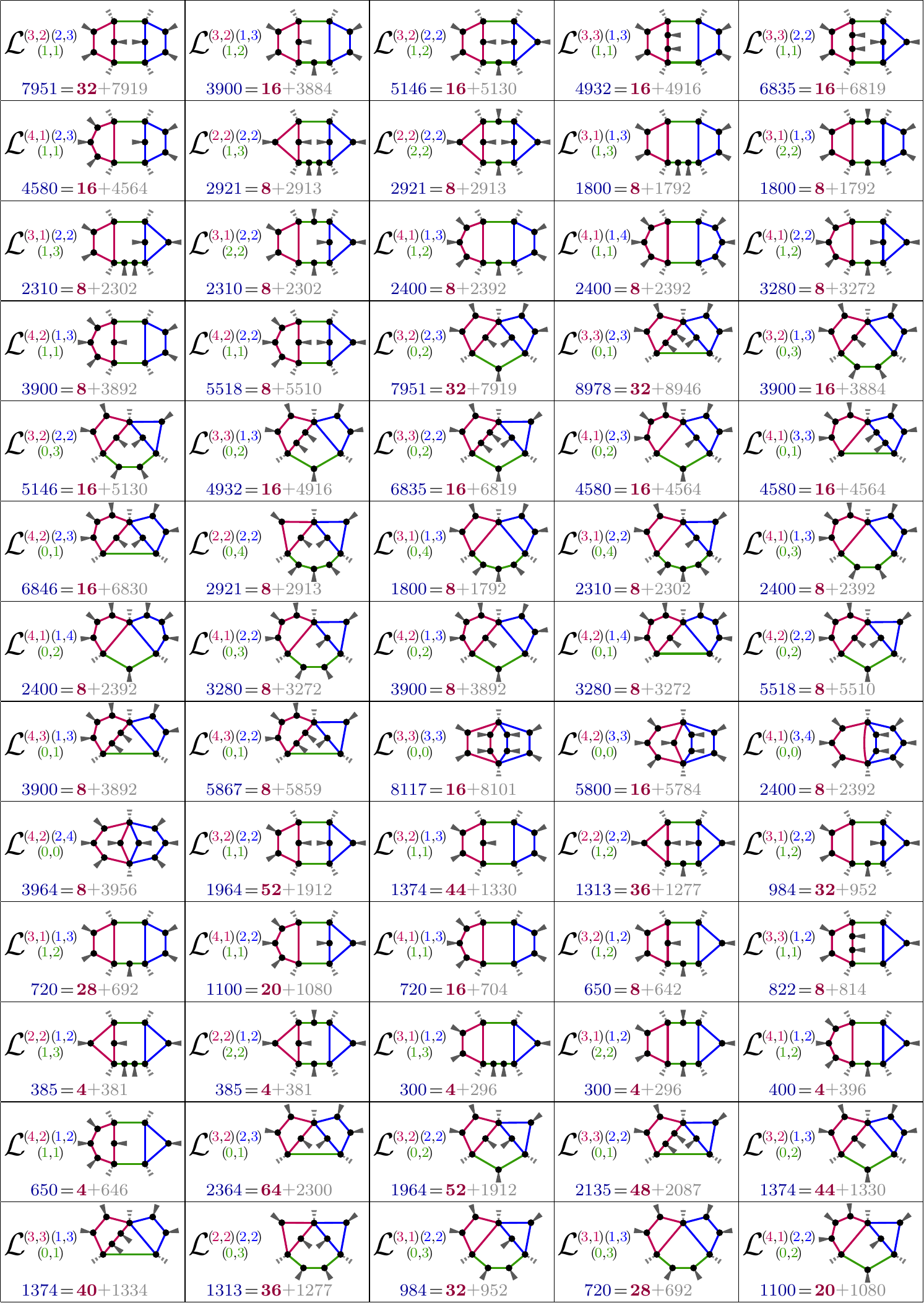}}}
\newcommand{\threeLoopLadderDataTableTwoPrecompiled}{\fwbox{0pt}{\fig{-291pt}{1}{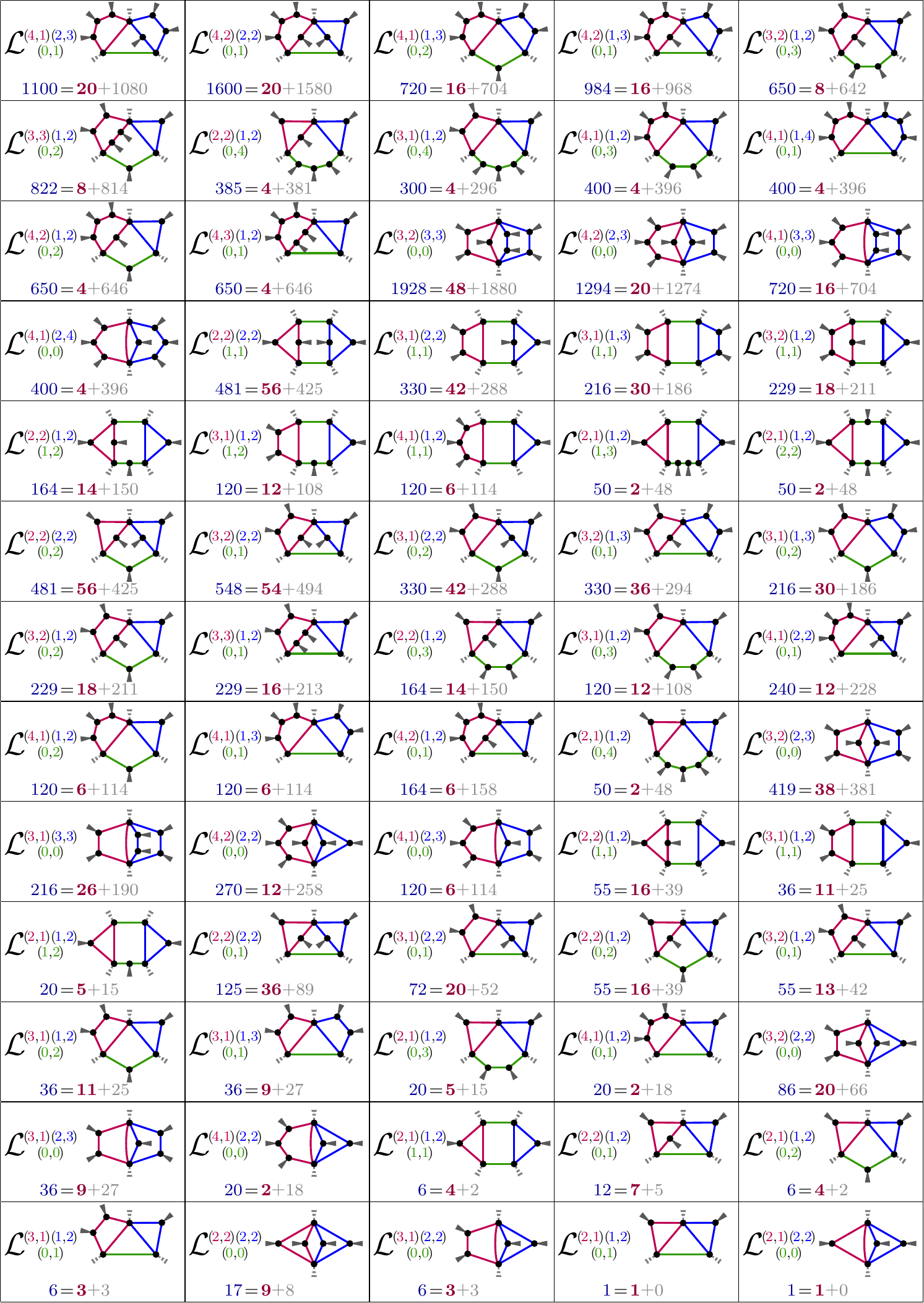}}}

\graphicspath{{figures/}}
\def\showTikzQ{1}

\title{~\\[-20pt]{\huge \mbox{Building Bases of Loop Integrands}}\\[-24pt]}
\author[a,b]{\vspace{-24pt}Jacob~L.~Bourjaily,}\emailAdd{bourjaily@psu.edu}
\author[c]{Enrico~Herrmann,}\emailAdd{eh10@stanford.edu}
\author[a,d]{Cameron~Langer,}\emailAdd{cklanger@ucdavis.edu}
\author[d]{Jaroslav~Trnka}\emailAdd{trnka@ucdavis.edu}
\affiliation[a]{Institute for Gravitation and the Cosmos, Department of Physics,\\Pennsylvania State University, University Park, PA 16892, USA}
\affiliation[b]{Niels Bohr International Academy and Discovery Center, Niels Bohr Institute,\\University of Copenhagen, Blegdamsvej 17, DK-2100, Copenhagen \O, Denmark}
\affiliation[c]{SLAC National Accelerator Laboratory, Stanford University, Stanford, CA 94039, USA}
\affiliation[d]{Center for Quantum Mathematics and Physics (QMAP),\\Department of Physics, University of California, Davis, CA 95616, USA}
\abstract{
We describe a systematic approach to the construction of loop-integrand bases at arbitrary loop-order, sufficient for the representation of general quantum field theories. We provide a graph-theoretic definition of `power-counting' for multi-loop integrands beyond the planar limit, and show how this can be used to organize bases according to ultraviolet behavior. This allows amplitude integrands to be constructed iteratively. We illustrate these ideas with concrete applications. In particular, we describe complete integrand bases at two loops sufficient to represent arbitrary-multiplicity amplitudes in four (or fewer) dimensions in any massless quantum field theory with the ultraviolet behavior of the Standard Model or better. We also comment on possible extensions of our framework to arbitrary (including regulated) numbers of dimensions, and to theories with arbitrary mass spectra and charges. At three loops, we describe a basis sufficient to capture all `leading-\mbox{(transcendental-)}weight' contributions of \emph{any} four-dimensional quantum theory; for maximally supersymmetric Yang-Mills theory, this basis should be sufficient to represent \emph{all} scattering amplitude integrands in the theory---for generic helicities and arbitrary multiplicity. 
}
\preprint{}

\begin{document}
\maketitle%\thispagestyle{empty}
\pagenumbering{roman}\clearpage
%================================================================================================================
%    1. Introduction 
%         
%================================================================================================================
\setcounter{page}{1}\vspace{-4pt}\setcounter{section}{-1}%
\pagenumbering{arabic}
\vspace{-6pt}\section{Introduction and Overview}\label{sec:introduction}\vspace{-0pt}
%================================================================================================================ 

Recent decades have been witness to breathtaking progress in our understanding of and our ability to represent and evaluate scattering amplitudes in perturbative quantum field theory. Much of this progress can be traced to the development of unitarity-based methods at the end of the last century \cite{Bern:1994zx,Bern:1994cg,Britto:2004nc}. Indeed, the heroic work of \emph{e.g.}~\cite{Bern:1992em,Bern:1996je,Bern:1996fj,Bern:1996ja,Bern:1998xc,Bidder:2005ri} at \emph{one-loop} level would lead to the discovery of \emph{tree-level} recursion \cite{Britto:2004ap,Britto:2005fq} among many other things (see \emph{e.g.}~\cite{Drummond:2006rz,Alday:2007hr,Drummond:2008vq,Beisert:2010jr,ArkaniHamed:2012nw,Arkani-Hamed:2013jha}). 

Although not originally described in this language, one way to characterize the principal insight of generalized unitarity is that loop \emph{integrands}---roughly `the sum of Feynman diagrams' \emph{prior to loop integration}---are meaningful quantities of interest when viewed as differential forms on the space of loop momenta. In any sufficiently well-behaved local and unitary quantum field theory, these integrands should be determined by their `cuts' (residues) in terms of lower-loop, and ultimately tree-level information. Provided a sufficiently large basis of loop integrands $\basis$ (viewed as a vector-space of rational functions), the coefficients $a_i$ of $\mathfrak{b}^i\!\in\!\basis$ of any loop amplitude $\mathcal{A}$ representable within this basis, 
\eq{\mathcal{A}=\sum_{\mathfrak{b}^i\in\basis}a_{i}\,\mathfrak{b}^i\,,\label{schematic_unitarity_construction}}
can be determined by linear algebra---by matching all cuts. How and under which conditions this works in detail is beyond the scope of this present work, but we refer the reader to \emph{e.g.}~\cite{Bern:1995db,Bern:2011qt,Ellis:2011cr} for more detailed discussions. 

We have been fairly schematic in (\ref{schematic_unitarity_construction}) for an important reason. Fixing any \emph{particular} scattering amplitude in any particular theory, the amplitude integrand itself (however it is found) could be viewed as a single basis element whose coefficient would be 1. This is tautological, and not especially useful or interesting. What is extremely interesting is that for a very wide class of quantum field theories (including all renormalizable ones), there exists a \emph{\textbf{finite}-dimensional basis} $\basis$ at any loop order in which \emph{all scattering amplitudes of that theory can be represented}---involving \emph{arbitrary} numbers and species of external states.
Examples of such bases suitable for representing all amplitudes in the Standard Model at one and two loops can be found in \cite{Ossola:2006us,Mastrolia:2012wf,Mastrolia:2013kca,Kleiss:2012yv} and \cite{Feng:2012bm}, respectively. 

Importantly, it is often possible to build a basis (and therefore represent amplitudes) iteratively. What we mean by this is that we can stratify an integrand basis $\basis$ into subspaces according to some notion of `power-counting' $p$---schematically, one partitions a basis $\basis$ into parts according to
\eq{\basis=\bigoplus_{p=0}^{\infty}\basis_p\label{schematic_basis_stratification}}
from which we can construct the coefficients of amplitudes iteratively:
\eq{\mathcal{A}=\sum_p\mathcal{A}_p\quad\text{with}\quad\mathcal{A}_p\equivR\mathcal{A}\Bigger{\cap}\basis_p\equivR \sum_{\mathfrak{b}_p^i\in\basis_p}\!\!a_i\,\mathfrak{b}^i_p\,.}
This work is principally concerned with the systematic enumeration and construction of integrand bases suitable for representing amplitudes in generic quantum field theories, and to define for non-planar theories how integrand bases can be carved up into subspaces as in (\ref{schematic_basis_stratification}) according to some proxy for their ultraviolet (`UV') behavior. An example of such a partitioning at one loop is the division of integrands by na\"ive power-counting---that is, the leading polynomial degree in $\ell^\mu$ as $\ell\!\to\!\infty$. We will elaborate on this in \mbox{section \ref{sec:one_loop}}, but roughly speaking, we may say that a one-loop integrand behaves `like a $p$-gon [at infinity]' if
\eq{%\mathcal{I}(\ell)\in\basis_p\quad\text{iff}\quad 
\lim_{\ell\to\infty}\!\!\big[\mathcal{I}(\ell)\big]=\frac{1}{(\ell^2)^p}\Big[1+\mathcal{O}(1/\ell^2)\Big]\,.\label{naive_one_loop_power_counting}}

As we will see later, the construction of {\it a} basis of loop integrands large enough to represent all amplitudes in the Standard Model, for example, to all-multiplicity and at any loop order turns out to be relatively easy. What is considerably less trivial is the fact that in any integer (or within $\epsilon$ of an integer) number of dimensions, this basis is finite-dimensional and upper-triangular in cuts and hence, can be made `prescriptive' in the sense of \cite{Bourjaily:2017wjl}.

A much more important result, however, is the fact that the integrands in a basis constructed in this way can be partitioned into non-overlapping sets according to their `power-counting'. This is a concept that requires considerable care in its definition---and it will occupy most of our attention below. The real problem---and the source of subtlety and potential confusion---is that most measures of power-counting have relied explicitly on aspects of loop integrands such as how the loop momenta are `\hyperlink{routing}{routed}' which, rather than being intrinsically graph-theoretic, are instead subject to the whims of the physicist who wrote down some particular list of rational loop integrands. We will propose one such partitioning defined in purely graph-theoretic terms and discuss the limitations of this definition in the conclusions.\\

In this work, we will be interested in {\it integrands} of perturbative scattering amplitudes\,\!---the rational differential forms over the space of loop momenta obtained by the Feynman expansion. Careful readers should object to our use of `the' in the previous sentence, as this implies a certain degree of uniqueness which does not exist for many quantum field theories. This non-uniqueness comes in at least two forms. 

The first source of non-uniqueness of `the' integrand arises from the fact that terms in the Feynman expansion do not come pre-equipped with any preferential choice of origins for the loop momenta, or how the loop momenta should be routed through the Feynman graph. For planar field theories (such as Yang-Mills in the planar limit), Feynman diagrams do come pre-equipped with an arguably preferential \hyperlink{routing}{routing} associated with the planar-dual of each Feynman graph (provided the labels of these dual variables are symmetrized across all loop momenta in all graphs). But such a choice is immediately absent for non-planar theories: for example, there is no natural sense in which the following Feynman diagrams encode any particular rational function of any single loop momentum variable:
\eq{\fig{-27.25pt}{0.6}{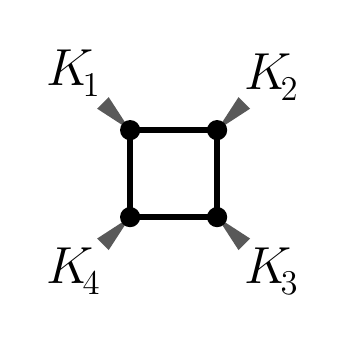}+\fig{-27.25pt}{0.6}{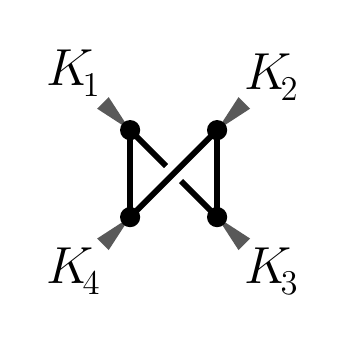}+\fig{-27.25pt}{0.6}{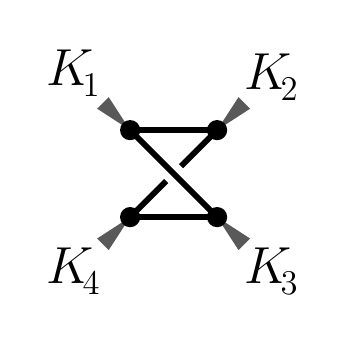}\,.}
There is no natural way to align the origins (in loop-momentum space) of the loop momenta across the three diagrams above. This non-uniqueness has historically been a source of much confusion. But the resolution is fairly obvious: there is simply no {\it particular} integrand to discuss. When we speak of {\it the} loop integrand, we merely mean a {\it representative integrand} of the equivalence class of rational functions generated by translations of all the internal loop momenta. Indeed, we will almost never have any need to choose any particular representative (equivalently, any choice of loop-momentum origins): our integrands will (almost) always be defined and discussed graph-theoretically.\footnote{In the representation of amplitudes, however, it is worth bearing in mind that the coefficients of integrands (for the representation of a particular amplitude) may themselves require that particular loop-momentum routings be chosen for the elements in the basis.}

It is worth mentioning that there has been a lot of progress in defining the non-planar loop integrand more rigorously \cite{Tourkine:2019ukp,Ben-Israel:2018ckc}, at least in the context of certain quantum field theories. Furthermore, it was observed in \cite{Herrmann:2016qea,Herrmann:2018dja,Edison:2019ovj} that even without a properly defined integrand there are very non-trivial properties of \emph{e.g.}~gravity amplitudes which have an imprint at the `integrand' level---as peculiar cancellations between diagrams---and which give a strong hint that a satisfactory definition of non-planar integrands should indeed exist. However, these observations are indirect and not immediately useful for defining a single rational integrand function.

The second source of non-uniqueness of loop integrands is considerably more important and surprisingly subtle: `the sum of Feynman diagrams' is not, in fact, guaranteed to be physically meaningful. Indeed, loop integrands of pure Yang-Mills theory need not be gauge invariant. This ambiguity is not hard to understand and is closely related to the fact that the forward limits of tree-amplitudes in non-supersymmetric gauge theories are ill-defined as not all diagrams that would be included in the forward limit of trees should be included in loops (as they become divergent). This partitioning of diagrams---to `throw out' those tree-diagrams which will diverge in the forward limit---is not gauge invariant. This is a real problem, and one that requires considerably more discussion than we will have room for here. The standard solution to this problem is to use \mbox{(\emph{e.g.}~dimensionally-)}regulated Feynman rules. In dimensional regularization \cite{tHooft:1972tcz}, for example, the loop integrand {\it would be} gauge-invariant as the problematic terms described above can be meaningfully said to vanish; but then the integrand is not regulator independent. This is a wholly acceptable viewpoint, and arguably a very powerful one considering the state of our tools for doing loop integration in dimensional regularization. However, it is not one that we insist upon.\\ 

Of course, loop integrands should be integrated. One consequence of the claims above is that for any fixed loop-order, the number of integrands we need to integrate is finite. We should be clear that this is a very different statement compared to the finiteness of the basis of master {\it integrals} \cite{Smirnov:2010hn} in dimensional regularization. In some ways, the finiteness we are describing is much stronger: as the statement about master integrals is dimensionally agnostic, the number of Feynman integrand topologies that must be considered will grow arbitrarily with multiplicity. In another sense, however, the finiteness of loop integrands is fairly weak; the number of independent {\it integrals} is always less than the number of independent integrands. We will not say more about integral-level identities, but emphasize that an independent set of integrals can always be chosen as a subset of independent integrands. 

We close this introduction with one small provocative comment. It is an empirical fact (often encountered in the evaluation of master integrals via differential equations \cite{Kotikov:1990kg,Bern:1992em,Gehrmann:1999as,Henn:2013pwa}) that the difference between an `easy' integral and a `hard' integral is immeasurable: that hard integrals are {\it so hard} that they are essentially technically intractable using current algorithms/technology; on the other hand, there are many `easy' integrands which are near-trivial to integrate. The cost of choosing a `wrong' basis of master integrals often far exceeds any savings from IBP reductions. As such, it is useful to consider the integrand-level representation of amplitudes very carefully.

%================================================================================================================
\vspace{-0pt}\paragraph*{The \emph{Stratification} of Perturbative Quantum Field Theories}\label{subsec:qft_stratifications}\vspace{-0pt}~\\[-12pt]
%================================================================================================================

\noindent 
Quantum field theories can be partially ordered according to the variety of loop integrands required to represent their amplitudes---that is, by the size of the smallest basis in which all their amplitudes can expressed according to (\ref{schematic_unitarity_construction}). Specifically, we will say that $[\text{theory A}]\succ[\text{theory B}]$ if a suitable basis for theory B exists which is a subset of such a basis for theory A. 

To illustrate this hierarchy of theories consider, for example, that
\eq{[\text{Standard Model}]\bigger{\succ}[\text{(Standard Model\textbackslash Higgs})]\bigger{\succ}[\text{QCD}]\bigger{\succ}[\text{Yang-Mills}]\ldots\,.}
This partial ordering is not hard to understand: each successive theory above involves a strict subset of the preceding theory's Feynman diagrams. What is much less trivial---and considerably more interesting---is that this partial ordering need not have anything to do with the number or kinds of Feynman diagrams for a theory. For example, there is a strict sense in which 
\eq{[\text{(pure) Yang-Mills}]\bigger{\succ}[\text{$\mathcal{N}=2$ super-Yang-Mills}]\bigger{\succ}[\text{$\mathcal{N}=4$ super-Yang-Mills}]\,.}
From a Feynman-diagrammatic point of view, adding supersymmetry to pure Yang-Mills means more fields and more Feynman diagrams. And so in what sense can it be that supersymmetric Yang-Mills theory (sYM) is `smaller' than non-supersymmetric (`pure') Yang-Mills? The answer is that for any multiplicity and any helicity configuration there exists bases of loop integrands for these theories such that
\eq{\basis^{\text{YM}}\,\bigger{\supsetneq}\,\basis^{\text{sYM}}\,.}

The fact that amplitudes in sYM are simpler than those in pure YM is closely related to the ameliorating effects of supersymmetry in the ultraviolet. The exceptionally good ultraviolet behavior of amplitudes in maximally supersymmetric ($\mathcal{N}\!=\!4$) Yang-Mills theory has been the subject of a great deal of interest. In the planar limit, this good behavior is tied to the dual-conformal symmetry of the theory \cite{Drummond:2006rz,Alday:2007hr,Drummond:2008vq}, but there is an increasingly sharp sense in which amplitudes outside the planar limit are expected \cite{Arkani-Hamed:2014via,Bern:2014kca,Bern:2015ple} or are known to have similarly good behavior at `infinite loop momenta' \cite{Bourjaily:2018omh}.

%================================================================================================================
\vspace{-0pt}\paragraph*{Problems with (Labeling) non-Planar Feynman Integrals}\label{subsec:non_planar_problems}\vspace{-0pt}~\\[-12pt]
%================================================================================================================

\noindent 
For planar\footnote{For the more mathematically minded reader, we should perhaps define precisely what we mean by a {\it planar} loop integrand. This is a loop integrand which admits a plane embedding once all `\hyperlink{externalEdges}{external}' propagators (those attached to monovalent vertices) are attached together at a single node `at infinity' on the compactified plane.} theories, there is a natural way to generalize the na\"ive power-counting defined at one loop by (\ref{naive_one_loop_power_counting}): namely, one can assign internal loop momenta according to each plane Feynman diagram's dual graph; symmetrizing with respect to these variables allows us to define an unambiguous rational loop integrand with loop-momentum variables that agree over all contributions. In terms of such labeling, one can discuss subsets integrands that behave `like a $p$-gon at infinity' with respect to all of the (now unambiguous) loop momenta by direct analogy to one loop. 

For non-planar loop integrands, there is no preferential way to route the loop momenta through a graph, making any definition analogous to (\ref{naive_one_loop_power_counting}) highly suspect. For example, consider the the three-loop `wheel' integral in scalar $\phi^4$-theory:
\eq{\fig{-38pt}{1.5}{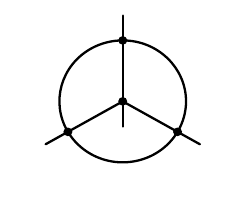}\label{scalar_wheel_111111}}
This Feynman integral has the topology of a tetrahedron, with loop momentum flowing around its edges. There are two seemingly natural (or at least, highly symmetric) choices for how to represent the dependence of this Feynman integrand on its three internal loop momenta: 
\eq{\fig{-38pt}{1.5}{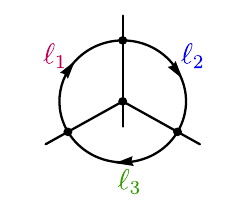}\quad\text{or}\quad\fig{-38pt}{1.5}{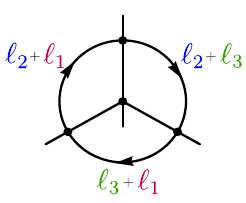}\,.\label{two_routings_of_wheel_111111}}
In the first case, each loop momentum flows through three propagators, suggesting that the integral behaves like a `triangle' at infinity; in the second case, each momentum flows through four propagators, suggesting that we view the integral as behaving like three `boxes' at infinity. Why is one choice preferred over the other? 

One may immediately think of several convincing reasons to prefer one routing loop momenta over another; but we predict that this rule would not lead to a \emph{unique} choice of labeling for complicated, generic graphs. Indeed, we strongly suspect that no preferential choice exists. As this example illustrates, not having  some preferential way to assign loop momentum variables to the propagators of a graph is already a problem for the description of integrands consisting exclusively of scalar propagators. The problem is compounded when integrands with loop-dependent numerators (required for theories with higher spin fields) are considered. 

One way to deal with the problems raised by this example would be to simply \emph{assign loop momentum labels} to all propagators and organize integrands in some way similar to the na\"ive power-counting defined for one loop integrals. This will indeed lead to \emph{some} notional hierarchy of integrands, and even allow one to discover that amplitudes in some theories are simpler than others (being expressible within better-behaved strata of bases). However pragmatic this approach may be, it nevertheless relies heavily on an obviously artificial choice of loop-momentum labeling. To avoid such burdensome complications, we must think more graph theoretically.

%================================================================================================================
\vspace{-0pt}\paragraph*{Overview of Our Main Results}\label{subsec:main_results_summary}\vspace{-0pt}~\\[-12pt]
%================================================================================================================

\noindent 
The lack of any canonical `\hyperlink{routing}{routing}' of loop momenta through a non-planar Feynman graph forces us to construct loop integrand bases more graph theoretically. Among the principal results of this work is a graph-(and representation-)theoretic description of loop integrands and a description of how integrand bases can be divided up according to (some notion of) `power-counting' similar to (\ref{schematic_basis_stratification}), such that this partitioning of the basis (loosely) tracks ultraviolet behavior, (some notion of) transcendentality, etc. 

A crucial ingredient in these results is a translationally-invariant description of vector-spaces of loop-dependent numerators of integrands. The formalism we describe is valid in any number of spacetime dimensions (including those of dimensional regularization); but there is much to be learned by specializing to a particular number of spacetime dimensions. In four dimensions, we reproduce the one-loop basis described by OPP in \cite{Ossola:2006us}, and its two loop extension discussed in \cite{Mastrolia:2012wf,Mastrolia:2013kca,Kleiss:2012yv}. What is more interesting, perhaps, is that we reproduce these examples in a way that renders them essentially representation-theoretic and naturally generalizes to arbitrary numbers of spacetime dimensions. 

Constructing a finite-dimensional $L$-loop integrand basis suitable for representing all amplitudes in the Standard Model in $4-2\epsilon$ dimensions turns out to be surprisingly easy. What is much more subtle and interesting is how such a basis may be divided up into strata like in (\ref{schematic_basis_stratification}) in any meaningful or useful way. By this we mean that it is not terribly difficult to define a graph-theoretic stratification of integrand bases, but it turns out to be surprisingly difficult to define a non-trivial substratum in which the amplitudes of maximally supersymmetric Yang-Mills may be represented, say. The stratification we propose here will not achieve this seemingly simple goal beyond some relatively high loop order (beyond at least 7 loops). 

In addition to these general ideas, we provide a complete description of two-loop integrand bases sufficient for representing all amplitudes in renormalizable quantum field theories in four or fewer spacetime dimensions. At three loops, we describe a complete basis of integrands sufficient to represent the leading transcendental weight part of any amplitude in four dimensions---which should suffice to represent all amplitudes in fully color-dressed maximally supersymmetric Yang-Mills theory.

%================================================================================================================
\vspace{-0pt}\subsection{Organization and Outline}\label{subsec:organization_and_outline}\vspace{-0pt}%~\\[-12pt]
%================================================================================================================

This work is organized as follows: In section \ref{sec:one_loop} we introduce all necessary ingredients and tools using the example of one-loop amplitudes for the sake of familiarity and concreteness. This includes the discussion of basic notational ideas, functional building blocks and a familiar (but generalizable) definition of power-counting. The final result of this section will be the list of topologies and the number of degrees of freedom for a given power-counting. Our main results lie in section \ref{sec:two_loops}, where we discuss in detail how to construct and stratify bases of integrands suitable for general two-loop amplitudes; we give a combinatorial rule for constructing and counting the numerator degrees of freedom for any integrand topology and power-counting. In section \ref{sec:three_loops}, we apply these ideas to three loops, and illustrate the results by providing a complete basis of triangle power-counting integrands in four dimensions.  Section \ref{sec:discussion_and_conclusions} is an extended discussion of caveats, open problems, and directions for further research. Several graph-theoretic definitions and notation is relegated to \mbox{appendix \ref{appendix:graph_theory}}. Finally, in the ancillary files of this work's submission to the \texttt{arXiv}, we give a \textsc{Mathematica} notebook which encapsulates our results for three-loop integrand bases in four spacetime dimensions and with triangle power-counting.

%================================================================================================================
%    1. One Loop Redux 
%         
%================================================================================================================
\newpage\vspace{-6pt}\section[\mbox{Review (and Redux): Building Bases of Integrands at One Loop}]{\mbox{Review (and Redux): Bases of Integrands at One Loop}}\label{sec:one_loop}\vspace{-2pt}
%================================================================================================================
%
The construction of \emph{integrand} bases at one loop is well-trodden territory, with a rich history of developments. The original ideas (in a modern form) go at least as far back as to Melrose, Passarino, and Veltman \cite{Melrose:1965kb,Passarino:1978jh}, and include the more recent ideas of Bern, Dixon, Kosower \cite{Bern:1994zx,Bern:1994cg,Bern:1997sc}, Ossola, Papadopoulos, Pittau (OPP) \cite{Ossola:2006us,Mastrolia:2010nb}, among many others (see for example \cite{vanNeerven:1983vr,Denner:2002ii,Britto:2004nc,Denner:2005nn,Cachazo:2008vp,Badger:2012dv,Mastrolia:2012an,Badger:2013sta,Mastrolia:2016dhn,Ita:2015tya}). The insights gained from these developments include the discovery of tree-level recursion relations for amplitudes \cite{Britto:2004ap,Britto:2005fq}, dual-conformal symmetry \cite{Drummond:2006rz,Alday:2007hr,Drummond:2008vq}, the Yangian \cite{Drummond:2009fd} of $\mathcal{N}=4$ sYM, and were put to use in many powerful practical applications (see \emph{e.g.}~\cite{Berger:2008sj,Bourjaily:2010wh}). 

Although this material is quite well established (see \emph{e.g.}~the reviews \cite{Ellis:2011cr,Zhang:2016kfo} and references therein), there are two key reasons for us offering yet another exposition here. The first is purely pragmatic: it will allow us to introduce critical concepts, notation, and illustrative examples that will prove important in our work ahead. 

The second reason why we feel this material will be useful for us to review, however, is to distinguish \emph{integrand} reduction (discussed in this work) from \emph{integral} reduction. The latter makes critical use of integration-by-parts relations \cite{Chetyrkin:1981qh,Tkachov:1981wb} about which we will have nothing to say here. A great example of the distinction between the two is the (ir-)reducibility of a pentagon integrand/integral in four spacetime dimensions---a case that we hope will be made more clear through our discussions. 

We begin our review of integrands with basic notational ideas, making the case for how loop-dependent numerators should be described algebraically. We will introduce the meaning of bases of integrands with fixed, `$p$-gon' power-counting (something which involves no subtleties at one loop), and then apply these ideas to re-derive many familiar facts about loop integrand bases in four dimensions. Subsequently, we describe how this can be generalized to arbitrary integer dimensions, and also comment on extending our integer-dimensional construction to the realm of dimensional regularization \cite{tHooft:1972tcz}.\\

%================================================================================================================
\vspace{-0pt}\subsection{One-Loop Integrand Bases: Basic Building Blocks}\label{subsec:one_loop_reduction_and_stratification}\vspace{-0pt}
%================================================================================================================

The basic idea of integrand reduction at one loop is very easy to understand. Every Feynman diagram in any quantum field theory will give rise to a rational function of \emph{the} (unambiguous up to translation) loop momentum $\ell$, external kinematics and possibly internal quantum numbers (such as color factors in gauge theory). For any particular theory, these rational functions span a finite-dimensional vector-space. We are going to  say much more about the precise meaning of these statements momentarily, but let us start by considering the kinds of rational functions that arise.

%================================================================================================================
\newpage\vspace{-0pt}\subsubsection{Loop-Dependent Denominators: `Scalar' \texorpdfstring{$p$}{\emph{p}}-gon Integrands}\label{subsubsec:scalar_p_gons}\label{subsubsec:1loop_pgon_int}\vspace{-0pt}
%================================================================================================================

At one loop, the \emph{Feynman rules} of a theory give us a map of the form\footnote{We do not explicitly allow for eikonal propagators $1/(\ell\cdot Q)$ here, even though an extension to this case should be straight forward.}
\eq{\hspace{-70pt}\text{Feynman diagram}\mapsto\frac{\mathfrak{N}(\ell)}{((\ell-Q_1)^2-m_1^2)((\ell-Q_2)^2-m_2^2)\cdots((\ell-Q_p)^2-m_p^2)},\,\hspace{-50pt}\label{eg_one_loop_diagram}}
where $\ell$ is the `loop momentum' variable to be integrated over, each $Q_i$ is some constant momentum offset, each $m_i$ is the mass of a particle through which the loop momentum `propagates', $\mathfrak{N}(\ell)$ is some \textbf{polynomial} in $\ell$ with coefficients that depend rationally on the external particles' momenta, color labels, and polarizations. To be clear, we consider $\mathfrak{N}(\ell)\simeq\mathfrak{N}'(\ell)$ if $\mathfrak{N}(\ell)\!=\!f(\text{ext})\mathfrak{N}'(\ell)$ for any function $f(\text{ext})$ independent of $\ell$. In particular, $f(\text{ext})$ can involve many propagators: namely, all those fixed to be $\ell$-independent  by momentum-conservation at each vertex. Thus, for example
\eq{
\label{eq:graph_equivalence_ext_props}
\vspace{-0pt}\fig{-15pt}{1}{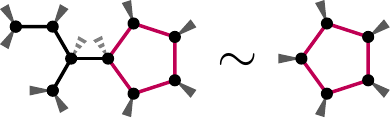}\vspace{-0pt}}
should be understood as equivalent as far as their {\it loop-dependence} is concerned. In this work, we will mostly be interested in the loop-momentum-\emph{dependent} parts of Feynman integrands---avoiding much (if any) discussion of the loop-independent factors\footnote{For some recent discussions on efficiently reconstructing the loop-independent functions that appear in a given basis, see \emph{e.g.}~\cite{Peraro:2016wsq,Abreu:2019odu,Peraro:2019svx}.} and therefore have no reason to consider loop-momentum-independent propagators that may be involved in any actual Feynman diagram. 

At one loop, we call an integrand with $p$ loop-dependent propagators a `$p$-gon'. The propagators of any $p$-gon have a natural ordering: namely, so that the offset momenta $Q_i$ appearing in the propagators of (\ref{eg_one_loop_diagram}) always differ by the momentum flowing into each of the vertices. Specifically, we can order the factors so that for each index $i$ (with cyclic labeling understood), \mbox{$P_i\equivL Q_{i+1}-Q_{i}$} is the sum of some subset of external momenta (which `flows into' the $i$th vertex). For example, 
\eq{\fig{-35.25pt}{1}{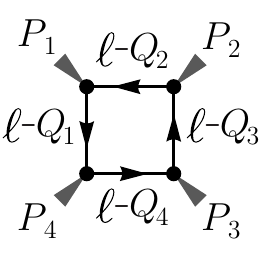}\,.\label{eg_one_loop_box}}
Notice that in (\ref{eg_one_loop_box}) we have used wedges to denote external momenta flowing into the graph. We use this notation to make it clear that we do not care how many \hyperlink{externalEdges}{external legs} are flowing into each vertex, whether they are massive or massless, or what other quantum numbers the external states may carry. Further details concerning our notation can also be found in appendix \ref{appendix:graph_theory}. In the following, we often suppress the external edge labels and write
\eq{\fig{-35.25pt}{1}{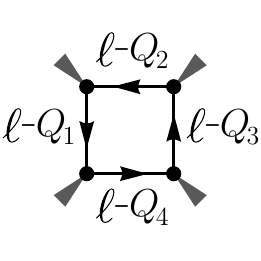}\;\;\bigger{\leftrightarrow\;}\,\frac{1}{\x{\ell}{Q_1}\x{\ell}{Q_2}\x{\ell}{Q_3}\x{\ell}{Q_4}}\,,\label{eg_box_int}}
where we have introduced the convenient shorthand notation
\eq{\x{\ell}{Q}\equivR (\ell\mi Q)^2\label{bracket_defined}\,,}
and where $Q$ denotes {\it any} momentum---{\it including} $\vec{0}$. The Feynman $i\epsilon$-prescription will be left implicit throughout our work. The propagator definition (\ref{bracket_defined}) supposes a massless spectrum. Indeed, in this work we show explicit results for massless theories only, even though a generalization to the massive case does not pose any conceptual difficulties. If we were to include massive propagators, we could define
\eq{\x{\ell}{Q}_m\equivR(\ell\mi Q)^2\mi m^2\,,\qquad\text{with}\qquad\x{\ell}{Q}_0\equivL\x{\ell}{Q}\,,\label{massive_bracket_defined}}
but we would need to consider all graphs such as (\ref{eg_box_int}) to carry additional labels to account for all the possible distribution of \hyperlink{internalEdges}{\emph{internal} edges'} particle masses. 

In eq.~(\ref{eg_box_int}), the $Q_i$'s should be understood to conserve momentum. Why have we not written them explicitly? The reason is that any explicit formula for the $Q_i$'s would require that we eliminate or trivialize the translational invariance of the loop momentum $\ell$. This could be done by going to dual-momentum coordinates, or more simply by choosing any one of the $Q_i$'s to be $\vec{0}$ (that is, picking an origin for $\ell$). We are leaving such a choice to the reader, because we do not want to give the appearance that anything we say in what follows depends on whether or how translational invariance is eliminated (or otherwise trivialized). 

At one loop, momentum conservation requires that all propagators involving the internal loop momentum $\ell$ must form a single, closed cycle. We define an integrand with exactly $p$ propagators involving the loop momentum $\ell$ to be a `$p$-gon' integrand. These will be denoted diagrammatically by a polygon with $p$ sides with $p$ \hyperlink{externalEdges}{`external' legs} denoting inflowing momentum; for example,
\eq{\vspace{-0pt}\fig{-15pt}{1}{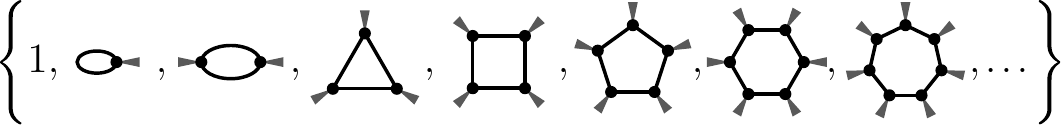}\vspace{-0pt}\label{example_one_loop_graphs}}
would be described as $0$-gon, $1$-gon, $\ldots$, $7$-gon integrands, respectively. In practice (and even here), some of these are rarely named in this way: the first integrands in (\ref{example_one_loop_graphs}) are more often referred to as `constant', `tadpole', `bubble', `triangle', and `box' integrands, respectively.

%================================================================================================================
\vspace{-0pt}\subsubsection{Loop-Dependent Numerators: Notational Biases for Bases}\label{subsubsec:one_loop_tensor_numerators}\vspace{-0pt}
%================================================================================================================

The real motivation for introducing notation such as (\ref{bracket_defined}) for propagators is that {\it inverse propagators} should be viewed as {fundamental, irreducible objects}. They are the right language in which to describe loop-dependent numerators exactly because of their ubiquitous role in the construction of loop-dependent denominators of any Feynman integral. Our attitude here is somewhat at odds with the more familiar approach to integrand reduction which treats Lorentz invariant scalar products as the primary monomials of consideration. For integrand reduction, one then distinguishes monomials that can be written in terms of inverse propagators and external kinematic invariants from irreducible scalar products (ISP)'s~\cite{Johansson:2013sda} (starting at two loops). Thus, it is worthwhile to compare and contrast these frameworks in detail.

We suggest that all polynomials involving loop momenta should be expressed in the space of sums of products of (generalized) inverse propagators\footnote{For practitioners, using inverse `propagators' not necessarily appearing in a Feynman graph should not seem so strange: such vector-spaces are also needed for \emph{integral} reduction, as implemented in public codes such as \texttt{FIRE}~\cite{Smirnov:2008iw}.}. In particular, we use
\eq{\ellKd{}{d}\equivR\Span{Q}\{\x{\ell}{Q}\}\quad\text{for}\quad Q\in\mathbb{R}^d\,\label{definition_of_single_tensor}}
to denote the vector-space generated by the $d$-dimensional \emph{translates} of inverse propagators involving momentum $\ell$. To be clear, $1/\x{\ell}{Q}$ need not be a propagator in any Feynman graph which relates to our description as `generalized' inverse propagators (that we occasionally drop in our discussions below). Provided that $\ell\!\in\!\mathbb{R}^d\,$,\footnote{Throughout this work, we will take $d\!\in\!\mathbb{N}_+$; when discussing dimensional regularization, we consider $\hat{e}_{{-}2\epsilon}$ to be an additional basis element used for translations.} then it is not hard to see that
\eq{[\squeezeD \ell\squeezeD]_d\simeq\mathrm{span}\Big\{\text{`}1\text{'},\underbrace{\ell\!\cdot\! \hat{e}_1,\ldots,\ell\!\cdot\!\hat{e}_d}_{\fwbox{0pt}{\text{components `$\ell^i$'$\!:=\!\ell\!\cdot\!\hat{e}_i$}}},\ell^2\Big\}\,,\label{naive_basis_of_single_inverse_props}}
where $\hat{e}_i$ are $d$-dimensional basis vectors for $\mathbb{R}^d$ and `$1$' signifies any $\ell$-independent monomial which should be understand to carry the scaling dimension of mass-squared. We will soon stop flagging this fact with scare quotes when writing `$1$'. To see the equivalence between (\ref{definition_of_single_tensor}) and (\ref{naive_basis_of_single_inverse_props}), one can confirm that 
\eq{\fwbox{0pt}{\hspace{-20pt}\text{`}1\text{'}\simeq\,Q^2\!=\frac{1}{2}\big[\x{\ell}{Q}\pl\x{\ell}{\mi Q}\mi2\x{\ell}{0}\big]\,,\quad
\ell\!\cdot\! \hat{e}_i=\frac{1}{4}\big[\x{\ell}{\mi \hat{e}_i}\mi\x{\ell}{\hat{e}_i}\big]\,,\quad\ell^2=\x{\ell}{0}\,.}}
From (\ref{naive_basis_of_single_inverse_props}), it is easy to determine the dimensionality of $\ellKd{}{d}$
\eq{\mathrm{rank}\big(\ellKd{}{d}\big)=d\pl2\,.\label{eq:ell_d_dimensionality}}
A key reason for using translates of inverse propagators to describe polynomial degrees of freedom in $\ell$ is that the definition (\ref{definition_of_single_tensor}) is translationally invariant:
\eq{[\squeezeD \ell\squeezeD]\simeq[\squeezeD \ell\pl Q\squeezeD];\quad\text{for any $Q$}\,.}
Thus, any propagators of a Feynman graph which differ by incoming external momenta will correspond to the same vector-space $\ellK{}$. Moreover, these spaces are identical for massive and massless propagators:
\eq{\Span{Q}\{\x{\ell}{Q}\}\simeq\,\,\Span{Q}\{\x{\ell}{Q}_m\}\,.}
Although $[\squeezeD \ell\squeezeD]$ includes the degree-two-in-components element $\ell^2$, it is important to emphasize that no other degree-two-in-components monomials are spanned by $[\squeezeD \ell\squeezeD]$. In particular, \mbox{$\ell^i\ell^j\!\notin\![\squeezeD \ell\squeezeD]$}; a corollary of this is that if we were to write $\ell^2\equivL\hat{\ell}^2\mi \mu^2$, then neither $\hat{\ell}^2$ nor $\mu^2$ would be in $[\squeezeD \ell\squeezeD]$---although, the combination $\ell^2\!(\!\equivL\hat{\ell}^2\mi \mu^2)\!\in\![\squeezeD \ell\squeezeD]$. We will have more to say about how degrees of freedom should be encoded for dimensional regularization below in section \ref{subsubsec:one_loop_for_arbitrary_regulated_integer_dim}.

%================================================================================================================
\vspace{-0pt}\subsection{Integrand Bases with `\texorpdfstring{$p$}{p}-gon' Power-Counting}\label{subsec:p_gon_power_counting_bases}\vspace{-0pt}
%================================================================================================================
%
Let us now describe how we may construct and stratify bases of one-loop integrands according to their ultraviolet behavior---or, more colloquially, their `power-counting'. As we will see in the next section, it turns out that one loop is a deceptively simple case in this regard. Roughly speaking, we say that $\mathcal{I}$ has `$p$-gon power-counting' if it scales like $p$ or more propagators as $\ell\!\to\!\infty$:
\eq{\lim_{\ell\to\infty}\!\!\big(\mathcal{I}\big)=\frac{1}{{(\ell^2)^p}}\big[1\pl\mathcal{O}(1/\ell^2)\big]\,.\label{naive_one_loop_uv_scaling}}
As we will argue in \ref{subsubsec:one_loop_contact_term_decomposition}, this definition turns out to not be entirely satisfactory for our purposes. To see why, we first introduce the relevant language to discuss numerators with multiple loop-momentum insertions of the fundamental building blocks defined in section \ref{subsubsec:one_loop_tensor_numerators}.

%================================================================================================================
\vspace{-0pt}\subsubsection{Vector-spaces of Loop-Dependent Numerators}\label{subsubsec:one_loop_higher_tensor_numerators}\vspace{-0pt}
%================================================================================================================
%
To allow for higher polynomial degrees in $\ell$, we may simply consider spaces constructed from monomials built from products of inverse propagators as straight forward generalization of our basic building blocks defined in eq.~(\ref{definition_of_single_tensor}). Let
\eq{\ellK{p}_d\equivR\Span{\oplus_iQ_i}\!\Big\{\x{\ell}{Q_1}\cdots\x{\ell}{Q_p}\!\Big\}\quad\text{for}\quad Q_i\!\in\!\mathbb{R}^d\,.}
From the embedding space perspective \cite{Weinberg:2010fx,SimmonsDuffin:2012uy}, it is easy to see that the vector-space $\ellK{p}_d$ is a $p$-fold symmetric, traceless product of $(d\pl2)$-dimensional vectors of $\mathfrak{so}_{d+2}$. As such, 
\eq{\mathrm{rank}\big(\ellKd{p}{d}\big)=\binom{d+p}{d}+\binom{d+p-1}{d}\,. \label{pnum_ddim_rank}}

To better understand these vector-spaces, consider the first non-trivial case of $\ellK{2}$. Following the definition above, we could start with an over-complete vector-space spanned by
\eq{[\squeezeD \ell\squeezeD]^2_d=\mathrm{span}\Big\{\x{\ell}{Q_1}^2,\ldots,\x{\ell}{Q_{d+2}}^2,\x{\ell}{Q_1}\x{\ell}{Q_2},\ldots,\x{\ell}{Q_{d+1}}\x{\ell}{Q_{d+2}}\Big\}\,,}
for some set of displacements $Q_i\!\in\!\mathbb{R}^d$; from this one may at first expect the rank to be $(d\pl2)+\binom{d\pl2}{2}$; but this over-counts the dimension of the space by (in this case) one non-trivial relation. To see this one relation most simply, consider the na\"{i}ve basis of monomials as we did in (\ref{naive_basis_of_single_inverse_props}); simply taking all pairs, we'd find the (still over-complete) generators:
\eq{[\squeezeD \ell\squeezeD]^2\simeq\mathrm{span}\Big\{1\!\cdot\!1,1\!\cdot\!\ell^i,{\color{hblue}1\!\cdot\!\ell^2},{\color{hred}(\ell^i)^2},\ell^i\!\cdot\!\ell^j,\ell^i\!\cdot\!\ell^2,(\ell^2)^2\Big\}\,.\label{naive_basis_of_double_inverse_props_v0}}
Here, the over-completeness is more manifest: ${\color{hblue}1\!\cdot\!\ell^2}\!\in\mathrm{span}_{\hspace{-5pt}\raisebox{-4pt}{{\scriptsize$i$}}}\big\{{\color{hred}(\ell^i)^2}\big\}$. Eliminating this over-completeness would result in a non-redundant basis for $\ellK{2}$ of the form
\eq{[\squeezeD \ell\squeezeD]^2\simeq\mathrm{span}\Big\{1\!\cdot\!1,1\!\cdot\!\ell^i,{\color{hred}(\ell^i)^2},\ell^i\!\cdot\!\ell^j,\ell^i\!\cdot\!\ell^2,(\ell^2)^2\Big\}\,.\label{naive_basis_of_double_inverse_props}}
Although we are not including the monomial $\ell^2$ as a basis vector in (\ref{naive_basis_of_double_inverse_props}), it is important to never forget that $\ell^2\!\in\![\squeezeD \ell\squeezeD]^2$; and moreover, more generally, $\ellK{1}\!\subset\!\ellK{2}$. 

This is part of a more general observation: because multiplying by the loop independent monomial `$1$'$\!\in\![\squeezeD \ell\squeezeD]$ is an injective map from $\ellK{p}\hookrightarrow\ellK{p\pl1}$, we always have
\eq{\ellK{a}\!\subset\!\ellK{b}\quad \forall\,a\!<\!b\,.\label{eq:ell_inclusion}}
Whenever we have such sequences of inclusions, it is natural to \emph{stratify} these vector-spaces according to their complements. Specifically, let
\eq{\ellKhat{p}\equivR\ellK{p}\backslash\ellK{p-1}\quad\text{so that}\quad \ellK{p}\!=\ellKhat{0}\!\oplus\!\ellKhat{1}\!\oplus\!\cdots\!\oplus\!\ellKhat{p}\,;\label{nested_ellK_inclusions}}
that is, $\ellKhat{p}$ is the part of $\ellK{p}$ \emph{not} spanned by $\ellK{p-1}$. The rank of $\ellKhat{p}$ is easy to compute from (\ref{pnum_ddim_rank}):
\eq{\mathrm{rank}\big(\ellKhat{p}_{d}\big)
=\mathrm{rank}\big(\ellK{p}_{d}\big) -\mathrm{rank}\big(\ellK{p-1}_{d}\big) 
=\binom{d\pl p}{p}-\binom{d\pl p\mi2}{p\mi2} 
%= \frac{d\pl 2p\mi 1}{p}\binom{d\pl p\mi2}{p-1}
\,.}
Such a grading of loop-dependent polynomial degrees of freedom is closer in spirit to the way that scalar products are often organized in the literature, \emph{e.g.}~\cite{Kosower:2011ty,Badger:2013gxa}. An important distinction, however, is that what we call $\ellKhat{p}$ consists of some (but not all) polynomials in the components of $\ell$ of degree $2p$ or degree $2p\mi1$. We will see below that this grading of vector-spaces plays an important role in the `power-counting' stratification of loop-dependent degrees of freedom in section \ref{subsubsec:stratifying_one_loop_bases}.

%================================================================================================================
\newpage
\vspace{-0pt}\subsubsection{Organizing Loop-Dependent Numerators by Contact-Terms}\label{subsubsec:one_loop_contact_term_decomposition}\vspace{-0pt}
%================================================================================================================
%
We gave a vague diagnostic for `$p$-gon' power-counting integrands in eq.~(\ref{naive_one_loop_uv_scaling}) above. However, this definition turns out to be somewhat inadequate, because we would like to differentiate between numerators such as $\ell^2\!\in\!\ellK{}$ from those such as $\ell^i\ell^j\!\notin\!\ellK{}$. Recall that $\ell^i\ell^j\!\in\!\ellK{2}$. Our choice to write $\ell^i\ell^j$ in terms of two-fold products of inverse propagators would seem to unnecessarily worsen what we consider the UV behavior of an integrand with such a numerator. However, this is perfectly okay: because $1\!\in\!\ellK{}$, it is \emph{always} possible to express an integrand with some UV behavior in a space of integrands with worse UV behavior. At the end of the day, the question of how to carve up integrand bases (and how UV behavior should even be defined) can only be answered by how useful the resulting basis is for representing amplitudes.

One major advantage of choosing to write all numerator degrees of freedom directly in terms of products of generalized inverse propagators is that it trivializes the determination of the scaling of eq.~(\ref{naive_one_loop_uv_scaling}). Let
\eq{\mathfrak{b}_{q}^p\equivR\frac{\ellK{q-p}}{\x{\ell}{Q_1}\cdots\x{\ell}{Q_q}}\,\quad\text{for}\quad q\geq p\,,\label{eq:pgon_pc_q_gon_integrand_def_1L}}
be the vector-space of all $q$-gon integrals with $(q\mi p)$-fold products of inverse propagators in their numerators. It is trivial to see that 
\eq{\lim_{\ell\to\infty}\!\!\big(\mathcal{I}\big)=\frac{1}{{(\ell^2)^p}}\quad\text{for all}\;\;\; \mathcal{I}\!\in\!\mathfrak{b}_q^p\,.\label{def_one_loop_uv_scaling}}
Indeed, starting from the space $\mathfrak{b}_p^p$---so-called `scalar' $p$-gon integrands, with loop-independent numerators---one may also consider $(p\pl1)$-gon integrals with numerators chosen from $\ellK{}$, $\mathfrak{b}_{p+1}^p$; $(p\pl2)$-gons with numerators chosen from $\ellK{2}$, $\mathfrak{b}_{p+2}^p$; and so on. All these integrands trivially scale `like a $p$-gon' at infinite loop momentum.

%================================================================================================================
\phantomsection\markright{}\addcontentsline{toc}{paragraph}{\hspace{-4pt}Graphical Rules for Denoting Integrand Vector-Spaces}
%================================================================================================================
\vspace{-0pt}\paragraph*{Graphical Rules for Denoting Integrand Vector-Spaces}\label{para:graphical_rules_for_inverse_prop_spaces}\vspace{-0pt}~\\[-12pt]
%================================================================================================================

\noindent 
To better discuss these vector-spaces of integrands, it may be useful to introduce some graphical/diagrammatic notation. Let us denote the vector-space of translates of inverse propagators over a propagator as
\eq{\fwbox{65pt}{\tikzBox{\draw[int](0,0)--(1,0);\draw[markedEdgeR](0,0)--(1,0);\node[anchor=north] at (0.5,0) {$\vec{\ell}$};}}\equivR\frac{[\ell]}{\ell^2}\,.\label{vector_space_of_decorated_edge}}
This vector-space includes the Feynman-rule-propagators of scalar fields, Fermions, and spin-1 bosons (in any gauge). In particular, this means that (\ref{vector_space_of_decorated_edge}) includes as elements all the propagators involved in the Feynman expansion for the Standard Model. And because $\ell^2\in\ellK{}$, it also includes so-called \emph{contact-terms}. 

Consider, for example, the space $\mathfrak{b}_4^3$, which is represented diagrammatically by
\eq{\mathfrak{b}_4^3\bigger{\;\Leftrightarrow}\tikzBox{\oneLoopGraphElement[0]{4}\draw[markedEdgeR](a2)--(a3);}\,.}
Because all the propagators of any one-loop graph involve the same undetermined loop momentum up to translation, it does not matter where we put the decoration:
\eq{\tikzBox{\oneLoopGraphElement[0]{4}\draw[markedEdgeR](a2)--(a3);}
\sim\tikzBox{\oneLoopGraphElement[0]{4}\draw[markedEdgeR](a3)--(a4);}
\sim\tikzBox{\oneLoopGraphElement[0]{4}\draw[markedEdgeR](a4)--(a1);}
\sim\tikzBox{\oneLoopGraphElement[0]{4}\draw[markedEdgeR](a1)--(a2);}\,.}
The statement about contact-terms being included can be written graphically as, for example,
\eq{\fig{-18.pt}{0.2125}{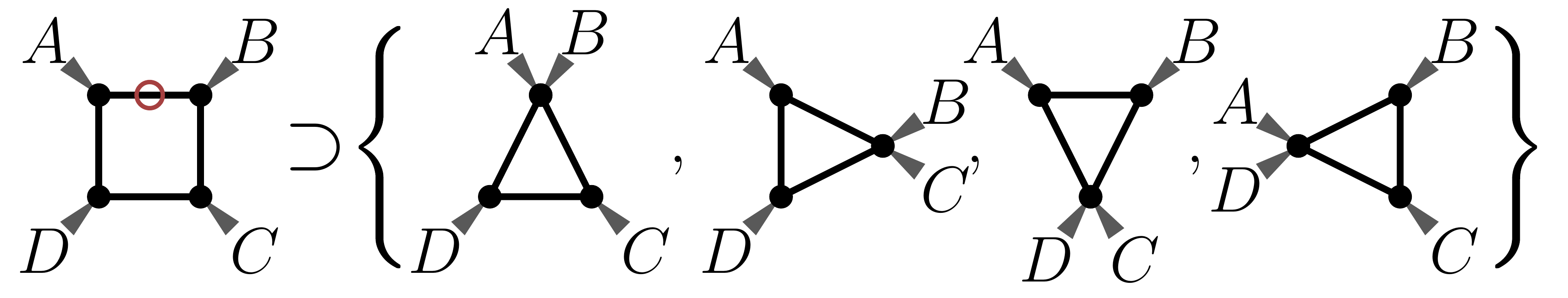}\,.\label{contact_term_inclusions}}
Notice that this shows that $\mathfrak{b}^3_3\!\subset\!\mathfrak{b}^3_{4}$. 

More generally, it is easy to see that $\mathfrak{b}^p_{q{-}1}\!\subset\!\mathfrak{b}^p_{q}$. It will be useful to talk about the complements of these inclusions, as we did for $\smash{\ellKhat{p}}$ defined in (\ref{nested_ellK_inclusions}). Let
\eq{\widehat{\fwbox{15pt}{{\color{topCount}\mathfrak{b}^p_q}}}\equivR\mathfrak{b}^p_{q}\backslash\mathfrak{b}^p_{q-1}\,,
\label{eq:contact_decomposition_abstract_1L}
}
denote the `{\color{topCount}top-level}' sub-space of ${\color{totalCount}\mathfrak{b}_q^p}$---the space \emph{not} spanned by so-called contact-terms. Returning to the example of $\mathfrak{b}_4^3$, it is natural to wonder the rank of $\smash{\widehat{\fwbox{15pt}{\rule{0pt}{8pt}\smash{\mathfrak{b}_4^{\raisebox{-5pt}{{\scriptsize3}}}}}}}$. Consider the case of $d\!=\!1$, for which $\mathrm{rank}\big(\mathfrak{b}_4^3\big)\!=\mathrm{rank}\big(\ellK{1}_{d=1}\big)\!=\!3$ and $\mathrm{rank}\big(\smash{\widehat{\fwbox{15pt}{\rule{0pt}{8pt}\smash{\mathfrak{b}_4^{\raisebox{-5pt}{{\scriptsize3}}}}}}}\big)\!=\!0$; the statement (\ref{contact_term_inclusions}) is still true, but not all of the four triangle contact-terms can be independent. For $d\!=\!2$, we similarly conclude that $\mathrm{rank}\big(\smash{\widehat{\fwbox{15pt}{\rule{0pt}{8pt}\smash{\mathfrak{b}_4^{\raisebox{-5pt}{{\scriptsize3}}}}}}}\big)\!=\!0$, but now the four scalar triangles in (\ref{contact_term_inclusions}) are independent. For $d\geq3$, we find that $\mathrm{rank}\big(\smash{\widehat{\fwbox{15pt}{\rule{0pt}{8pt}\smash{\mathfrak{b}_4^{\raisebox{-5pt}{{\scriptsize3}}}}}}}\big)\!=\!d\mi2$.\\

Let us define the `$p$-gon power-counting basis' of integrands at one loop to be
\eq{\mathfrak{B}_p\equivR\fwbox{22pt}{\mathfrak{b}_p^p}\bigger{\,\cup\,}\mathfrak{b}_{p+1}^p\bigger{\,\cup\,}\mathfrak{b}_{p+2}^p\bigger{\cup}\cdots\,=\fwbox{22pt}{\fwbox{15pt}{\mathfrak{b}^p_p}}\bigger{\oplus}\fwboxL{22pt}{\widehat{\fwboxL{15pt}{\mathfrak{b}^p_{p{+}1}}}}\bigger{\oplus}\fwboxL{22pt}{\widehat{\fwboxL{15pt}{\mathfrak{b}^p_{p{+}2}}}}\bigger{\oplus}\cdots\,.}
Graphically, we could write
\eq{\vspace{-0pt}\fig{-95pt}{1}{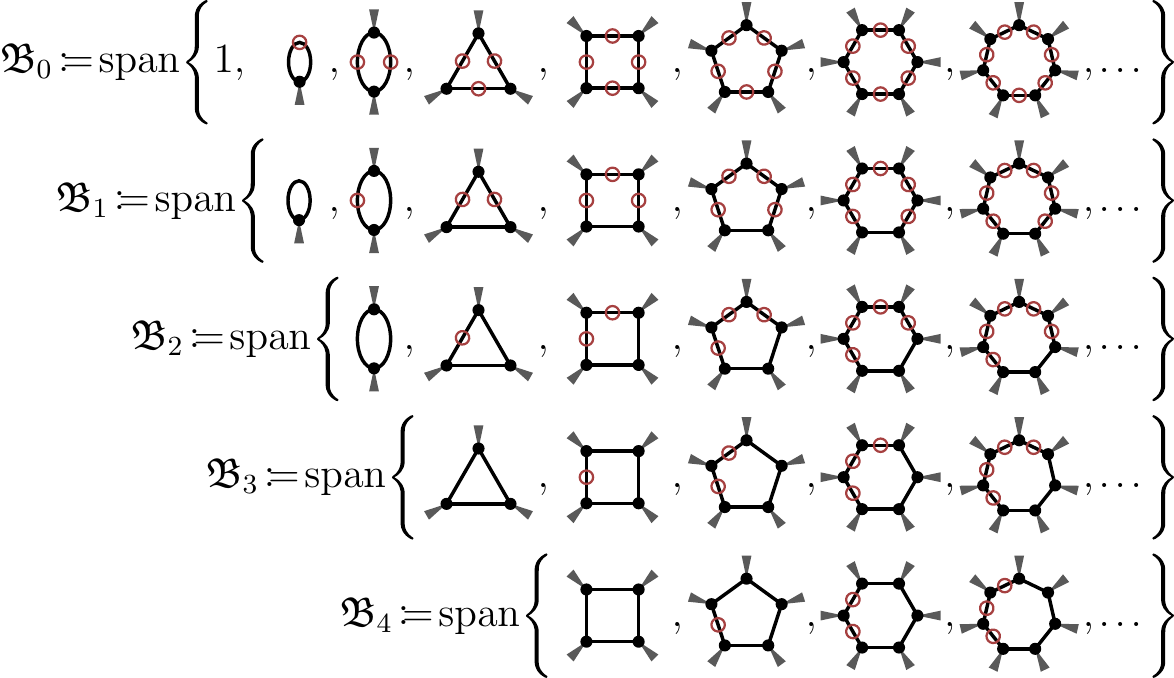}\vspace{-0pt}}
and so on. Although each of these spaces may appear to be infinite-dimensional, this is never the case. Indeed, for any fixed spacetime dimension, the rank of $\mathfrak{B}_p$ is bounded. The proof of this statement is originally due to Passarino and Veltman in \mbox{ref.\ \cite{Passarino:1978jh}}; let us see how we can understand this fact in our present formalism.  Let $\mathfrak{B}_p^{(d)}$ denote the space $\mathfrak{B}_p$ in spacetime dimension $d\!\in\!\mathbb{N}_+$. The easiest way to see that $\mathfrak{B}_p^{(d)}$ is finite-dimensional is to notice that the nested sequence of inclusions
\eq{\mathfrak{b}_p^p\bigger{\,\subset\,}\mathfrak{b}_{p+1}^p\bigger{\,\subset\,}\mathfrak{b}_{p+2}^p\bigger{\subset}\cdots\,,}
turns around on itself eventually: $\mathfrak{b}_{q}^p\bigger{\supset}\mathfrak{b}_{q+1}^p$ for some $q$ in any dimension $d$ and power-counting $p$. This happens as soon as $\mathrm{rank}\big(\smash{\widehat{\fwbox{13pt}{\rule{0pt}{8pt}\smash{\mathfrak{b}_q^{\raisebox{-5pt}{{\scriptsize $p$}}}}}}}\big)=0$---that is, when the space of contact-terms spans everything (as we saw above for $\smash{\widehat{\fwbox{15pt}{\rule{0pt}{8pt}\smash{\mathfrak{b}_4^{\raisebox{-5pt}{{\scriptsize3}}}}}}}$ when $d\!\leq\!2$). When does this happen? There are three cases to consider: $p\!>\!d$, $p\!=\!d$, and $p\!<\!d$, which we discuss in turn. 

First, consider the case $p\!>\!d$; we will show that $\mathfrak{b}_p^p\!\supset\!\mathfrak{b}_{p+1}^p$, which implies that $\mathrm{rank}\!\big(\,\smash{\widehat{\fwbox{18pt}{\rule{0pt}{8pt}\smash{\mathfrak{b}_{p{+}1}^{\raisebox{-5pt}{{\scriptsize $p$}}}}}}}\big)\!=\!0$ and that $\mathfrak{B}^{(d\!<\!p)}_p\!\!=\mathfrak{b}_p^p$. To prove this, we merely need to note that 
\eq{\begin{split}\mathrm{rank}\big(\smash{\fwboxL{22pt}{\widehat{\fwboxL{15pt}{\rule{0pt}{8pt}\smash{\mathfrak{b}_{p+1}^{\raisebox{-5pt}{{\scriptsize $p$}}}}}}}}\big)&=\mathrm{rank}\big(\mathfrak{b}_{p+1}^p\backslash\mathfrak{b}^p_{p}\big)\\
&=\mathrm{rank}\big(\ellK{}_d\big)-\mathrm{rank}\big(\mathrm{span}\big\{\x{\ell}{Q_1}\mathfrak{b}_p^p,\ldots,\x{\ell}{Q_{p+1}}\mathfrak{b}_p^p\big\}\big)=0\,.\end{split}}
The first term above is $d\pl2\!=\!\mathrm{rank}\big(\ellK{}_d\big)$, and the second term has rank at least $d\pl2$ as $p\pl1\!\geq\!d\pl2$. Thus, $q\!=\!p$. 

The next case to consider is when $p\!=\!d$. For this, it is not hard to show (by direct construction) that  $\mathfrak{b}_{d+2}^d\!\subset\!\mathfrak{b}_{d+1}^d$, so that $\mathrm{rank}\big(\fwboxL{22pt}{\smash{\widehat{\fwboxL{18pt}{\rule{0pt}{8pt}\smash{\mathfrak{b}_{d+2}^{\raisebox{-5pt}{{\scriptsize $d$}}}}}}}}\big)\!=\!0$. As such, the $d$-gon power-counting basis would consist of $\mathfrak{B}_d^{(d)}\!\!=\!\mathfrak{b}_d^d\bigger{\oplus}\fwboxL{22pt}{\smash{\widehat{\fwboxL{15pt}{\rule{0pt}{8pt}\smash{\mathfrak{b}_{d+1}^{\raisebox{-5pt}{{\scriptsize $d$}}}}}}}}$, with each of these vector-spaces carrying a single degree of freedom in their numerators. Thus, $q\!=\!p\pl1$. Actually, this is a bit of an overstatement: as we will discuss at greater length in section~\ref{subsubsec:conformality_one_loop} below, the span of $(d\pl1)$-gons in $\fwboxL{22pt}{\smash{\widehat{\fwboxL{15pt}{\rule{0pt}{8pt}\smash{\mathfrak{b}_{d+1}^{\raisebox{-5pt}{{\scriptsize $d$}}}}}}}}$ is over-complete: any choice of $(d\pl2)$ of these integrands whose union consists of $d\pl2$ propagators will satisfy a single relation. 
 
The final case to consider is that of $p\!<\!d$, for which we will show that $\mathrm{rank}\big(\fwboxL{15pt}{\smash{\widehat{\fwboxL{13pt}{\rule{0pt}{8pt}\smash{\mathfrak{b}_{q}^{\raisebox{-5pt}{{\scriptsize $p$}}}}}}}}\big)\!=\!0$ for all $q\!>\!d$, and hence $q\!=\!d$. From this, it follows that 
\eq{\mathfrak{B}_p^{(d)}=\!\!\fwbox{22pt}{\fwbox{15pt}{\mathfrak{b}^p_p}}\bigger{\oplus}\fwboxL{22pt}{\widehat{\fwboxL{15pt}{\mathfrak{b}^p_{p{+}1}}}}\bigger{\oplus}\cdots\bigger{\oplus}\fwboxL{22pt}{\widehat{\fwboxL{15pt}{\mathfrak{b}^p_{d-1}}}}\bigger{\oplus}\fwboxL{22pt}{\widehat{\fwboxL{15pt}{\mathfrak{b}^p_{d}}}}\,\quad\text{for}\quad p\!<\!d\,. }
We can prove this claim by simply constructing a general formula for  $\mathrm{rank}\big(\fwboxL{15pt}{\smash{\widehat{\fwboxL{13pt}{\rule{0pt}{8pt}\smash{\mathfrak{b}_{q}^{\raisebox{-5pt}{{\scriptsize $p$}}}}}}}}\big)$. 
\eq{
\label{eq:q_gon_pgonPC_rank_decomp}
\begin{split}
{\color{totalCount}\mathfrak{d}_d^p[q]}&\equivR\mathrm{rank}\big(\mathfrak{b}_{q}^p\big)=\mathrm{rank}\big(\ellK{q-p}_d\big)\,,\\
&\equivL\underbrace{\fwbox{0pt}{\phantom{\sum_{i>0}}}{\color{topCount}\widehat{\mathfrak{d}}_d^p[q]}}_{\text{{\color{topCount}top rank}}}+\underbrace{\sum_{i>0}\binom{q}{i}\widehat{\mathfrak{d}}_d^p[q-i]}_{\text{{\color{black}contact-term rank}}}\;,\quad\text{with}\;\;\;\widehat{\mathfrak{d}}_d^p[p]\equivR1\,,\;\; \widehat{\mathfrak{d}}_d^p[q\!<\!p]\equivR0\,,
\end{split}}
where we have recursively defined
\eq{{\color{topCount}\widehat{\mathfrak{d}}_d^p[q]}\equivR\mathrm{rank}\big(\fwboxL{15pt}{\smash{\widehat{\fwboxL{15pt}{\rule{0pt}{8pt}\smash{\mathfrak{b}_{q}^{\raisebox{-5pt}{{\scriptsize $p$}}}}}}}}\big)\,.}
This formula can be viewed as one for constructing particular bases for subspaces of $\mathfrak{b}_{q}^p$ for its contact-terms---elements involving some subsets of the $q$ propagators of the $q$-gon---assuming that these subspaces are all independent, which ensures the veracity of (\ref{eq:q_gon_pgonPC_rank_decomp}) as a statement about ranks. It turns out to be relatively straight forward to solve the recurrence relation (\ref{eq:q_gon_pgonPC_rank_decomp}) for $\widehat{\mathfrak{d}}_d^p[q]$; doing so, we find
\eq{{\color{topCount}\widehat{\mathfrak{d}}_d^p[q]}=\binom{d\mi p}{d\mi q}+\binom{d\mi p\mi 1}{d\mi q} = \frac{d\pl q\mi 2p}{q\mi p}\binom{d\mi p\mi 1}{d\mi q}\,\quad\text{for}\quad p\!\leq\!q\!<\!d\,\text{ or }\,p\!<\!q\!\leq\!d.\label{binomial_formula_for_hatted_ranks}}
%

%================================================================================================================
\vspace{-0pt}\subsubsection{Stratifying Bases of One-Loop Integrands}\label{subsubsec:stratifying_one_loop_bases}\vspace{-0pt}
%================================================================================================================

There is one more extremely useful set of inclusions to consider related to these integrand bases: $\mathfrak{B}_{p}\!\subset\!\mathfrak{B}_{p-1}$ for all $p$. Moreover, we can see that this holds true topology-by-topology: $\mathfrak{b}_{q}^p\!\subset\!\mathfrak{b}_q^{p-1}$ in general. This follows trivially from the fact that $\ellK{a}\!\subset\!\ellK{a+1}$ (see eq.~\ref{eq:ell_inclusion}), and has the interpretation that we are always able to express any integrand with good ultraviolet behavior (high $p$) in terms of those with worse ultraviolet behavior (lower $p$). 

To discuss the \emph{new} degrees of freedom as the power-counting of the basis worsens (as $p$ decreases), we may define
\eq{\fwboxL{14pt}{\smash{\widehat{\rule{0pt}{9.5pt}\smash{\widehat{\fwboxL{14pt}{\rule{0pt}{8pt}\smash{\mathfrak{b}_{q}^{\raisebox{-5pt}{{\scriptsize $p$}}}}}}}}}}\equivR\fwboxL{14pt}{\smash{\widehat{\fwboxL{14pt}{\rule{0pt}{9.5pt}\smash{\mathfrak{b}_{q}^{\raisebox{-5pt}{{\scriptsize $p$}}}}}}}}\backslash
\fwboxL{14pt}{\smash{\widehat{\fwboxL{20pt}{\rule{0pt}{9.5pt}\smash{\mathfrak{b}_{q}^{\raisebox{-5pt}{{\scriptsize $p{+}1$}}}}}}}}\,\hspace{5pt},
}
---the vector-space of (contact-free) degrees of freedom with $p$-gon power-counting not expressible by numerators with $(p\pl1)$-gon power-counting. Similarly, we can discuss the ranks of these spaces by defining
\eq{\fwboxL{15pt}{\smash{\widehat{\rule{0pt}{7.5pt}\smash{\widehat{\fwboxL{12pt}{\rule{0pt}{6pt}\smash{\hspace{4pt}\mathfrak{d}_{d}^{\raisebox{-10pt}{{\scriptsize $p$}}}}}}}}}}[q]\equivR\mathrm{rank}\big(\fwboxL{14pt}{\smash{\widehat{\rule{0pt}{9.5pt}\smash{\widehat{\fwboxL{14pt}{\rule{0pt}{8pt}\smash{\mathfrak{b}_{q}^{\raisebox{-5pt}{{\scriptsize $p$}}}}}}}}}}\big)={\color{topCount}\widehat{\mathfrak{d}}_d^p[q]}\mi{\color{topCount}\widehat{\mathfrak{d}}_d^{p+1}[q]}\,
}
where the last equality follows from (\ref{binomial_formula_for_hatted_ranks}) and requires that $p\!\leq\!q\!<\!d$, or $p\!<\!q\!\leq\!d$. Interestingly, it is not hard to see that
\eq{\fwboxL{15pt}{\smash{\widehat{\rule{0pt}{7.5pt}\smash{\widehat{\fwboxL{12pt}{\rule{0pt}{6pt}\smash{\hspace{4pt}\mathfrak{d}_{d}^{\raisebox{-10pt}{{\scriptsize $p$}}}}}}}}}}[q]={\color{topCount}\widehat{\mathfrak{d}}_{d{-}1}^p[q]}\,,}
which illustrates how integrand bases can be constructed by iteratively shifting the spacetime dimension upward.

%================================================================================================================
\newpage\vspace{-0pt}\subsubsection{The Vices and Virtues of Conformality \texorpdfstring{\newline(`$d$-gon' Power-Counting in $d$ Dimensions)}{\emph{d}-gon Power-Counting in \emph{d} Dimensions}}\label{subsubsec:conformality_one_loop}\vspace{-0pt}
%================================================================================================================
%
In section \ref{subsubsec:one_loop_contact_term_decomposition}, we have alluded to the fact that something special happens in the decomposition of integrand bases into contact-terms when the power-counting parameter $p$ coincides with the spacetime dimension---when $p\!=\!d$. In this case, the loop integrands (when combined with the measure $d^d\ell$) become invariant under the rescaling of loop momenta $\ell\!\to\!\alpha\,\ell$ and we refer to this property as `conformality'. From the integrand perspective,  we would like to explain in a little more detail why we now have to include basis elements with $d\pl1$ loop dependent propagators which were completely reducible when $p\!<\!d$. 

Before discussing the general case for arbitrary integer $d$, it is perhaps instructive to consider the concrete case of $d\!=\!4$ spacetime dimensions. Following our one-loop discussion for the vector-space of $q$-gon integrals with $(q\mi p)$ numerator insertions of generalized inverse propagators in eq.~(\ref{eq:pgon_pc_q_gon_integrand_def_1L}), box (or 4-gon) power-counting implies that all integrands with five propagators can have a single loop-momentum dependent numerator insertion schematically written as
\eq{\mathfrak{b}^{4}_{5} = \frac{[\squeezeD \ell\squeezeD]_4}{(\ell|Q_1)(\ell|Q_2)(\ell|Q_3)(\ell|Q_4)(\ell|Q_5)}\,,} 
where the numerator $[\squeezeD \ell\squeezeD]_4$ contains {\color{totalCount}6} total degrees of freedom. Of course, any choice of this six dimensional basis is na\"ively as good as any other. However, as suggested in eq.~(\ref{eq:contact_decomposition_abstract_1L}), we find it most desirable to decompose the basis according to propagator topologies. Such a basis decomposition is extremely natural in conjunction with generalized unitarity and is also employed in the OPP setup \cite{Ossola:2006us,Mastrolia:2010nb}. We are led to the following basis choice
\eq{[\squeezeD \ell\squeezeD]_4= \text{span} \big\{ {\color{topCount}\epsilon(\ell, Q_1,Q_2,Q_3,Q_4,Q_5)},{\color{dim}(\ell|Q_1)},{\color{dim}(\ell|Q_2)},{\color{dim}(\ell|Q_3)},{\color{dim}(\ell|Q_4)},{\color{dim}(\ell|Q_5)}\big\} \label{1loop_box_pc_parity_basis}}
which contains the five inverse propagators $(\ell|Q_i)$ together with the dual of these five generated by the six-dimensional epsilon symbol. Even though the explicit form of this numerator is not relevant for us here, note that ${\color{topCount}\epsilon(\ell, Q_1,Q_2,Q_3,Q_4,Q_5)}$ can either be conveniently evaluated by going to embedding space \cite{Weinberg:2010fx,SimmonsDuffin:2012uy} or by writing an expansion in inverse propagators with additional (complex) momenta. The numerators that are proportional to the inverse propagators $(\ell|Q_i)$ give rise to scalar box contact term topologies, \emph{i.e.}~integrands that can be obtained from the pentagon by pinching one of the propagators. One advantage of the basis choice in (\ref{1loop_box_pc_parity_basis}) is that spacetime parity of each element is manifest: the scalar boxes give rise to parity-even integrands and the pentagon with $\epsilon$-insertion is parity-odd. Since the parity-odd pentagon vanishes upon integration over the parity-even Feynman contour, this element is often neglected if one is interested in properties of integrated scattering amplitudes.\\

There are a few advantages of the box power-counting basis in four dimensions at one loop. In the context of maximally supersymmetric ($\mathcal{N}\!=\!4$) Yang-Mills theory (sYM) \cite{Brink:1976bc,Gliozzi:1976qd} in the planar limit (taking $N\!\to\!\infty$ for an $\mathfrak{su}_{N}$ gauge theory), box power-counting makes dual conformal invariance \cite{Drummond:2006rz,Alday:2007hr,Drummond:2008vq} manifest at the integrand-level and displays the inherently good UV behavior of the theory. At one loop, there is a similar statement about the surprisingly good UV behavior of supergravity theories (dubbed the `no-triangle hypothesis' \mbox{\cite{Bern:2006kd,BjerrumBohr:2006yw,BjerrumBohr:2008ji}}) which, however, fails at high enough loop-order or for sufficiently many external particles (starting at two loops \cite{Bourjaily:2018omh}). From an integration perspective, box power-counting implies a low numerator polynomial rank which is advantageous for \emph{integral reduction} and direct integration as well (see \emph{e.g.}~\mbox{\cite{Bourjaily:2013mma,Bourjaily:2018aeq,Bourjaily:2019igt,Bourjaily:2019jrk,Bourjaily:2019vby}}).

On the other hand, a substantial downside of the box power-counting basis in (\ref{1loop_box_pc_parity_basis}) is that it is \emph{over-complete}. For details of the argument, see section 2 of \cite{Bourjaily:2017wjl}; but the bottom line is that the parity-odd pentagons are not all independent, but satisfy \emph{integrand}-level relations that must be eliminated in order to specify a complete and not over-complete basis. 

There are many ways to see this redundancy among pentagon integrands. One way would be analogous to what we saw for the decomposition of $3$-gon power-counting box integrands for $d\!=\!1$ below eq.~(\ref{eq:contact_decomposition_abstract_1L}) above. Specifically, we may consider the decomposition of a hexagon integrand with box power-counting in four dimensions. Our analysis above (see eqs.~(\ref{eq:q_gon_pgonPC_rank_decomp}) and (\ref{pnum_ddim_rank})) shows that the total rank of the numerators for an integrand in the space $\mathfrak{b}_6^4$ is ${\color{totalCount}\mathfrak{d}_4^4[6]}\!=\!\mathrm{rank}(\ellK{2}_4)\!=\!{\color{totalCount}20}$. Decomposing this space into contact-terms would give us $\smash{\binom{6}{5}}\!\times\!{\color{topCount}1}$ {\color{topCount}top-level} degrees of freedom for contact-term pentagons, and $\smash{\binom{6}{2}}\!\times\!{\color{topCount}1}$ {\color{topCount}top-level} degrees of freedom for contact-term boxes; if these contact-term contributions to the hexagon numerators were all independent, they would span a $6\pl15\!=\!21$-dimensional space. Thus, there must be one redundancy (and it is easy to see that the scalar box integrands are always independent in $d\!=\!4$). Thus, the hexagon's six contact-term, box-power-counting-pentagon integrands must satisfy one algebraic relation (in order to span a merely five-dimensional space).

Besides eliminating the redundancy by picking an arbitrary subset of independent parity-odd pentagons, there is another way to proceed that is also most relevant to represent more general quantum field theories at one-loop and has already led to numerous fruitful results for higher-loop amplitudes even in sYM \cite{Bourjaily:2019iqr,Bourjaily:2019gqu}. As we have argued in section \ref{subsubsec:one_loop_contact_term_decomposition}, increasing the power-counting from box (4-gon) to triangle (3-gon) scaling is sufficient to eliminate (at \emph{integrand} level) the parity-odd pentagon entirely. In these cases, one is left with triangle-power-counting, chiral boxes and scalar triangle integrands.\\

\newpage
Our four-dimensional discussion can be easily generalized to any integer dimension $d$ where we run into exactly the same issue. In $d$ dimensions, the $d$-gon power-counting numerator of a $(d\pl1)$-gon involves exactly one insertion of generalized inverse propagators. By eq.~(\ref{eq:ell_d_dimensionality}), this space is $(d\pl2)$-dimensional and can be spanned by the $(d\pl1)$ inverse propagators of the $(d\pl1)$-gon together with one parity-odd integrand dual to these. At higher points, these parity-odd integrands always satisfy linear relations analogous to what we described above. To specify a complete and non-redundant basis would require us to eliminate certain permutations of the parity-odd integrands, which is undesirable from a symmetry point of view. 

%================================================================================================================
\vspace{-0pt}\subsection{One-Loop Integrands for Theories in Various Dimensions}\label{subsec:arbitrary_dim_version_of_one_loop}\vspace{-0pt}
%================================================================================================================
%
In this part of our work, we would like to take a moment to illustrate the more abstract definitions of subsections \ref{subsec:one_loop_reduction_and_stratification} and \ref{subsec:p_gon_power_counting_bases} with some concrete examples. In particular, we first discuss four-dimensional integrand bases at one loop in subsection \ref{subsec:one_loop_four_dim_redux} and generalize this setup in two ways. First, we extend the four-dimensional analysis to arbitrary integer spacetime dimension $d$ in \ref{subsubsec:one_loop_for_arbitrary_integer_dim}. Second, more nontrivially, we extend our discussion to the realm of dimensional regularization---to build integrand bases within $\epsilon$ of an integer dimension $d$ in \ref{subsubsec:one_loop_for_arbitrary_regulated_integer_dim} where it becomes important how we define the relevant extra-dimensional components of the loop momenta (\emph{i.e.}~the precise definition of so-called $\mu$-terms). 

%================================================================================================================
\vspace{-0pt}\subsubsection{Bases for Theories Defined in Four Spacetime Dimensions}\label{subsec:one_loop_four_dim_redux}\vspace{-0pt}
%================================================================================================================
%

In this subsection, we briefly discuss the specific and particularly relevant case of $d\!=\!4$-dimensional integrands to make some of the general statements above more concrete. As mentioned above, all these results are well-known \cite{Passarino:1978jh,Ossola:2006us,Mastrolia:2010nb} and have long been put to use in the generalized unitarity program \cite{Bern:1994zx,Bern:1994cg,Britto:2004nc}. Nonetheless, we find it valuable to review here. (Our discussion here follows an earlier exposition in sections 2 and 4 of ref.~\cite{Bourjaily:2017wjl}.) We start by specializing eqs.~(\ref{definition_of_single_tensor}) and (\ref{naive_basis_of_single_inverse_props}) to $d=4$\footnote{In the four dimensions, there are a number of alternate ways (such as momentum twistors \cite{Hodges:2009hk} or the embedding space formalism \cite{Weinberg:2010fx,SimmonsDuffin:2012uy}) to represent this setup; but we prefer to use the general notation introduced in this section.}
\eq{[\squeezeD \ell\squeezeD]_{4} = \Span{Q\in\mathbb{R}^4}\{\x{\ell}{Q}\} \simeq \mathrm{span}\{1,\ell\!\cdot\! \hat{e}_1,\ldots,\ell\!\cdot\!\hat{e}_4,\ell^2\}\,, \label{single_tensor_basis_equality}}
which forms a six-dimensional vector-space spanned by (generalized) inverse propagators. When discussing four-dimensional numerators, we often drop the explicit indication of dimensionality. One consequence of (\ref{single_tensor_basis_equality})---in line with the general discussion above---is that any monomial $\{1,\ell\!\cdot\! \hat{e}_1,\ldots,\ell\!\cdot\!\hat{e}_4,\ell^2\}$ can be expanded in the six-dimensional space of inverse propagators. Importantly, this implies that (independent of the field-theory power-counting) any integrand with six and more propagators is expressible in terms of integrands with five and fewer propagators.

%================================================================================================================
\newpage\vspace{-0pt}\subsubsection*{Organizing Four-Dimensional Integrand Bases by Contact-Terms}\label{subsubsec:p_gon_power_counting_bases_contact_decomp_in_four_dim}\vspace{-0pt}
%================================================================================================================

To illustrate the structure of higher power-counting bases at one loop in four spacetime dimensions, we follow the discussion of section 4 of \cite{Bourjaily:2017wjl}. The simplest extension of the box power-counting basis is to allow one additional loop-dependent numerator insertion so that all basis elements scale as $\sim 1/(\ell^2)^3$ at large $\ell$. Thus, we may start with the following three structures $\mathfrak{B}_3^{(4)}\!=\!\text{span}\big\{\mathfrak{b}^3_3\!\oplus\!\mathfrak{b}^3_4\!\oplus\!\mathfrak{b}^3_5\big\}$; in the notation of eq.~(\ref{eq:pgon_pc_q_gon_integrand_def_1L}):
\begin{align*}
\left\{ \frac{[\squeezeD \ell\squeezeD]^0_4}{(\ell|Q_1)(\ell|Q_2)(\ell|Q_3)}\,, 
\frac{[\squeezeD \ell\squeezeD]^1_4}{(\ell|Q_1)(\ell|Q_2)(\ell|Q_3)(\ell|Q_4)}\,, 
\frac{[\squeezeD \ell\squeezeD]^2_4}{(\ell|Q_1)(\ell|Q_2)(\ell|Q_3)(\ell|Q_4)(\ell|Q_5)}
\right\}\,.
\end{align*}

As explained above (see also \cite{Bourjaily:2017wjl}), it turns out that the pentagon \emph{integrands} are not independent of the box- and triangle integrands and one is able to eliminate all topologies with five or more propagators. This is due to the fact that in four dimensions, the rank of the space of two loop-momentum insertions according to (\ref{pnum_ddim_rank}) and (\ref{eq:q_gon_pgonPC_rank_decomp}) is ${\color{totalCount}\mathfrak{d}^{3}_4[5]}\!=\!\mathrm{rank}\big(\ellKd{2}{4}\big)\!=\!{\color{totalCount}20}$; this {\color{totalCount}20}-dimensional space can be fully spanned by contact-terms---namely, $\smash{\binom{5}{1}}\!\times\!{\color{topCount}2}$ degrees of freedom from contact-term box integrands and $\smash{\binom{5}{2}}\!\times\!{\color{topCount}1}$ degrees of freedom from contact-term triangles. The full triangle-power-counting basis can therefore be spanned by box integrands (with non-trivial numerators) as well as scalar triangle integrands. 

A particularly convenient choice of basis for the 3-gon power-counting boxes are the so-called `chiral' numerators (which turn out to be incredibly useful for matching unitarity cuts). Choosing this particular basis, we attribute the {\color{topCount}2} {\color{topCount}top-level} degrees of freedom to each box integrand according to:
\begin{align}
\begin{split}
   \mathfrak{b}^3_4 & = \frac{[\ell]}{(\ell|Q_1)(\ell|Q_2)(\ell|Q_3)(\ell|Q_4)}\,, 
    \\
   \fwboxR{0pt}{\text{with}\hspace{10pt}} [\ell]_4 &= \text{span}\big\{
    {\color{topCount}(\ell|\ell^\ast_{1})},{\color{topCount}(\ell|\ell^\ast_{2})},
    {\color{dim}(\ell|Q_1)},{\color{dim}(\ell|Q_2)},{\color{dim}(\ell|Q_3)},{\color{dim}(\ell|Q_4)}\big\}\,,
\end{split}    
\end{align}
where $\ell^\ast_{i}$ are the two solutions to the quadruple-cut equations $(\ell|Q_1){=}{\cdots}{=}(\ell|Q_4){=}0$.\\

Having discussed the 3-gon power-counting basis, we can easily extend the analysis to higher numerator ranks, including $0$-gon power-counting. Since bases with lower power-counting are always subsets of bases with higher power-counting (in the notation of section \ref{subsubsec:stratifying_one_loop_bases}; $\mathfrak{B}_{p+1}\!\subset\!\mathfrak{B}_p$ for any $p$) we now know that for any $p\!<\!4$-gon power-counting, the bases of one-loop integrands are spanned by box-, triangle-, bubble-, tadpole topologies and potentially polynomial terms without propagators. (Polynomial terms do not play a role for integrated amplitudes in dimensional regularization as they would give rise to power-divergences which are set to zero, see \emph{e.g.}~\cite{Collins:1984xc}.) We can therefore simply list the dimensionality of the bases of integrands for a given power-counting in strictly four spacetime dimensions; this can be found in Table~\ref{tab:one_loop_4d_bases}.

\begin{table}[ht!]\vspace{-10pt}$$\hspace{-1pt}\vspace{-0pt}\fig{-100pt}{1}{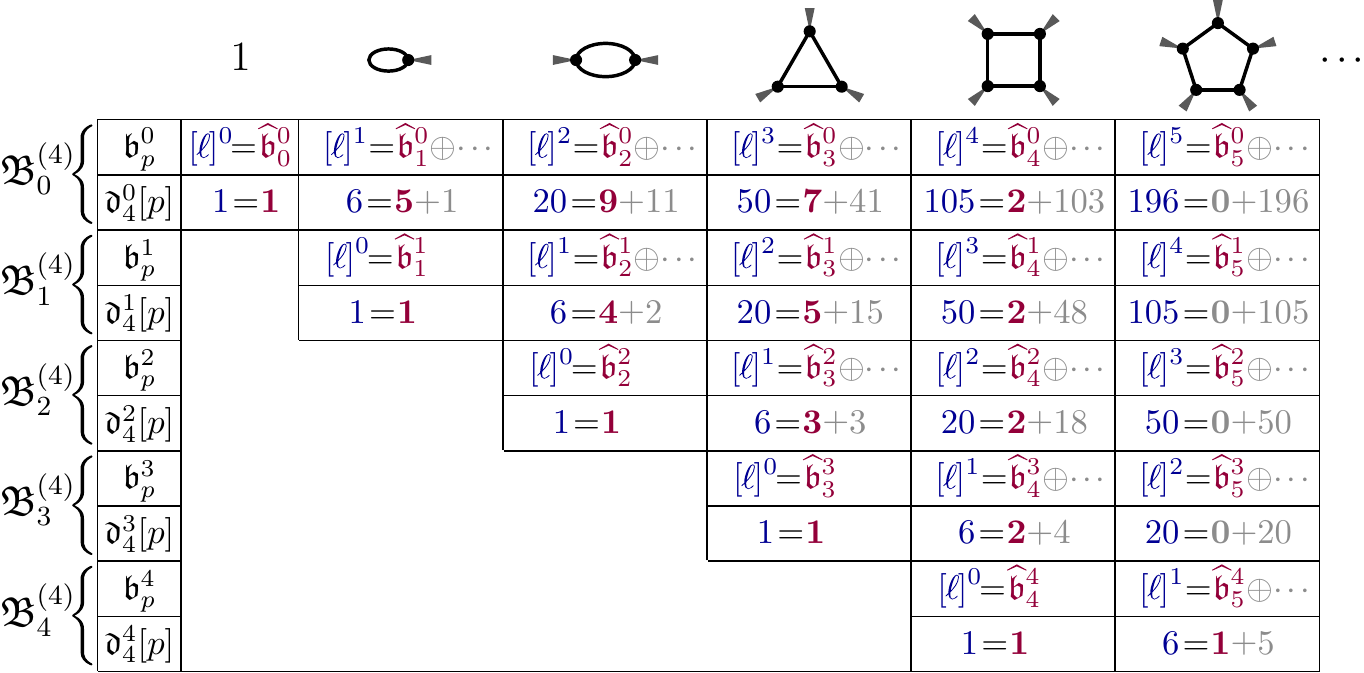}\vspace{-20pt}$$
\caption{One-loop integrand bases in 4 dimensions with various $p$-gon power-counting.\label{tab:one_loop_4d_bases}}\vspace{-0pt}\end{table}

Note that all numbers in Table~\ref{tab:one_loop_4d_bases} are dictated by the general formula for the rank of symmetric traceless tensors of $\mathfrak{so}_6$, consistent with eq.~(\ref{pnum_ddim_rank}). As indicated in the table, we can split the overall ranks associated to a given basis topology according to `\hyperlink{parent}{parent}' and `\hyperlink{daughter}{daughter}' (or `contact-term') degrees of freedom as has been advocated for in eq.~(\ref{eq:q_gon_pgonPC_rank_decomp}). In this split, by parent degrees of freedom, we mean integrand basis elements that can be fixed in a generalized unitarity setup at the level of maximal cuts \cite{Bern:2007ct} of the corresponding topology. Likewise, `daughter' or `contact-term' degrees of freedom correspond to integrand basis elements associated to certain pinched topologies whose coefficients can be fixed using unitarity cuts of the pinched topology. For more details, we refer the interested reader to our original exposition of the split of integrand degrees of freedom in section 4 of \cite{Bourjaily:2017wjl}.

%================================================================================================================
\vspace{-0pt}\subsubsection{Bases for Theories Defined in Various Spacetime Dimensions}\label{subsubsec:one_loop_for_arbitrary_integer_dim}\vspace{-0pt}
%================================================================================================================

Based on our 4-dimensional discussion in \ref{subsec:one_loop_four_dim_redux}, it should be quite clear how to extend the basis construction to arbitrary integer $d$ spacetime dimensions. Let us therefore only briefly summarize the corresponding results. Having fixed the spacetime dimension, there is one additional figure of merit required to define a basis of integrands, namely the desired power-counting. 

We can then look at different power-counting bases. Similar to the 4-dimensional case with $4$-gon power-counting (see eq.~(\ref{1loop_box_pc_parity_basis}) we discussed in section \ref{subsubsec:conformality_one_loop}, there are irreducible $(d\pl1)$-gon integrands that can be chosen to be parity-odd together with parity-even $d$-gon integrands). Again, na\"ively this forms an \emph{over-complete} basis and one is forced to eliminate linear relations between different parity-odd $(d\pl1)$-gons.

Exactly like the 4-dimensional case, boosting the power-counting to $(d\mi1)$-gon power-counting and higher (more loop-momentum dependence in the numerator) allows us to eliminate this redundancy and one is left with a complete (and not over-complete) basis of integrands where each integrand basis element involves up to $d$ propagators. We can equally well split the resulting integrand degrees of freedom into `\hyperlink{parent}{parent}' and `\hyperlink{daughter}{daughter}' basis elements associated to the maximal number of propagator topologies and various pinched topologies. The relevant data of the size of the $d$-dimensional integrand bases of $p$-gon power-counting is dictated by eq.~(\ref{pnum_ddim_rank}) and we summarize the relevant results in Table~\ref{tab:power_counting_table_1L}.\\~\\~\\\vspace{50pt}~

\begin{table}[h!]\vspace{-10pt}$$\hspace{-1pt}\vspace{-0pt}\fig{-100pt}{1}{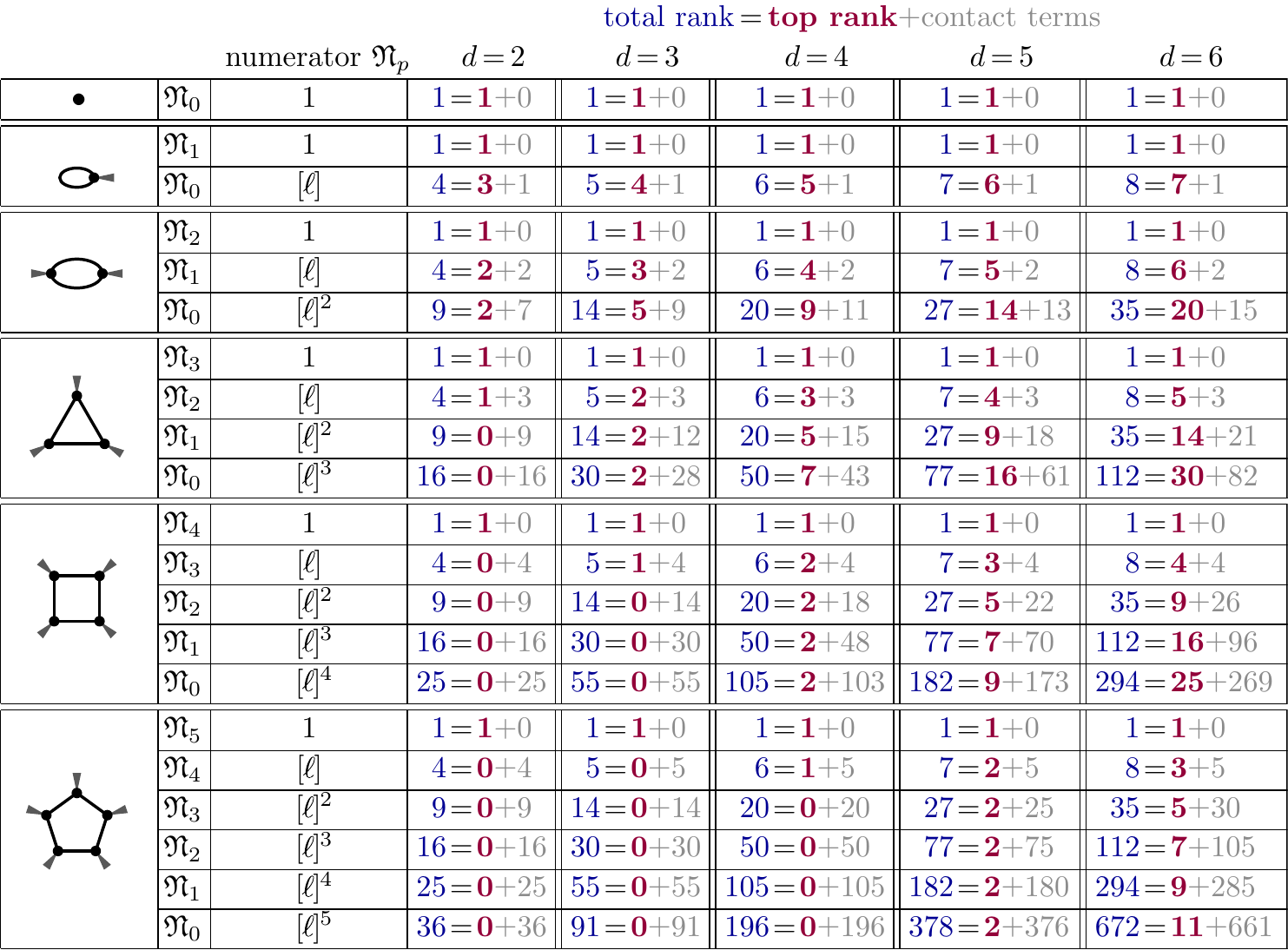}\vspace{-20pt}$$
\caption{One-loop degrees of freedom for $p$-gon power-counting in 2--6 dimensions.\label{tab:power_counting_table_1L}}\vspace{-10pt}\end{table}

%================================================================================================================
\newpage\vspace{-0pt}\subsubsection{Bases for Theories Defined within \texorpdfstring{$\epsilon$}{eps} of an Integer Dimension}\label{subsubsec:one_loop_for_arbitrary_regulated_integer_dim}\vspace{-0pt}
%================================================================================================================
%
So far we have been discussing one-loop integrand bases in integer dimensions, $d\!\in\!\mathbb{N}_+$. Even though this integer-dimensional basis counting is interesting in its own right, from a practical point of view it may be desirable to have an analogous construction in the context of dimensional regularization.\footnote{Note that we work in a dimensional regularization scheme where the external states must lie in strictly-integer-$d$ spacetime dimensions. Examples of such schemes are the `t Hooft Veltman scheme \cite{tHooft:1972tcz}, the dimensional reduction scheme \cite{Siegel:1979wq}, and the four-dimensional helicity scheme \cite{Bern:2002zk}, but \emph{not} the conventional dimensional regularization scheme \cite{Collins:1984xc}, where all momenta and states are continued away from an integer dimension.} To this end, the first goal will be to upgrade the definitions of eqs.~(\ref{definition_of_single_tensor}) and (\ref{naive_basis_of_single_inverse_props}) to include the extra-dimensional loop-momentum components. In particular, we are interested in a situation, where loop momenta are defined in $d\mi2\epsilon$ spacetime dimensions and all external particle momenta and polarizations are defined in strictly-integer-$d$ spaceitme dimensions (see \emph{e.g.}~\cite{Bern:2000dn}). Therefore, we schematically decompose each loop momentum $\ell_i$ (with an eye towards possible generalizations to higher loops) according to 
\eq{\ell_i \equivL \hat{\ell}_{i} + \vec{\mu}_i\qquad\text{with }\quad\hat{\ell}_i\!\in\!\mathbb{R}^d\;\;\;\text{and}\;\;\;\vec{\mu}_i\!\in\!\mathbb{R}^{-2\epsilon}}
where the extra-dimensional components are orthogonal to the integer-dimensional loop momentum, $\hat{\ell}_{i}\!\cdot\!\vec{\mu}_j\!=\!0$ for all $i,j$. Moreover, requiring that the external states live in the integer-dimensional space, we have that $\vec{\mu}_i\!\cdot\!p_a\!=\!0$ for all external momenta (and similarly for external polarizations). Thus, the only new, extra-dimensional Lorentz invariants that appear in integrand construction would be the so-called `$\mu$-terms' defined via
\eq{ \vec{\mu}_i\!\cdot\!\vec{\mu}_j \equivL -\mu^2_{ij}\,. \label{eq:mu_def_l_dot_l}}
(When discussing one loop, we will drop these indices---leaving only $\mu^2$.)

In the context of generalized unitarity, it has been known for some time (see \emph{e.g.}~the discussion in \cite{Badger:2013gxa}) that the extra-dimensional pieces of amplitude integrands can be obtained from unitarity cuts in higher, integer-dimensional spacetimes. This of course raises the question of how many extra dimensions are required; but as we will see, it also raises important questions about how to organize integrands by power-counting in order to represent integrands in various quantum field theories. 

To make the stakes clear, we may ask: if it were known that a theory's unregulated amplitude integrands are representable in terms of $p$-gon power-counting integrands in $d$ dimensions, is it true that, when regulated in dim-reg, this theory's amplitude integrands are \emph{still} representable in a $p$-gon power-counting basis? For example, it is known that unregulated amplitude integrands in (for the present argument, say planar) sYM are representable in terms of integrands with box power-counting; once regulated, is this still true?\\

For the sake of concreteness and illustration, let us first restrict ourselves to the case of one loop. For integrands in $d\mi2\epsilon$ dimensions the $\mu$-terms can be obtained from a ($d\pl1$)-dimensional analysis by noting that $\vec{\mu}$ must span a \emph{single}-dimensional space spanned by the basis element `$\hat{e}_{\mi2\epsilon}$': for one loop, we may identify $\vec{\mu}\equivR\ell\!\cdot\!\hat{e}_{-2\epsilon}$.  Notice that this is structurally identical to merely assuming that $\hat{e}_{\mi2\epsilon}$ spans one additional spacetime direction. Thus, we know that any $d$-dimensional theory's amplitudes may be represented in dimensional regularization in terms of a basis of $(d\pl1)$-dimensional loop integrands. Are the regulated amplitudes of sYM representable in terms of a $p\!=\!4$ power-counting basis in $5$-dimensions? 

If we were to use the na\"ive definition of power-counting of eq.~(\ref{naive_one_loop_uv_scaling}), then the Lorentz-invariant monomial $\mu^2$ would scale identically as $\ell^2\!=\!\hat{\ell}^2\mi\mu^2$. However, as argued at the end of \mbox{section \ref{subsubsec:one_loop_tensor_numerators}}, our space would treat these two cases differently---as $\mu^2\!\notin\!\ellK{}$ while $(\hat{\ell}^2\mi\mu^2)\!\in\!\ellK{}$. In our setup, $\mu^2\!\in\!\ellK{2}$, so that we would declare integrands involving this bare monomial to have worse power-counting. The fact that $\mu^2$ should be considered to have worse power-counting than $(\hat{\ell}^2\mi\mu^2)$ is not merely semantic: the space $\ellK{}_{d\pl1}$ involves $\mu^2$, but \emph{only} in a very specific (and very precise) way: in combination with $\hat{\ell}^2$. 

All existing representations of dimensionally-regulated one-loop amplitude integrands in sYM specifically make use of the bare monomial $\mu^2$  (see \emph{e.g.}~\cite{Bern:1992em,Bern:1996je,Bern:1996ja}) times pentagons (and higher). As such, these representations do not satisfy our notion of having $p\!=\!4$ power-counting in $5$ dimensions. It would be highly non-trivial (and somewhat surprising) if our more restrictive definition of box power-counting could still be used to represent these amplitudes. In a forthcoming work \cite{1LtoAppear}, it will be shown that this is in fact the case: all regulated amplitude integrands of sYM \emph{can} be represented in the more restrictive space of $4$-gon power-counting integrands in 5 dimensions, as we have defined them here.\\

Beyond one loop, the new Lorentz invariants (\ref{eq:mu_def_l_dot_l}) would appear to require some $\smash{\binom{L\pl1}{2}}$ new degrees of freedom introduced into loop integrands. This could be achieved by simply constructing integrands in $d\pl\smash{\binom{L\pl1}{2}}$ spacetime dimensions. But surely this is overkill! Is it obvious that we cannot do better? It remains an important open---and directly answerable---question of what spacetime dimension is required to represent dimensionally-regulated amplitudes in any given theory. It would be very worthwhile to address this question using some of the known expressions for regulated amplitude integrands that exist in the literature for sYM \cite{Bern:1997nh,Bern:2005iz,Bern:2007hh,Bern:2008pv,Bern:2010tq,Carrasco:2011mn,Bern:2012uc}, supergravity \mbox{\cite{Bern:1997nh,Bern:2007hh,Bern:2008pv,Bern:2010tq,Carrasco:2011mn,Bern:2017ucb,Bern:2018jmv}}, or even the recently studied two-loop four-graviton amplitude in pure GR \cite{Abreu:2020lyk}.

%================================================================================================================
%    2. Two Loop Bases
%         
%================================================================================================================
\newpage\vspace{-6pt}\section{Building Bases of Integrands at Two Loops}
\label{sec:two_loops}\vspace{-0pt}
%================================================================================================================

The construction of loop integrand bases at two loops turns out to be considerably more subtle and interesting than at one loop. The principal reason for this is that there are fundamentally non-planar integrands, and hence no obviously preferential `routing' of loop momenta.\footnote{Recall that the \hyperlink{routing}{routing} of loop momenta corresponds to a particular solution to momentum conservation at every vertex, expressed as a function of $L$ loop momentum variables $\ell_i$.} Moreover, different choices of loop momentum routing can have a severe effect on how integrand bases would be stratified by more pragmatic approaches. We will see how this works in detail below.

In this section, we clarify our general basis-building strategy and apply our ideas to the case of two-loop bases built for theories defined in various spacetime dimensions. We will start in \mbox{section \ref{subsec:two_loop_general_structure}} with a generalization of our loop integrand formalism to two loops; in particular, we introduce notation for describing vector-spaces of generic two-loop Feynman integrals with loop-dependent numerators constructed from inverse propagators. In \mbox{section \ref{subsec:two_loop_opp_in_d_dimensions}} we apply these ideas to describe the `$0$-gon power-counting basis' of two loop integrands in $d$ dimensions, and we will show how to better understand and generalize the results of \cite{Feng:2012bm}.

%================================================================================================================
\vspace{-0pt}\subsection{Two-Loop Integrand Bases: Basic Building Blocks}\label{subsec:two_loop_general_structure}\vspace{-0pt}
%================================================================================================================
In this subsection, we first introduce the labeling of two-loop graphs in terms of their propagator structure, before examining the basic numerator building blocks in \ref{subsubsec:two_loop_tensor_structure}, both of which generalize our one-loop discussions above.

%================================================================================================================
\vspace{-0pt}\subsubsection{Loop-Dependent Denominators: Vacuum/Skeleton Graphs}\label{subsubsec:two_loop_vacuum_graphs}\vspace{-0pt}
%================================================================================================================

The topology of any two-loop Feynman graph can be characterized by three numbers:
\eq{\bigger{\twoLoopLadderNameText{a}{b}{c}\;\Leftrightarrow\;\;} \fig{-26pt}{0.3}{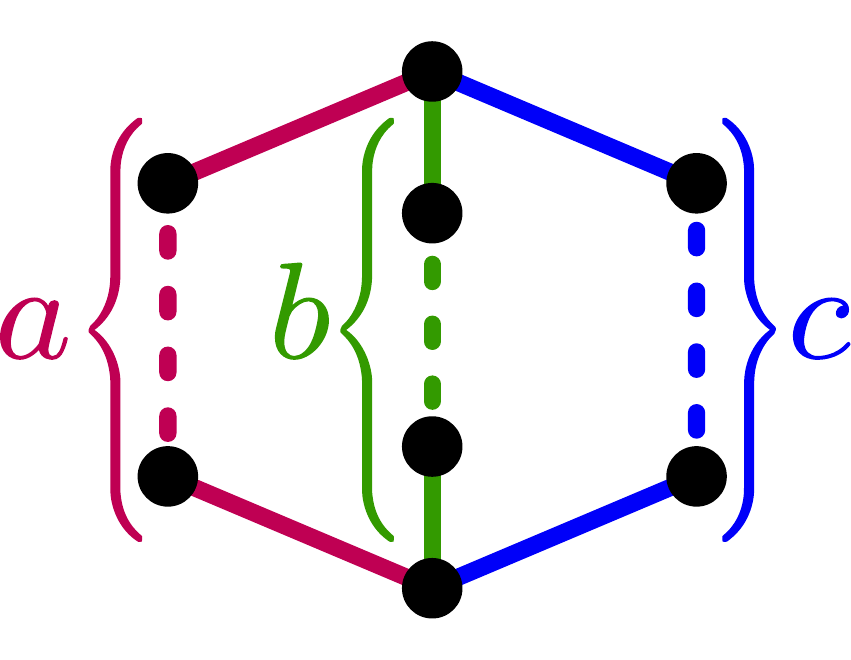}}
The numbers $\{{\color{emphA}a},{\color{emphB}b},{\color{emphC}c}\}$ indicate how many (loop-momentum-dependent-)propagators differ only by external momentum flowing into the graph. Just as there is no preferential {routing} for the undetermined loop momenta, these labels have no natural ordering. That being said, we choose to write representative graphs using the convention that ${\color{emphB}b}\leq{\color{emphC}c}\leq{\color{emphA}a}$, where the indices have been colored merely to direct the eye. None of the counting we perform in this section depends on this choice, it just gives us a convenient way to label graphs.  

It is not hard to see that these indices fully characterize the loop-dependent propagator structure of any two loop Feynman diagram, and similar conventions for such graphs have been used by others, see \emph{e.g.}~\cite{Gluza:2010ws,Badger:2013gxa}. To be clear, we allow any of these indices to vanish---corresponding to product topologies. Some simple examples of these include
\eq{
\twoLoopLadderNameText{1}{1}{1}\;\bigger{\;\Leftrightarrow\!\!\hspace{-9pt}}\twoLoopTableFigure{1}{1}{1}
\qquad
\twoLoopLadderNameText{4}{0}{4}\;\bigger{\;\Leftrightarrow\;\;}\twoLoopTableFigure{4}{0}{4}
\qquad
\twoLoopLadderNameText{4}{1}{3}\;\bigger{\;\Leftrightarrow\!\!}\twoLoopTableFigure{4}{1}{3}}
which have been called elsewhere `sunrise', `kissing-boxes', and `penta-box' integrands, respectively. 

In order to write a rational expression for $\twoLoopLadderNameText{a}{b}{c}$, we would require either (modest) redundancy or some (ephemeral) choice of loop momentum routing. Redundantly, we could choose to write
\eq{
\label{eq:two_loop_prop_labeling}
\twoLoopLadderNameText{a}{b}{c}\;\;\bigger{\Leftrightarrow}\frac{1}{\x{{\color{emphA}\ell_A}}{Q_{\color{emphA}1}}\cdots\x{{\color{emphA}\ell_A}}{Q_{\color{emphA}a}}\x{{\color{emphB}\ell_B}}{R_{\color{emphB}1}}\cdots\x{{\color{emphB}\ell_B}}{R_{\color{emphB}b}}\x{{\color{emphC}\ell_C}}{S_{\color{emphC}1}}\cdots\x{{\color{emphC}\ell_C}}{S_{\color{emphC}c}}}}
subject to the constraint that $\cca{\ell_A}+\ccb{\ell_B}+\ccc{\ell_C}\in\mathbb{R}^d$; or we may solve this condition of momentum conservation and eliminate one of the three classes of loop momenta. When required in the following, we choose to solve momentum conservation and associate $\cca{\ell_A}=\cca{\ell_1}$ with $\cca{a}$-type propagators, $\ccc{\ell_C}=\ccc{\ell_2}$ with $\ccc{c}$-type propagators, and $\ccb{\ell_B}=\ccb{\ell_1{-}\ell_2}$ with $\ccb{b}$-type propagators. Furthermore, unless otherwise specified, we assume all $Q_i$, $R_j$ and $S_k$ in eq.~(\ref{eq:two_loop_prop_labeling}) to be distinct; in a later discussion, however, we will drop this requirement and also allow for `doubled-propagator' graphs which can be relevant for two-loop amplitudes---depending on renormalization scheme (see \emph{e.g.}~\cite{Baumeister:2019rmh} for the absence of certain residues for integrals with doubled-propagators in the on-shell scheme). 

In $d$ spacetime dimensions, it should be clear from our one-loop discussion that we mostly need to consider integrand topologies with $d{+}1$ propagators of either ${\color{emphA}a},\ {\color{emphB}b}$, or ${\color{emphC}c}$ type, as any topology with more propagators can be trivially reduced by one-loop methods (at the cost of worsening the power-counting). As we will see shortly, for $p\!<\!d$-gon power-counting we can reduce the number of topologies relevant for two-loop integrands even further. In order to make this discussion more transparent, let us first mention the relevant building blocks for the numerator structures.

%================================================================================================================
\vspace{-0pt}\subsubsection{Loop-Dependent Numerators: Notation and Biases for Bases}\label{subsubsec:two_loop_tensor_structure}\vspace{-0pt}
%================================================================================================================

In order to discuss the general structure of two-loop numerators, we generalize our initial discussion of the fundamental one-loop numerator objects from section \ref{subsubsec:one_loop_tensor_numerators}. We have argued that it is most natural to express loop-dependent numerators in terms of generalized inverse propagators
\eq{[\ell]_d = \text{span}_Q\left\{(\ell|Q)\right\}\,, \qquad Q\in \mathbb{R}^d} 
with loop-momentum independent coefficients. Following our discussion of the two-loop propagator structure in the previous subsection, it is natural to define the associated two-loop numerator building blocks
\begin{align}
\begin{split}
[{\color{emphA}\ell_A}]_d &\equivR \text{span}_{Q}\!\left\{(\cca{\ell_A}|Q)\right\} 
            (\simeq \text{span}\{1,\cca{\ell_A}\!\cdot\!\hat{e}_1,\ldots, 
                                \cca{\ell_A}\!\cdot\! \hat{e}_d, \cca{\ell^2_A}\})\,, \\
[{\color{emphB}\ell_B}]_d &\equivR \text{span}_{R}\left\{(\ccb{\ell_B}|R)\right\} 
            (\simeq \text{span}\{1,\ccb{\ell_B}\!\cdot\!\hat{e}_1,\ldots,
                                \ccb{\ell_B}\!\cdot\!\hat{e}_d, \ccb{\ell^2_B}\})\,, \\
[{\color{emphC}\ell_C}]_d &\equivR \text{span}_{S\,}\left\{(\ccc{\ell_C}|S)\right\} 
            (\simeq \text{span}\{1,\ccc{\ell_C}\!\cdot\!\hat{e}_1,\ldots,
                                \ccc{\ell_C}\!\cdot\!\hat{e}_d, \ccc{\ell^2_C}\}) \,\,.   
\end{split}            
\end{align}
Since individual numerator polynomials $[\ell_i]_d^k$ are basically one-loop objects, we may refer back to our one-loop discussions for more details on their properties. 

In the following section, we will discuss integrand bases (with `$0$-gon power-counting') consisting of graphs $\twoLoopLadderNameText{a}{b}{c}$ with numerator-spaces constructed directly as products of these factors---for example,  
\eq{\emphEdges
\twoLoopLadderNameText{3}{1}{3} \Leftrightarrow 
\begin{tikzpicture}[scale=\figScale,baseline=-3.05]\twoLoopGraph{3}{1}{3}{1.25};
\end{tikzpicture}\quad\text{with numerator  }\mathfrak{N}_0\!\big(\!\twoLoopLadderNameText{3}{1}{3}\big)=[\cca{\ell_A}]^3[\ccb{\ell_B}]^1[\ccc{\ell_C}]^3\,.
}
However, one novelty that will arise when we consider bases with better power-counting is that we must also consider vector-spaces of numerators constructed as sums of these objects. For example, we may consider an integrand 
\eq{\emphEdges
\twoLoopLadderNameText{3}{1}{3} \Leftrightarrow 
\begin{tikzpicture}[scale=\figScale,baseline=-3.05]\twoLoopGraph{3}{1}{3}{1.25};
\end{tikzpicture}\quad\text{with numerator  }\mathfrak{N}_3\!\big(\!\twoLoopLadderNameText{3}{1}{3}\big)=[\cca{\ell_A}][\ccc{\ell_C}]\!\!\oplus\!\![\ccb{\ell_B}]
}
---which would be appropriate for 3-gon power-counting, as we will define below. Such sums of vector-spaces never appear at one loop, for the simple reason that \mbox{$[\ell]^j\!\!\oplus\![\ell]^k\!\subset[\ell]^{\mathrm{max}(j,k)}$.} 

Consider for example the sum of two vector-spaces
\eq{\text{rank}\left([\ell_i]_d\!\oplus\![\ell_j]_d\right), \quad \text{for } i\neq j,\ i,j\!\in\{A,B,C\}\,.\label{eg_space_sum}}
Each space individually has $\text{rank}[\ell_i]_d = (d+2)$. However, the rank of the combination (\ref{eg_space_sum}) is a little less trivial. One might think that the linear span of the two spaces would add independently resulting in a combined rank of $2(d\pl2)$ for (\ref{eg_space_sum}); however, this would double-count the constant `1' shared by both pieces. Eliminating this over-counting, one can easily verify that the above $\text{rank}\left([\ell_i]_d \oplus [\ell_j]_d\right) = 2(d\pl2)\mi1$. 

Going one step further, we can consider the combined rank of all three numerator factors,
\eq{ \text{rank}\!\big([\cca{\ell_A}]_d\!\oplus\![\ccb{\ell_B}]_d\!\oplus\![\ccc{\ell_C}]_d\big) =\,\, ?}
Besides over-counting the constant term, we now encounter a subtlety with momentum conservation. As mentioned above, the three loop-momenta $\cca{\ell_A},\ \ccb{\ell_B}$, and $\ccc{\ell_C}$ are not all independent but satisfy a $d$-dimensional momentum conservation constraint. In fact, adding this third numerator factor only adds a single basis element:
\eq{ \text{rank}\!\big([\cca{\ell_A}]_d\!\oplus [\ccb{\ell_B}]_d\!\oplus\![\ccc{\ell_C}]_d \big)=2 (d+2)\,.}
In the particular choice of \hyperlink{routing}{routing} introduced above, this counting can be understood as the combination of $\ell_i^2$ for $i\!=\!1,2$, the $2\!\times\!d$ angles of the two independent loop momenta $\ell_1\!\cdot\!\hat{e}_1,\ldots,\ell_1\!\cdot\!\hat{e}_d, \ \ell_2\!\cdot\!\hat{e}_1,\ldots, \ell_2\!\cdot\!\hat{e}_d$, the constant term `1', as well as the angle between the two loop-momenta, which is proportional to $\ell_1\!\cdot\!\ell_2$. Note that, just as in our one-loop discussion, besides $\ell_i^2$, the only new degree-two-in-components element spanned by $[\ell_1]\!\oplus\![\ell_1\mi\ell_2]\!\oplus\![\ell_2]$ is $\ell_1\!\cdot\!\ell_2$; all other other linear combination of terms $(\ell_1\!\cdot\!\hat{e}_i)(\ell_2\!\cdot\!\hat{e}_j)$ would be assigned to numerators with higher power-counting.

%================================================================================================================
\newpage\vspace{-0pt}\subsection{Integrand Bases with `0-gon' Power-Counting in \texorpdfstring{$d$}{\emph{d}} Dimensions}\label{subsec:two_loop_opp_in_d_dimensions}\vspace{-0pt}
%================================================================================================================

Before diving into the general discussion and some of the subtleties with defining the two-loop version of $p$-gon power-counting in the following subsection \ref{subsec:power_counting_defined_at_two_loops}, we would like to discuss the conceptually straightforward case of $0$-gon power-counting that is relevant for theories such as the Standard Model. Here, we restrict our discussion to an integer-dimensional setup, but we expect that a suitable generalization of our one-loop implementation of a dimensional regularization friendly basis including $\mu$-terms should be possible as well.

%================================================================================================================
\vspace{-0pt}\subsubsection{Vector-Spaces of Loop-Dependent Numerators}\label{subsubsec:two_loop_opp_numerator_descriptions}\vspace{-0pt}
%================================================================================================================
As alluded to above, the rule for writing down the relevant numerator space for 0-gon power-counting is extremely simple to state: for any propagator structure characterized by the three indices $\cca{a},\ \ccb{b}$, and $\ccc{c}$ in $\twoLoopLadderNameText{a}{b}{c}$, we write exactly the same number of generalized inverse propagators in the numerator. As such, the numerator is given by a monomial in terms of $[\cca{\ell}_{\cca{A}}][\ccb{\ell}_{\ccb{B}}][\cca{\ell}_{\ccc{C}}]$
\eq{
\twoLoopLadderNameText{a}{b}{c} \leftrightarrow \mathfrak{N}_0\!\big(\!\twoLoopLadderNameText{a}{b}{c}\big) \equivR [\cca{\ell}_{\cca{A}}]^{\ca}_{}[\ccb{\ell}_{\ccb{B}}]^{\cb}_{}[\ccc{\ell}_{\ccc{C}}]^{\cc}_{}\,.
\label{0gon_rule_at_two_loops}}
Let us give one concrete example to make the above definition abundantly clear,
\eq{
\twoLoopLadderNameText{4}{2}{2}\;\bigger{\;\Leftrightarrow\!\!}\twoLoopTableFigure{4}{2}{2} \hspace{-3pt}\leftrightarrow \mathfrak{N}_0\!\big(\!\twoLoopLadderNameText{4}{2}{2}\big)\equivR [\cca\ell_{\cca{A}}]^{\cca{4}}_{d}[\ccb\ell_{\ccb{B}}]^{\ccb{2}}_{d}[\ccc\ell_{\ccc{C}}]^{\ccc{2}}_{d}\,.
}

The logic behind this rule should be apparent: as any propagator involving $\ell_i$ approaches infinity, every numerator in the vector-space (\ref{0gon_rule_at_two_loops}) will cancel all the propagators involving $\ell_i$. Thus, at infinite loop momentum, these integrands scale like a constant---or, a $0$-gon.

A key advantage of discussing $0$-gon power-counting is the extremely simple graph-theoretic rule on how to construct the associated numerator space to start with. The next non-trivial part of this analysis is to determine how many of the resulting numerators are linearly independent. As we have seen in our simple linear example in the previous subsection \ref{subsubsec:two_loop_tensor_structure}, there are intricate dependencies that only become more involved due to various completeness relations. Of course, computing the rank of linearly independent basis vectors is a `simple' linear algebra problem that can, however, become prohibitive for sufficiently large graphs in higher dimensions as the ranks of the associated matrices grow.  

In the particular case of 0-gon power-counting, where the numerator ansatz is a simple monomial, we were able to find closed form formulae for the ranks $\mathfrak{f}_d^0(\ca,\cb,\cc)\equivR \text{rank}\left([\cca\ell_{\cca{A}}]^{\ca}_{d}[\ccb\ell_{\ccb{B}}]^{\cb}_{d}[\ccc\ell_{\ccc{C}}]^{\cc}_{d}\right)$ of the integrand spaces in integer dimensions $d\!\leq\!4$ for arbitrary values of $\ca,\cb$ and $\cc$.\footnote{We thank Andrew McLeod for help with finding these formulae.} 
\begin{align}
\label{eq:closed_form_rank_formula_2loop}
\begin{split}
  \mathfrak{f}_1^0({\color{emphA}a},{\color{emphB}b},{\color{emphC}c}) = 
  &\, 1\pl 2 s_1\pl 4 s_2\,, \\
  \mathfrak{f}_2^0({\color{emphA}a},{\color{emphB}b},{\color{emphC}c}) = 
  &\, (1\pl s_1\pl s_2)^2\!\!\,, \\
  \mathfrak{f}_3^0({\color{emphA}a},{\color{emphB}b},{\color{emphC}c}) = 
  &\, \frac{1}{36} \Big[36\pl12 s_1^3\pl29 s_2^2\pl 4 s_2^3\pl s_1^2 (54\pl 26 s_2)\pl 3 s_3 (9\pl4 s_3)  \\
  & \hspace{80pt}\pl s_2 (61\pl 18 s_3)\pl s_1 (78\pl 9 s_2 (9\pl 2 s_2)\pl 22 s_3)\Big] \,,   \\
  \mathfrak{f}_4^0({\color{emphA}a},{\color{emphB}b},{\color{emphC}c}) = 
  & \frac{1}{144}\Big[(1\pl s_1\pl s_2) (4\pl 2 s_1\pl s_2)^2 (9\pl 3 s_1\pl s_2)  \\
  & \hspace{40pt}\pl 2 (4\pl 2 s_1\pl s_2) (7 s_1\pl 4 (3\pl s_2)) s_3\pl (15\pl 8 s_1\pl 4 s_2) s_3^2\Big]\,,
\end{split}
\end{align}
written in terms of symmetric polynomials
\eq{
s_1\equivR(\ca\pl\cb\pl\cc)\,,
\quad 
s_2\equivR(\ca \cb\pl\cb\cc\pl\cc\ca)\,,
\quad 
s_3\equivR \ca\cb \cc\,.
}
These expressions were obtained by writing a polynomial ansatz in $\cca{a},\ccb{b},\ccc{c}$ and matching this with the boundary one-loop-square cases when one of the indices is zero. Furthermore to fix the remaining ambiguity, we matched the ansatz with explicitly calculated off-shell ranks. Leftover rank `data' was then used as a nontrivial cross-check of eqs.~(\ref{eq:closed_form_rank_formula_2loop}). 

As written, it is not clear that the ranks eqs.~(\ref{eq:closed_form_rank_formula_2loop}) are integer, but it turns out that they indeed are. It would be interesting to find a group-theoretic interpretation of these formulae analogous to the interpretation of the one-loop ranks as the dimensions of symmetric traceless tensors of $\mathfrak{so}_{d+2}$.

%================================================================================================================
\vspace{-0pt}\subsubsection{Organizing Loop-Dependent Numerators by Contact Topologies}\label{subsubsec:top_and_contact_divisions}\vspace{-0pt}
%================================================================================================================

Having counted the dimensionality of the relevant numerator spaces, it is straightforward to write down an arbitrary representative basis that fills up the full rank space. In the following, however, we will argue that there is again a natural organization of the numerators in terms of contact-terms and top-level degrees of freedom analogous to eqs.~(\ref{eq:contact_decomposition_abstract_1L}) and (\ref{eq:q_gon_pgonPC_rank_decomp}):
\eq{\begin{split}\mathfrak{f}_d^{\,p}({\color{emphA}a},{\color{emphB}b},{\color{emphC}c})&\equivR\mathrm{rank}\Big[\mathfrak{N}_p\big(\twoLoopLadderNameText{a}{b}{c}\big)\Big]\\
&\equivL\underbrace{\fwbox{0pt}{\phantom{\sum_{(i,j,k)>\vec{0}}}}{\color{topCount}\widehat{\mathfrak{f}}_d^{\,p}({\color{emphA}a},{\color{emphB}b},{\color{emphC}c})}}_{\text{{\color{topCount}top rank}}}+\underbrace{\sum_{(i,j,k)>\vec{0}}\binom{\ca}{i}\binom{\cb}{j}\binom{\cc}{k}\,\,\widehat{\mathfrak{f}}_d^{\,p}({\color{emphA}a-i},{\color{emphB}b-j},{\color{emphC}c-k})}_{\text{{\color{black}contact-term rank}}}\,.
\end{split}\label{contact_term_rank_decomposition}}
This is very similar to the case at one loop. And as with one loop, this formula requires boundary data to solve---namely, which integrands are given numerators `$1$'. As we will see, this boundary data is provided by our definition of `scalar' $p$-gon power-counting integrands discussed in the following subsection. 

The recursive rank formula (\ref{contact_term_rank_decomposition}) can be interpreted directly as giving us a rule for \emph{constructing} the corresponding vector-spaces of numerators
\begin{align}\mathfrak{N}_{p}\!\big(\!\twoLoopLadderNameText{a}{b}{c}\big)
&\equivL\underbrace{\fwbox{0pt}{\phantom{}}{\color{topCount}\widehat{\mathfrak{N}}_{p}\!{\color{black}\big(}\!\twoLoopLadderNameText{a}{b}{c}{\color{black}\big)}}}_{\text{{\color{topCount}top-level numerators}}}\label{contact_term_decomposition}\\
&\hspace{-34pt}\underbrace{\bigoplus_{(i,j,k)>\vec{0}}\hspace{-10pt}\big[\x{\cca{\ell_A}}{Q_{a_{1}}}\cdots\x{\cca{\ell_A}}{Q_{a_{i}}}\big]\!\!\big[\x{\ccb{\ell_B}}{Q_{b_{1}}}\cdots\x{\ccb{\ell_B}}{Q_{b_{j}}}\big]\!\!\big[\x{\ccc{\ell_C}}{S_{c_{1}}}\cdots\x{\ccc{\ell_C}}{S_{c_{k}}}\big]\,\,\!\!\widehat{\mathfrak{N}}_{p}\!{\color{black}\big(}\!\twoLoopLadderNameText{a{\color{black}-i}}{b{\color{black}-j}}{c{\color{black}-k}}\hspace{32pt}{\color{black}\big)}}_{\text{{\color{black}contact-term numerators}}}\nonumber
\end{align}
This makes it clear that (\ref{contact_term_rank_decomposition}) requires that the vector-spaces appearing in (\ref{contact_term_decomposition}) are all mutually independent. This will be true whenever $p\!<\!d$.  

In order to stratify the relevant loop-dependent numerators for all two-loop integrand topologies, we propose a bottom-up strategy: first identify  graphs with `scalar' numerators---those with the fewest propagators for a given $p$---and work our way upwards. In doing so, we have full control over the relevant numerator spaces of all contact-terms of more complicated graphs. Together with our general counting formulae in eq.~(\ref{eq:closed_form_rank_formula_2loop}) it is then easy to compute the {\color{topCount}top-level} degrees of freedom of a given graph. Alternatively, one can also compute the {\color{topCount}top-level} rank of a given numerator by evaluating the span of numerators on the maximal cut \cite{Bern:2007ct} surface of a given topology. In $d\!\leq\!4$, we have pursued both strategies and independently confirmed the various numerator ranks, which also serves as a nontrivial cross-check of our closed form expressions (\ref{eq:closed_form_rank_formula_2loop}).\\

After these general considerations, it is perhaps instructive to return to $0$-gon power-counting and demonstrate how integrand decomposition works for a few concrete examples. Recall that for our definition of $0$-gon power-counting, the basic integrand topology has \emph{no} propagator and is solely given by a loop-momentum independent normalization. (Of course, in dimensional regularization all these topologies correspond to power-divergent integrals that integrate to zero and therefore are usually not considered. However, for building bases of integrands, these topologies are relevant.) Therefore, we assign
\eq{
 \twoLoopLadderNameText{0}{0}{0}  \Leftrightarrow
 \bullet 
 \quad\text{with}\quad
 \mathfrak{N}_0\!\big(\! \twoLoopLadderNameText{0}{0}{0} \big)\equivR 1
}
a single degree of freedom to this topology. Going up in the number of propagators, the next topology to consider is a tadpole, where one loop is completely pinched and the other loop has a single propagator. In this case, we write down a single numerator factor
\begin{align}\emphEdges
\twoLoopLadderNameText{1}{0}{0}  \;\Leftrightarrow \hspace{-10pt}
%\oneLoopGraph[90]{1} 
\begin{tikzpicture}[scale=\figScale,baseline=-3.05]\twoLoopGraph{1}{0}{0}{1.25};
\end{tikzpicture}
\quad\text{with}\quad\mathfrak{N}_0\!\big(\!\twoLoopLadderNameText{1}{0}{0} \big)\equivR [\cca{\ell_A}]^{\cca{1}}\,.
\end{align}
Specifying to $d\!=\!4$ for concreteness, we have $\mathfrak{f}_4^0(\cca{1},\ccb{0},\ccc{0})=6$. $\twoLoopLadderNameText{1}{0}{0}$ is of course a one-loop graph and we know how to decompose its {\color{totalCount}6} total numerator degrees of freedom: ${\color{totalCount}{6}}\!=\!{\color{topCount}{5}}{\color{dim}\pl1}$ represents {\color{topCount}5} {top-level} degrees of freedom and 1 contact term.

Next we consider the tadpole$\times$tadpole graph whose numerator is given by the two one-loop numerator factors
\begin{align}
\emphEdges
\twoLoopLadderNameText{1}{0}{1} 
 \;\Leftrightarrow \hspace{-10pt}
\begin{tikzpicture}[scale=\figScale,baseline=-3.05]\twoLoopGraph{1}{0}{1}{1.25};
\end{tikzpicture}
%\oneLoopGraph[-90]{1}\hspace{-2cm}\oneLoopGraph[90]{1}
\text{with}\quad\mathfrak{N}_0\!\big(\!\twoLoopLadderNameText{1}{0}{1} \big)\equivR  [\cca{\ell_A}]^{\cca{1}}[\ccc{\ell_C}]^{\ccc{1}}
\end{align}%\resetGraphDefaults
In $d\!=\!4$, this numerator has $\mathfrak{f}_4^0(\cca{1},\ccb{0},\ccc{1})\!=\!{\color{topCount}36}$ total degrees of freedom that can be decomposed into \mbox{${\color{totalCount}{36}} = {\color{topCount}{25}}{\color{dim}\,\pl2\!\times\!5\pl1}$}: {\color{topCount}25} top-level degrees of freedom together with 11 contact-terms that we have identified with degrees of freedom for its \hyperlink{daughter}{daughters}' $\twoLoopLadderNameText{1}{0}{0}$ and $\twoLoopLadderNameText{0}{0}{0}$.

With these arguably trivial one-loop type examples in hand, we can now discuss an honest irreducible two-loop graph 
\eq{
\twoLoopLadderNameText{1}{1}{1}
 \;\Leftrightarrow\hspace{-14pt}
\twoLoopTableFigure{1}{1}{1} 
\text{with}\quad\mathfrak{N}_0\!\big(\!\twoLoopLadderNameText{1}{1}{1} \big)\equivR  [\cca{\ell_A}]^{\cca{1}}[\ccb{\ell_B}]^{\ccb{1}}[\ccc{\ell_C}]^{\ccc{1}}\,.
}
In four dimensions, \twoLoopLadderNameText{1}{1}{1} has $\mathfrak{f}_4^0(\cca{1},\ccb{1},\ccc{1})\!=\!{\color{totalCount}181}$ total degrees of freedom that are decomposable into $3\times 25$ degrees of freedom of \twoLoopLadderNameText{1}{0}{1} topologies, $3\times5$ \twoLoopLadderNameText{1}{0}{0} and 1 degree of freedom obtained by pinching all three propagators leading to a single \twoLoopLadderNameText{0}{0}{0}. Adding up all the contact degrees of freedom constitutes $91$ of the ${\color{totalCount}181}$ degrees of freedom in \twoLoopLadderNameText{1}{1}{1}, leaving {\color{topCount}90} top-level degrees of freedom for the sunrise integral:
\eq{
\mathfrak{f}_4^0(\cca{1},\ccb{1},\ccb{1})=\underset{d=4}{\text{rank}}\Big[\mathfrak{N}_0\!\big(\!\twoLoopLadderNameText{1}{1}{1}\big)\Big]=\text{rank}\Big\{[\cca{\ell_A}]^{\cca{1}}_4[\ccb{\ell_B}]^{\ccb{1}}_4[\ccc{\ell_C}]^{\ccc{1}}_4\Big\}=
\tensorDecomp{181}{90}{91}\,.
}
As mentioned above, the same number of top-level degrees of freedom can alternatively be obtained by evaluating the numerators $\mathfrak{N}_{[\cca{1},\ccb{1},\ccc{1}]}$ on the triple-cut surface $\cca{\ell^2_{\cca{A}}}=\ccb{\ell^2_{\ccb{B}}}=\ccc{\ell^2_{\ccc{C}}}=0$ and checking the remaining matrix rank.   

Proceeding in a similar fashion, we can explicitly stratify the bases of integrands for all other two-loop topologies. As one might guess from our iterative description, this algorithm is extremely suitable for automation in available computer-algebra systems to allow for an efficient rank counting. We will not give the complete answer to the counting problem here, but defer a detailed presentation of our results to Table~\ref{tab:two_loop_counting_table}, which also includes similar results for different degrees of power-counting in various dimensions.

Like in our one-loop discussion, where we had found that all pentagon integrands (and integrands with more propagators) become reducible in $d\!=\!4$ beyond 4-gon power-counting, we find that the basis of integrands of two-loop topologies is completely spanned by contact-terms beyond some number of propagators for any given power-counting. Whenever this happens, we deem such a topology reducible and we do not discuss it further. In particular, for the 0-gon power-counting under consideration in this section, this appears in $d\!=\!4$ for integrands with more than 8 propagators or whenever a single loop-momentum $\ca,\ \cb$, or $\cc$ appears in more than 4 propagators. More generally, we find that (for $p\!<\!d$-gon power-counting), all graphs with more than $d\!\times\!L$ propagators are reducible. (Notice that the number of propagators at two loops is simply $\ca\pl\cb\pl\cc$.) There is one notable exception to this rule involving propagator-renormalization graphs that we discuss separately in \ref{para:exceptional_cases}.

%================================================================================================================
\vspace{-0pt}\subsection{Defining `\texorpdfstring{$p$}{\emph{p}}-gon Power-Counting' at Two Loops}\label{subsec:power_counting_defined_at_two_loops}\vspace{-0pt}
%================================================================================================================

As with one loop, we will start our analysis of two-loop integrands without any restriction to the dimension of spacetime. One striking difference between one and two (or more) loops is that it will no longer be so obvious to describe the `power-counting' of an integrand beyond $p\!=\!0$: due to the lack of a natural origin in loop momentum space, the na\"ive definition given for one loop is no longer sufficient. 

Consider, for example, the two-loop graph
\emphEdges
\eq{%
\begin{tikzpicture}[scale=\figScale,baseline=-3.05]\twoLoopGraph{3}{1}{2}{1.25};
\end{tikzpicture}\,.%
\label{example_graph_for_routing_problems}}
For this integrand, how many propagators involve the loop momenta $\ell_1,\ell_2$? Possible answers include $\{4,3\}, \{5,4\},$ or $\{5,3\}$, as can be seen by re-drawing the graph in three ways: 
\eq{%
\begin{tikzpicture}[scale=\figScale,baseline=-3.05]\twoLoopGraph{3}{1}{2}{1.25};
\end{tikzpicture}\qquad%
\begin{tikzpicture}[scale=\figScale,baseline=-3.05]\twoLoopGraph{2}{3}{1}{1.25};
\end{tikzpicture}\quad\text{or}\quad%
\begin{tikzpicture}[scale=\figScale,baseline=-3.05]\twoLoopGraph{3}{2}{1}{1.25};
\end{tikzpicture}\;.
\label{example_graph_for_routing_problems_explained}}
This problem seems artificial in the planar case, if only because there is always a (seemingly) natural prescription for how to route the loop momenta of a planar graph---namely, according to the faces of the plane\footnote{Recall that a {\it planar} graph is one which admits a plane embedding. A {\it plane} graph is one endowed with a particular embedding. If a graph is planar, its plane embedding is unique provided the graph's edge-connectivity is 3 or greater. The easiest example of a planar graph which admits multiple plane embeddings is a (2-edge-connected) graph which includes as a sub-diagram a loop-correction to a propagator.} graph. To be clear: we define the \hyperlink{routing}{{\it routing}} of loop momenta of an $L$ loop diagram to be a choice of $L$ \hyperlink{simpleCycle}{simple cycles} whose union encompasses all edges of the graph. 

When we consider diagrams that are not planar, we are forced to reckon with the fact that there is no intrinsic (or even obviously preferential) choice of routing. We are using the example (\ref{example_graph_for_routing_problems}) in order to emphasize that this problem affects planar diagrams as well. 

Minimally, this indicates that more care is required to discuss the `power-counting' of an integrand. Another example which will help illustrate our point would be  the following:
\eq{%
\begin{tikzpicture}[scale=\figScale,baseline=-3.05]\twoLoopGraph{3}{1}{1}{1.25};
\end{tikzpicture}\hspace{3pt}\leftrightarrow\hspace{-2pt} \tikzBox[0.7]{\twoLoopGraph{1}{3}{1}{1.25}}\,.%
\label{example_graph_for_routing_problems2}}
While this example could be interpreted as four propagators involving $\ell_1$ and four propagators involving $\ell_2$, no one would overlook the fact that this integral contains a bubble! As such, it would seem absurd to assign this integral a power-counting of two box-integrals. One solution to this problem (especially to combat the example in (\ref{example_graph_for_routing_problems2})) would be to declare that any {routing} must consist of {cycles} of `minimal' length---by some metric of ordering on cycle sets. This feels like a dangerous approach to us, as at asymptotically large loop order, there would seem to be no natural way to ordering the `minimality' of large collections of loops (and very little reason to believe that, choosing some way to order choices for {routing}, that degeneracies of choice would behave similarly). 

Thus, we are forced to face the problem that there would seem to be no intrinsically obvious way to assign `power-counting' to a multi-loop (especially non-planar) Feynman integrand. However, in the following, we will put forward one suggestion of an intrinsically graph-theoretic way to define power-counting.

%================================================================================================================
\vspace{-0pt}\subsubsection{`Scalar' \texorpdfstring{$p$}{\emph{p}}-gon Integrands at Two Loops}\label{subsubsec:scalar_integrals}\vspace{-0pt}
%================================================================================================================
Our basic strategy for defining integrands $p$-gon power-counting follows from the recursive definition of how integrand numerators get stratified by contact-terms according to (\ref{contact_term_decomposition}). That is, we will define a vector-space of numerators for a given graph \emph{relative to} its contact-terms. For example, suppose that we wanted to construct a space of integrands that all behave exactly like the `scalar' integrand
\eq{\begin{tikzpicture}[scale=\figScale,baseline=-3.05]\twoLoopGraph{2}{1}{2}{1.25};
\end{tikzpicture}\,.\label{example_scalar_integrand212}}
Adding any new propagators \emph{together with} each new propagator's vector-space of inverse propagators in the numerator will clearly result in an integrand that behaves like (\ref{example_scalar_integrand212}). Graphically, integrands that `{scale like} \twoLoopLadderNameText{2}{0}{2} (at infinity)' would include
\eq{\left\{\begin{tikzpicture}[scale=\figScale,baseline=-3.05]\twoLoopGraph{2}{2}{2}{1.25};\draw[markedEdgeR](a3)--(b2);
\end{tikzpicture}\,,
\begin{tikzpicture}[scale=\figScale,baseline=-3.05]\twoLoopGraph{2}{3}{2}{1.25};\draw[markedEdgeR](a3)--(b3);\draw[markedEdgeR](b2)--(v0);
\end{tikzpicture}\,,
\begin{tikzpicture}[scale=\figScale,baseline=-3.05]\twoLoopGraph{3}{1}{2}{1.25};\draw[markedEdgeR](a1)--(a2);
\end{tikzpicture}\,,
\begin{tikzpicture}[scale=\figScale,baseline=-3.05]\twoLoopGraph{4}{1}{2}{1.25};\draw[markedEdgeR](a0)--(a1);\draw[markedEdgeR](a3)--(a4);
\end{tikzpicture}\,,
\begin{tikzpicture}[scale=\figScale,baseline=-3.05]\twoLoopGraph{4}{2}{2}{1.25};\draw[markedEdgeR](a0)--(a1);\draw[markedEdgeR](a3)--(a4);\draw[markedEdgeR](a4)--(b2);
\end{tikzpicture}\,,\ldots\right\}.}
It is easy to see that this defines numerators for all integrands that contain  \twoLoopLadderNameText{2}{0}{2} as a contact-term. This may or may not be a \emph{useful} vector-space of integrands to define, as there may be no interesting quantum field theories whose amplitudes are expressible in this space.

This rule can easily be generalized to construct numerators for integrands that scale like one or more of a list of `scalar' integrands (at infinity): we simply add the vector-spaces of loop-dependent monomials assigned to each integrand as dictated by each of a given set of the scalars (which are obtainable as edge contractions relative to the graph) from a given list. Thus, we may take any subset of graphs to be given `scalar' numerators `$1$', and thereby define an infinite set of integrands with more propagators which scale like one or more of the graphs from this list. The only missing ingredient is to define the space of integrands to be taken to be scalar. It should also be clear that this rule generalizes to arbitrary loop order.

Our proposal for integrands assigned scalar numerators for $p$-gon power-counting is as follows. 
\begin{verse}
\textbf{Definition:} a \emph{scalar} $p$-gon is any integrand having \hyperlink{girth}{girth} $p$, such that all its daughters---graphs obtained by single edge-contractions---have girth $<\!p$.
\end{verse}
We denote the space of $p$-gon power-counting scalars at $L$ loops by $\mathfrak{S}_p^L$. Recall that the \emph{girth} of a graph is the length of its shortest cycle. This definition clearly generalizes to any loop-order. 

At two loops, it is quite easy to list the scalar $p$-gons for any $p$. For example, 
\eq{\begin{split}
\hspace{-10pt}\text{1-gon power-counting scalars $\mathfrak{S}_1^2\equivR$\hspace{-10pt}} & \quad
\left\{\rule{0pt}{24pt}\hspace{-10pt}\begin{tikzpicture}[scale=\figScale,baseline=-3.05]\twoLoopGraph{1}{0}{1}{1.25};
\end{tikzpicture}\hspace{-10pt}\right\}\,\,\hspace{-30pt}\\
\hspace{-10pt}\text{2-gon power-counting scalars $\mathfrak{S}_2^2\equivR$\hspace{-10pt}} & \quad
\left\{\rule{0pt}{24pt}\begin{tikzpicture}[scale=\figScale,baseline=-3.05]\twoLoopGraph{2}{0}{2}{1.25};
\end{tikzpicture}\,, \quad
\hspace{-7pt}\begin{tikzpicture}[scale=\figScale,baseline=-3.05,rotate=90]\twoLoopGraph{1}{1}{1}{1.25};
\end{tikzpicture}\hspace{7.75pt}\right\}\,\,\hspace{-30pt}\\
\hspace{-10pt}\text{3-gon power-counting scalars $\mathfrak{S}_3^2\equivR$\hspace{-10pt}} & \quad
\left\{\begin{tikzpicture}[scale=\figScale,baseline=-3.05]\twoLoopGraph{3}{0}{3}{1.25};
\end{tikzpicture}\,, \quad
\hspace{-7pt}\begin{tikzpicture}[scale=\figScale,baseline=-3.05]\twoLoopGraph{2}{1}{2}{1.25};
\end{tikzpicture}\right\}\,\hspace{-30pt}\\
\hspace{-10pt}\text{4-gon power-counting scalars $\mathfrak{S}_4^2\equivR$\hspace{-10pt}} &\quad
\left\{\begin{tikzpicture}[scale=\figScale,baseline=-3.05]\twoLoopGraph{4}{0}{4}{1.25};
\end{tikzpicture}\,, \quad
\hspace{-10.5pt}\begin{tikzpicture}[scale=\figScale,baseline=-3.05]\twoLoopGraph{3}{1}{3}{1.25};
\end{tikzpicture}\,, \quad
\hspace{-7pt}\begin{tikzpicture}[scale=\figScale,baseline=-3.05]\twoLoopGraph{2}{2}{2}{1.25};
\end{tikzpicture}\right\}\,\hspace{-30pt}\\
\hspace{-10pt}\text{5-gon power-counting scalars  $\mathfrak{S}_5^2\equivR$\hspace{-10pt}} & \quad
\left\{\begin{tikzpicture}[scale=\figScale,baseline=-3.05]\twoLoopGraph{5}{0}{5}{1.25};
\end{tikzpicture}\,, \quad
\hspace{-7pt}\begin{tikzpicture}[scale=\figScale,baseline=-3.05]\twoLoopGraph{4}{1}{4}{1.25};
\end{tikzpicture}\,, \quad
\hspace{-7pt}\begin{tikzpicture}[scale=\figScale,baseline=-3.05]\twoLoopGraph{3}{2}{3}{1.25};
\end{tikzpicture}\hspace{2.5pt}\right\}\,\hspace{-30pt}
\label{pgon_power-counting_graphs_egs}
\end{split}}
and so on. Assigning `scalar' numerators `$1$' to each of these integrands allows us to define spaces of integrands that behave asymptotically like one or more of these. 

It is clear that this definition is purely graph theoretic, and provides us with a precise rule for assigning vector-spaces of numerators to every graph that contains one or more of the scalar graphs as contact-terms. Moreover, it is clear that these vector-spaces are defined without respect to the dimension of spacetime (although the sizes of these vector-spaces, and the degree to which these spaces are spanned by contact-terms will, of course, depend strongly on the dimension of spacetime).

%================================================================================================================
\newpage\vspace{-0pt}\subsubsection{Two-Loop Numerators \emph{Relative to} Scalar \texorpdfstring{$p$}{\emph{p}}-gon Integrands}\label{subsubsec:p_gon_tensor_reduction}\vspace{-0pt}
%================================================================================================================
%
Given this definition of scalar $p$-gon integrands, we may follow the rule described above to construct the vector-space of $p$-gon power-counting integrands---those integrands which are constructed explicitly to scale like one or more of the scalar $p$-gons at infinite loop momentum. That is, we assign any graph that is a \hyperlink{parent}{parent} of a scalar graph, $\Gamma\!\succ\!\Sigma\!\in\!\mathfrak{S}_p^L$, a numerator consisting of the inverse propagators associated with the edge-set in the quotient of $\Gamma$ relative to $\Sigma$; if there are multiple $\Sigma\!\in\!\mathfrak{S}_p^L$, we add the vector-spaces of loop-dependent numerators for each. Let us illustrate this rule with a number of concrete examples. 

Consider the graph \twoLoopLadderNameText{2}{1}{2}; what numerator would be assigned to this Feynman integrand for various power-countings? Because \twoLoopLadderNameText{2}{1}{2}$\,\in\!\mathfrak{S}^2_3$, it would be assigned the numerator `$1$' for 3-gon power-counting. For 2-gon (or `bubble') power-counting, it is easy to see that \mbox{\twoLoopLadderNameText{2}{1}{2}$\,\succ$\twoLoopLadderNameText{2}{0}{2}$\,\in\!\mathfrak{S}^2_2$} and also \mbox{\twoLoopLadderNameText{2}{1}{2}$\,\succ$\twoLoopLadderNameText{1}{1}{1}$\,\in\!\mathfrak{S}^2_2$}. Thus, 
\vspace{-6pt}\eq{\mathfrak{N}_2\!\big(\!\twoLoopLadderNameText{2}{1}{2}\big)=\mathrm{span}\Bigg\{\raisebox{-7.5pt}{$\underbrace{\begin{array}{@{}c@{}r@{}}
\tikzBox[0.7]{\twoLoopGraph{2}{1}{2}{1}\coordinate(b1)at(v0);\coordinate(b2)at(v1);
\decorateB\draw[markedEdgeR](b1)--(b2);
}\fwboxL{0pt}{\text{{\footnotesize$\hspace{5.25pt}\oplus\hspace{-10pt}$}}}&\\[-1pt]\fwboxR{0pt}{=\mathrm{span}\hspace{2.5pt}\big\{\hspace{12pt}}\text{{\footnotesize$[\ccb{\ell_B}]$}}&\fwboxL{0pt}{\text{{\footnotesize$\hspace{5.25pt}\oplus\hspace{-10pt}$}}}
\end{array}}_{\substack{\\[-3pt]\begin{turn}{90}$\prec$\end{turn} \\[-10pt] \tikzBox[0.7]{\twoLoopGraph{2}{0}{2}{1}}}}\hspace{15pt}
\underbrace{\begin{array}{@{}c@{}r@{}}
\tikzBox[0.7]{\twoLoopGraph{2}{1}{2}{1}
\decorateA\draw[markedEdgeR](a1)--(a2);\decorateC\draw[markedEdgeR](c1)--(c2);
}&\\[-1pt]\text{{\footnotesize$[\cca{\ell_A}][\ccc{\ell_C}]$}}&\fwboxL{0pt}{\hspace{0.5pt}\big\}\,.}
\end{array}}_{\substack{\\[-3pt]\begin{turn}{90}$\prec$\end{turn} \\[-10pt] \begin{turn}{90}\tikzBox[0.7]{\twoLoopGraph{1}{1}{1}{1}}\end{turn}}}
$}
\Bigg\}
\vspace{-14pt}}
Now, the {\color{totalCount}total rank} of this space varies with dimension, as does the breakdown of this vector-space into {\color{topCount}top-level} degrees of freedom and contact-terms. Specifically, we find
\vspace{-12pt}\eq{\begin{split}
\fwboxR{0pt}{\text{%$\mathfrak{N}_2\!\big(\!\twoLoopLadderNameText{2}{1}{2}\big)$ 
in $d\!=\!2:$}\hspace{10pt}}\mathrm{rank}\big([\ccb{\ell_B}]_2\!\!\oplus[\cca{\ell_A}]_2[\ccc{\ell_C}]_2\big)&={\color{totalCount}16}={\color{topCount}3}{\color{dim}+13}\,;\\
\fwboxR{0pt}{\text{ in $d\!=\!3:$}\hspace{10pt}}\mathrm{rank}\big([\ccb{\ell_B}]_3\!\!\oplus[\cca{\ell_A}]_3[\ccc{\ell_C}]_3\big)&={\color{totalCount}25}={\color{topCount}8}{\color{dim}+17}\,;\\
\fwboxR{0pt}{\text{ in $d\!=\!4:$}\hspace{10pt}}\mathrm{rank}\big([\ccb{\ell_B}]_4\!\!\oplus[\cca{\ell_A}]_4[\ccc{\ell_C}]_4\big)&={\color{totalCount}36}={\color{topCount}15}{\color{dim}+21}\,;
\end{split}\vspace{-14pt}}
and so-on. To be clear, the breakdown of $\mathfrak{N}_2\!\big(\!\twoLoopLadderNameText{2}{1}{2}\big)$ into {\color{topCount}top-level} degrees of freedom and contact-terms in any number of dimensions follows (recursively) via the definitions (\ref{contact_term_rank_decomposition}) and (\ref{contact_term_decomposition}). 

For $1$-gon power-counting, the numerators assigned to \twoLoopLadderNameText{2}{1}{2} would be given as the sum of three monomials:
\vspace{-10pt}\eq{\mathfrak{N}_1\!\big(\!\twoLoopLadderNameText{2}{1}{2}\big)=\mathrm{span}\Bigg\{\raisebox{-7.5pt}{$\underbrace{\begin{array}{@{}c@{}r@{}}
\tikzBox[0.7]{\twoLoopGraph{2}{1}{2}{1}\coordinate(b1)at(v0);\coordinate(b2)at(v1);
\decorateA\draw[markedEdgeR](a1)--(a2);\decorateC\draw[markedEdgeR](c1)--(c2);\decorateB\draw[markedEdgeR](b1)--(b2);
}\fwboxL{0pt}{\text{{\footnotesize$\hspace{5.25pt}\oplus\hspace{-10pt}$}}}&\\[-1pt]\fwboxR{0pt}{=\mathrm{span}\hspace{2.5pt}\big\{\hspace{0.75pt}}\text{{\footnotesize$[\cca{\ell_A}][\ccb{\ell_B}][\ccc{\ell_C}]$}}&\fwboxL{0pt}{\text{{\footnotesize$\hspace{1.25pt}\oplus\hspace{-10pt}$}}}
\end{array}\hspace{14pt}
\begin{array}{@{}c@{}r@{}}
\tikzBox[0.7]{\twoLoopGraph{2}{1}{2}{1}
\decorateA\draw[markedEdgeR](a1)--(a2);\draw[markedEdgeR](a1)--(a0);\decorateC\draw[markedEdgeR](c1)--(c2);
}\fwboxL{0pt}{\text{{\footnotesize$\hspace{5.25pt}\oplus\hspace{-10pt}$}}}&\\[-1pt]\text{{\footnotesize$[\cca{\ell_A}]^{\cca2}[\ccc{\ell_C}]$}}&\fwboxL{0pt}{\text{{\footnotesize$\hspace{5.25pt}\oplus\hspace{-10pt}$}}}%\fwboxL{0pt}{\hspace{0.5pt}\big\}\,.}
\end{array}
\hspace{18pt}
\begin{array}{@{}c@{}r@{}}
\tikzBox[0.7]{\twoLoopGraph{2}{1}{2}{1}
\decorateA\draw[markedEdgeR](a1)--(a2);\decorateC\draw[markedEdgeR](c1)--(c2);\draw[markedEdgeR](c1)--(c0);
}&\\[-1pt]\text{{\footnotesize$[\cca{\ell_A}][\ccc{\ell_C}]^{\ccc2}$}}&\fwboxL{0pt}{\hspace{1.5pt}\big\}\,.}
\end{array}}_{\substack{\\[-3pt]\begin{turn}{90}$\prec$\end{turn} \\[-10pt] \tikzBox[0.7]{\twoLoopGraph{1}{0}{1}{1}}}}\hspace{15pt}
$}\hspace{-14.25pt}
\Bigg\}
\vspace{-20pt}}
As before, we can easily decompose the ranks (and breakdowns) of these vector-spaces in various dimensions:
\vspace{-4pt}\eq{\begin{split}
\fwboxR{0pt}{\text{%$\mathfrak{N}_2\!\big(\!\twoLoopLadderNameText{2}{1}{2}\big)$ 
in $d\!=\!2:$}\hspace{10pt}}\mathrm{rank}\big([\cca{\ell_A}]_2[\ccb{\ell_B}]_2[\ccc{\ell_C}]_2\!\oplus\![\cca{\ell_A}]^{\cca2}_2[\ccc{\ell_C}]_2^{\ccc1}\!\oplus\![\cca{\ell_A}]^{\cca1}_2[\ccc{\ell_C}]_2^{\ccc2}\big)&={\color{totalCount}63}={\color{topCount}\mathbf{0}}{\color{dim}+63}\,;\\
\fwboxR{0pt}{\text{ in $d\!=\!3:$}\hspace{10pt}}\mathrm{rank}\big([\cca{\ell_A}]_3[\ccb{\ell_B}]_3[\ccc{\ell_C}]_3\!\oplus\![\cca{\ell_A}]^{\cca2}_3[\ccc{\ell_C}]_3^{\ccc1}\!\oplus\![\cca{\ell_A}]^{\cca1}_3[\ccc{\ell_C}]_3^{\ccc2}\big)&={\color{totalCount}131}={\color{topCount}16}{\color{dim}+115}\,;\\
\fwboxR{0pt}{\text{ in $d\!=\!4:$}\hspace{10pt}}\mathrm{rank}\big([\cca{\ell_A}]_4[\ccb{\ell_B}]_4[\ccc{\ell_C}]_4\!\oplus\![\cca{\ell_A}]^{\cca2}_4[\ccc{\ell_C}]_4^{\ccc1}\!\oplus\![\cca{\ell_A}]^{\cca1}_4[\ccc{\ell_C}]_4^{\ccc2}\big)&={\color{totalCount}229}={\color{topCount}49}{\color{dim}+180}\,.
\end{split}\vspace{-8pt}}

For one further illustration of how this works, consider the case of \twoLoopLadderNameText{3}{2}{3}. For $p$-gon power-counting with $p\!=\!0,\ldots,4$, we would have integrands built according to
\eq{\begin{split}
&\mathfrak{B}_0\!\supset\!\Bigg\{\tikzBox[0.7]{\twoLoopGraph{3}{2}{3}{1}\coordinate(b1)at(v0);\coordinate(b3)at(v1);
\decorateA\draw[markedEdgeR](a0)--(a1);\draw[markedEdgeR](a1)--(a2);\draw[markedEdgeR](a2)--(a3);
\decorateC\draw[markedEdgeR](c0)--(c1);\draw[markedEdgeR](c1)--(c2);\draw[markedEdgeR](c2)--(c3);
\decorateB\draw[markedEdgeR](b1)--(b2);\draw[markedEdgeR](b2)--(b3);}\decorateNo
\Bigg\}\\[-4pt]
&\mathfrak{B}_1\!\supset\!\Bigg\{
\tikzBox[0.75]{\twoLoopGraph{3}{2}{3}{1}\coordinate(b1)at(v0);\coordinate(b3)at(v1);
\decorateA\draw[markedEdgeR](a0)--(a1);\draw[markedEdgeR](a1)--(a2);\draw[markedEdgeR](a2)--(a3);
\decorateC\draw[markedEdgeR](c0)--(c1);\draw[markedEdgeR](c2)--(c3);
\decorateB\draw[markedEdgeR](b2)--(b3);}{\oplus}
\tikzBox[0.75]{\twoLoopGraph{3}{2}{3}{1}\coordinate(b1)at(v0);\coordinate(b3)at(v1);
\decorateA\draw[markedEdgeR](a0)--(a1);\draw[markedEdgeR](a2)--(a3);
\decorateC\draw[markedEdgeR](c0)--(c1);\draw[markedEdgeR](c1)--(c2);\draw[markedEdgeR](c2)--(c3);
\decorateB\draw[markedEdgeR](b2)--(b3);}{\oplus}
\tikzBox[0.75]{\twoLoopGraph{3}{2}{3}{1}\coordinate(b1)at(v0);\coordinate(b3)at(v1);
\decorateA\draw[markedEdgeR](a0)--(a1);\draw[markedEdgeR](a2)--(a3);
\decorateC\draw[markedEdgeR](c0)--(c1);\draw[markedEdgeR](c2)--(c3);
\decorateB\draw[markedEdgeR](b1)--(b2);\draw[markedEdgeR](b2)--(b3);}\decorateNo
\Bigg\}\\[-4pt]
&\mathfrak{B}_2\!\supset\!\Bigg\{
\tikzBox[0.75]{\twoLoopGraph{3}{2}{3}{1}\coordinate(b1)at(v0);\coordinate(b3)at(v1);
\decorateA\draw[markedEdgeR](a0)--(a1);\draw[markedEdgeR](a1)--(a2);\draw[markedEdgeR](a2)--(a3);
\decorateC\draw[markedEdgeR](c1)--(c2);
}{\oplus}
\tikzBox[0.75]{\twoLoopGraph{3}{2}{3}{1}\coordinate(b1)at(v0);\coordinate(b3)at(v1);
\decorateA\draw[markedEdgeR](a1)--(a2);
\decorateC\draw[markedEdgeR](c0)--(c1);\draw[markedEdgeR](c1)--(c2);\draw[markedEdgeR](c2)--(c3);
}{\oplus}
\tikzBox[0.75]{\twoLoopGraph{3}{2}{3}{1}\coordinate(b1)at(v0);\coordinate(b3)at(v1);
\decorateA\draw[markedEdgeR](a0)--(a1);\draw[markedEdgeR](a2)--(a3);
\decorateC\draw[markedEdgeR](c0)--(c1);\draw[markedEdgeR](c2)--(c3);
\decorateB\draw[markedEdgeR](b2)--(b3);}{\oplus}
\tikzBox[0.75]{\twoLoopGraph{3}{2}{3}{1}\coordinate(b1)at(v0);\coordinate(b3)at(v1);
\decorateA\draw[markedEdgeR](a1)--(a2);
\decorateC\draw[markedEdgeR](c1)--(c2);
\decorateB\draw[markedEdgeR](b1)--(b2);\draw[markedEdgeR](b2)--(b3);}\decorateNo
\Bigg\}\\[-4pt]
&\mathfrak{B}_3\!\supset\!\Bigg\{
\tikzBox[0.75]{\twoLoopGraph{3}{2}{3}{1}\coordinate(b1)at(v0);\coordinate(b3)at(v1);
\decorateB\draw[markedEdgeR](b1)--(b2);\draw[markedEdgeR](b2)--(b3);
}{\oplus}
\tikzBox[0.75]{\twoLoopGraph{3}{2}{3}{1}\coordinate(b1)at(v0);\coordinate(b3)at(v1);
\decorateA\draw[markedEdgeR](a0)--(a1);\draw[markedEdgeR](a2)--(a3);
\decorateC\draw[markedEdgeR](c1)--(c2);
}{\oplus}
\tikzBox[0.75]{\twoLoopGraph{3}{2}{3}{1}\coordinate(b1)at(v0);\coordinate(b3)at(v1);
\decorateA\draw[markedEdgeR](a1)--(a2);
\decorateC\draw[markedEdgeR](c0)--(c1);\draw[markedEdgeR](c2)--(c3);
}{\oplus}
\tikzBox[0.75]{\twoLoopGraph{3}{2}{3}{1}\coordinate(b1)at(v0);\coordinate(b3)at(v1);
\decorateA\draw[markedEdgeR](a1)--(a2);
\decorateC\draw[markedEdgeR](c1)--(c2);
\decorateB\draw[markedEdgeR](b2)--(b3);}\decorateNo
\Bigg\}\\[-4pt]
&\mathfrak{B}_4\!\supset\!\Bigg\{
\tikzBox[0.75]{\twoLoopGraph{3}{2}{3}{1}\coordinate(b1)at(v0);\coordinate(b3)at(v1);
\decorateB\draw[markedEdgeR](b2)--(b3);
}{\oplus}
\tikzBox[0.75]{\twoLoopGraph{3}{2}{3}{1}\coordinate(b1)at(v0);\coordinate(b3)at(v1);
\decorateA\draw[markedEdgeR](a1)--(a2);
\decorateC\draw[markedEdgeR](c1)--(c2);
}\decorateNo
\Bigg\}\,.
\end{split}}
These examples illustrate how for any power-counting $p$ we may decorate all graphs that are \hyperlink{parent}{parents} of subsets of $\mathfrak{S}_p^L$ with vector-spaces of loop-dependent numerators. 

%================================================================================================================
\vspace{-0pt}\subsection{Two-Loop Integrand Bases in Various Dimensions}\label{subsec:illustrations_of_bases_at_two_loops}\vspace{-0pt}
%================================================================================================================

%================================================================================================================
\vspace{-0pt}\subsubsection[\mbox{{\it Exempli gratia}: \texorpdfstring{$d$}{\emph{d}}-Dim, \texorpdfstring{$p$}{\emph{p}}-gon Power-Counting Bases for \texorpdfstring{$d\!\leq\!4$}{\emph{d}<=\emph{4}}}]{\mbox{{\it Exempli gratia}: $p$-gon Power-Counting Bases in $d$-Dimensions ($d\!\leq\!4$)}}\label{subsubsec:big_table_at_two_loops}\vspace{-0pt}
%================================================================================================================

In \mbox{Table \ref{tab:two_loop_counting_table}} we have summarized the bases for non-product topologies relevant for $d\!=\!2,3,4$ dimensions for $p$-gon power-counting through $p\!=\!4$. In \mbox{Table \ref{tab:two_loop_counting_table}}, we have written `$[1]$', `$[1{-}2]$', and `$[2]$' for $[\cca{\ell_A}],[\ccb{\ell_B}],$ and $[\ccc{\ell_C}]$, respectively. (We do not list product topologies, as their numerator-spaces are entirely dictated by the breakdown of one-loop degrees of freedom.)

As the reader will recall from one loop, the case of $p\!=\!d$ is a non-typical case. Thus, \mbox{Table \ref{tab:two_loop_counting_table}} does not include the complete list of topologies relevant for $4$-gon power-counting in 4 dimensions. In addition to the product topologies---\twoLoopLadderNameText{4}{0}{5} and \twoLoopLadderNameText{5}{0}{5}---whose degrees of freedom ({\color{topCount}}1 {top-level} degree of freedom for each) follow from one loop results, there are three additional integrand structures that need to be included to make the basis complete. These integrands are
\vspace{-4pt}\emphEdges\eq{\fwbox{0pt}{\hspace{-20pt}\begin{array}{@{}l@{}c@{$\;\quad\;$}c@{$\;\quad\;$}c@{}}
&\tikzBox[0.7]{\twoLoopGraph{4}{1}{4}{1}}&\tikzBox[0.7]{\twoLoopGraph{4}{2}{3}{1}}&\tikzBox[0.7]{\twoLoopGraph{3}{3}{3}{1}}\\
\mathfrak{N}_4\textbf{:}&[\cca{\ell_A}][\ccc{\ell_C}]\!\oplus\![\ccb{\ell_B}]&[\cca{\ell_A}]^{\cca2}[\ccc{\ell_C}]\!\oplus\![\cca{\ell_A}][\ccb{\ell_B}]&[\cca{\ell_A}][\ccb{\ell_B}][\ccc{\ell_C}]\!\oplus\![\cca{\ell_A}]^{\cca2}\!\!\oplus\![\ccb{\ell_B}]^{\ccb2}\!\!\oplus\![\ccc{\ell_C}]^{\ccc2}\\
\underset{d=4\hspace{4pt}~}{\text{\textbf{ranks: }}}&{\color{totalCount}36}\!=\!{\color{topCount}\mathbf{3}}{\color{dim}+33}&{\color{totalCount}120}\!=\!{\color{topCount}\mathbf{2}}{\color{dim}+118}&{\color{totalCount}181}\!=\!{\color{topCount}\mathbf{1}}{\color{dim}+180}
\end{array}}\label{additional_4gon_integrands_in_4d}}
(In each of these examples it is interesting to notice that the total numerator space is spanned by the first monomial: $[\ccb{\ell_B}]\!\subset[\cca{\ell_A}][\ccc{\ell_C}]$, \mbox{$[\cca{\ell_A}][\ccb{\ell_B}]\!\subset\![\cca{\ell_A}]^{\cca2}[\ccc{\ell_C}]$}, $[\cca{\ell_A}]^{\cca2}\!\!\subset\![\cca{\ell_A}][\ccb{\ell_B}][\ccc{\ell_C}]$, and so-on.)

%======  precompiled table.  ======= 
\begin{table}[ht!]\vspace{-20pt}$$\vspace{-0pt}\fig{-100pt}{1}{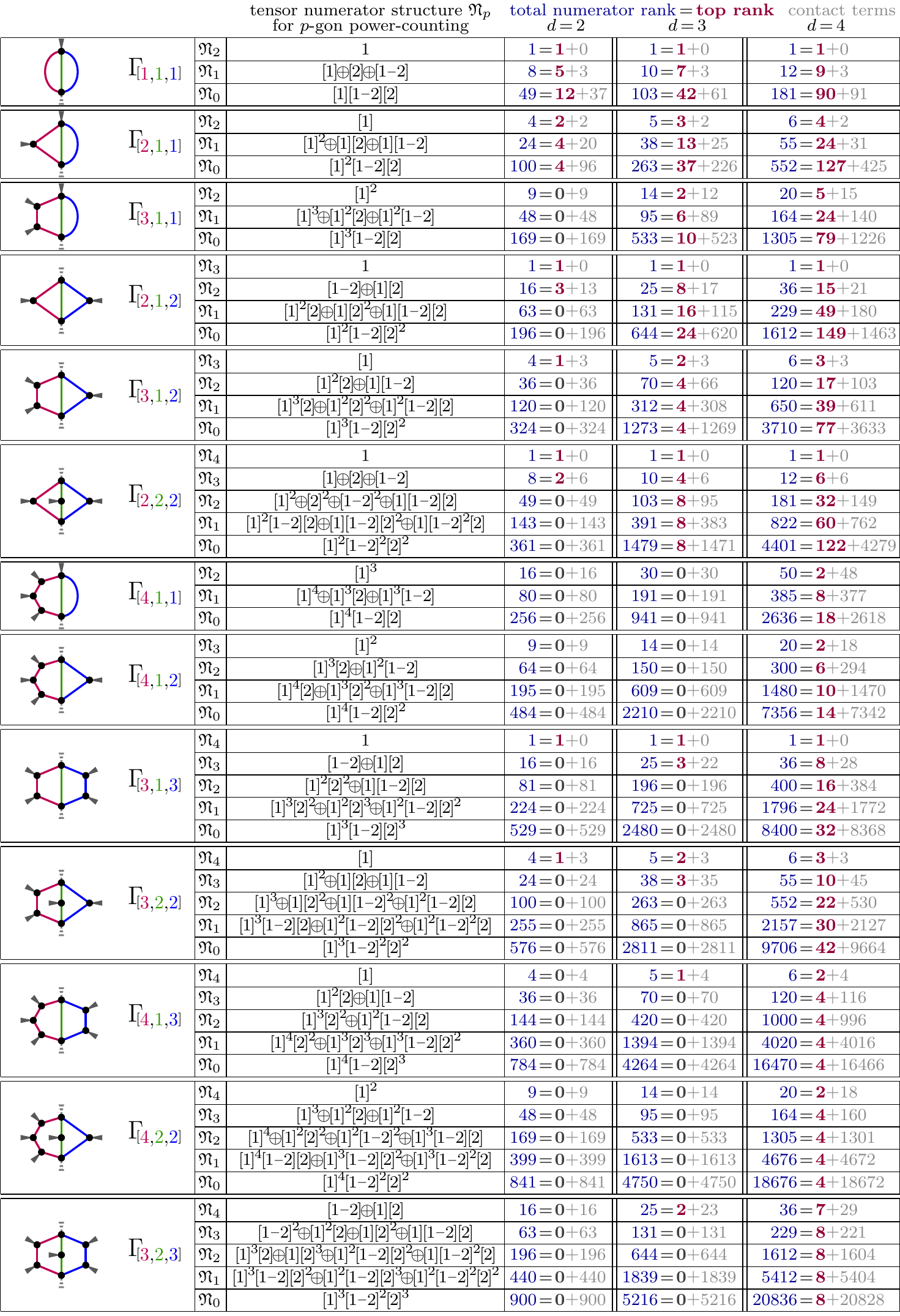}\vspace{-20pt}$$
\caption{
\label{tab:two_loop_counting_table}
Two-loop integrands' loop-dependent numerators $\mathfrak{N}_p$ for $p$-gon power-counting, and the breakdown of their ranks into {\color{hred}top rank} and {\color{deemph}contact-terms} in various dimensions.}\vspace{-40pt}\end{table}\vspace{0pt}
%======                      ======= 

\newpage~\newpage
These additional topologies required for 4-gon power-counting in $4$ dimensions suffer from the same problem we saw for the $4$-gon power-counting pentagons at one loop: they are topologically over-complete. By this we mean that when we include all graphs of these topologies we find that these integrands satisfy non-trivial relations among themselves. It is interesting to note how this over-completeness manifests itself (and can be cleverly avoided) in the case of the planar integrands at two loops. 

The degrees of freedom associated with the product topologies \twoLoopLadderNameText{4}{0}{4}, \twoLoopLadderNameText{4}{0}{5}, and \twoLoopLadderNameText{5}{0}{5} can all be re-cast as non-contact degrees of freedom attached to \twoLoopLadderNameText{4}{1}{4}. This can easily be understood from the viewpoint of cuts: these topologies can all be seen as necessary to match the 4 chiral solutions to the kissing-box cuts, and it is clear that we can match all four using just \twoLoopLadderNameText{4}{1}{4}, if this integrand were assigned {\color{topCount}4} top-level degrees of freedom. This is precisely what was done in ref.~\cite{ArkaniHamed:2010kv,ArkaniHamed:2010gh,Bourjaily:2015jna} at two loops: by excluding the product topologies from the basis, a topologically \emph{complete}---and importantly, \emph{not over-complete}---basis of planar, two-loop integrands with 4-gon power-counting was constructed (and used to represent all two-loop amplitude integrands of planar sYM). We do not know of any similarly clever choice of assigning non-planar degrees of freedom to two-loop integrands in (\ref{additional_4gon_integrands_in_4d}). \\

One final comment worth mentioning is that the breakdown of integrand numerators into {\color{topCount}top-level} degrees of freedom and contact-terms depends on there being generic (and non-zero) momentum flowing into every \hyperlink{necessaryEdges}{\emph{necessary} external edge} indicated (by a solid wedge attached to the vertex of a graph). When there are conspiracies amongst the momenta, or when some momenta vanish, some of the propagators can become doubled, and their corresponding degrees of freedom must change slightly. Dealing with this subtlety requires a small aside.

%================================================================================================================
\phantomsection\markright{}\addcontentsline{toc}{paragraph}{\hspace{-4pt} {\it Si Opus Sit}: Exceptional Cases Requiring Moderate Care}
%================================================================================================================
\vspace{-0pt}\paragraph*{{\it Si Opus Sit}: Exceptional Cases Requiring Modest Refinement}\label{para:exceptional_cases}\vspace{-0pt}~\\[-12pt]
%================================================================================================================

\noindent
In many renormalization schemes, amplitudes require integrands that involve loop-corrections to propagators---or otherwise integrands involving doubled propagators\footnote{Applying unitarity-based ideas to determine the coefficients of such integrands can be subtle, but some technology does exist to deal with this case; see for example refs.~\cite{Sogaard:2014ila} or \cite{Abreu:2017xsl}.} (see \emph{e.g.}~\cite{Baumeister:2019rmh} for possible exceptions). These integrands must be handled with some care, as our combinatorial rules discussed above relied upon an assumption of generality among the propagators of a given integrand. This rule is enforced by our conventions requiring that solid wedges of momenta flowing into a graph are all generic and non-vanishing, while dashed wedges of external momenta can be taken to be zero. Thus, all the propagators in 
\eq{\emphEdges\begin{tikzpicture}[scale=\figScale,baseline=-3.05]\twoLoopGraph{2}{1}{1}{1.25};
\end{tikzpicture}\;\,,\label{generic_211}}
for example, are distinct. This is true even if the momentum flowing into the bottom, `\hyperlink{optionalEdges}{optional}' external edge of the graph were taken to be zero. However, if the momentum flowing into the top of the graph were to become zero as well as in
\def\figScale{1}\eq{\emphEdges\begin{tikzpicture}[scale=\figScale,baseline=-3.05]\twoLoopVacGraph{2}{1}{1}{1.25}\leg{(a1)}{180};\emptyLeg{(v1)}{-90};\emptyLeg{(v0)}{90};\end{tikzpicture}\label{degenerate_211}}
then, as a rational function of loop momentum, the integrand would be indistinguishable from 
\def\figScale{1}\eq{\emphEdges\begin{tikzpicture}[scale=\figScale,baseline=-3.05]\twoLoopVacGraph{2}{1}{1}{1.25}\leg{(a1)}{180};\emptyLeg{(v1)}{-90};\emptyLeg{(v0)}{90};\end{tikzpicture}\sim\begin{tikzpicture}[scale=\figScale,baseline=-3.05]\twoLoopVacGraph{2}{1}{1}{1.25}\emptyLeg{(a1)}{180};\emptyLeg{(v1)}{-90};\leg{(v0)}{90};\end{tikzpicture}\;\,.}
This example may seem like a purely academic concern, but it affects the breakdown of {\color{totalCount}total} numerator degrees of freedom into {\color{topCount}top-level} and contact-terms---and has a knock-on effect for many graphs that include (\ref{degenerate_211}) among their contact-terms. Moreover, this can render otherwise reducible integrands suddenly irreducible. 

Consider for example case of bubble (2-gon) power-counting in $d\!=\!4$. In the generic case of \twoLoopLadderNameText{2}{1}{1} shown in (\ref{generic_211}), it is easy to see that its {\color{totalCount}6} total numerator degrees of freedom from $[\cca{\ell_A}]_4$ decompose into {\color{topCount}4} top-level degrees of freedom and 2 contact-terms. In the degenerate case, there is only a single contact term, leaving us now with {\color{topCount}5} top-level numerators. 

Consider now the degenerate case of \twoLoopLadderName{3}{1}{1}\!---also for bubble power-counting in $d\!=\!4$. It would be assigned a numerator of $[\cca{\ell_A}]^{\cca2}_4$ with {\color{totalCount}20} total degrees of freedom. In this case, \emph{some of} its contact-terms would be the non-degenerate (\ref{generic_211}), while others would be the degenerate case. A simple exercise shows that 
\eq{\underset{d=4}{\mathrm{rank}}\Bigg[\mathfrak{N}_2\Bigg(\emphEdges\begin{tikzpicture}[scale=\figScale,baseline=-3.05]\twoLoopVacGraph{3}{1}{1}{1.25}\leg{(a1)}{180-30};\leg{(a2)}{180+30};\emptyLeg{(v1)}{-90};\emptyLeg{(v0)}{90};\end{tikzpicture}\Bigg)\Bigg]={\color{totalCount}20}={\color{topCount}9}{\color{dim}+11}\,,}
instead of the usual breakdown of ${\color{totalCount}20}\!=\!{\color{topCount}5}{\color{dim}+15}$ of the generic case of \twoLoopLadderNameText{3}{1}{1}. Continuing in this manner, we would discover that
\eq{\underset{d=4}{\mathrm{rank}}\Bigg[\mathfrak{N}_2\Bigg(\emphEdges\begin{tikzpicture}[scale=\figScale,baseline=-3.05]\twoLoopVacGraph{5}{1}{1}{1.25}\leg{(a1)}{90+36};\leg{(a2)}{90+36*2};\leg{(a3)}{90+36*3};\leg{(a4)}{90+36*4};\optLeg{(v1)}{-90};\leg{(v0)}{90};\end{tikzpicture}\Bigg)\Bigg]={\color{totalCount}105}={\color{topCount}\mathbf{0}}{\color{dim}+105}\quad\text{while}\quad\underset{d=4}{\mathrm{rank}}\Bigg[\mathfrak{N}_2\Bigg(\emphEdges\begin{tikzpicture}[scale=\figScale,baseline=-3.05]\twoLoopVacGraph{5}{1}{1}{1.25}\leg{(a1)}{90+36};\leg{(a2)}{90+36*2};\leg{(a3)}{90+36*3};\leg{(a4)}{90+36*4};\emptyLeg{(v1)}{-90};\emptyLeg{(v0)}{90};\end{tikzpicture}\Bigg)\Bigg]={\color{totalCount}105}={\color{topCount}2}{\color{dim}+103}\,.\nonumber}
%

%================================================================================================================
%    3. Three Loops
%         
%================================================================================================================
\newpage
\vspace{-6pt}\section{Building Bases of Integrands at Three Loops}\label{sec:three_loops}\vspace{-0pt}
%================================================================================================================

Following such detailed one- and two-loop discussions above, we will be more telegraphic in our description of our main results at three loops. From what we have seen previously, it is fairly straightforward to generate a basis of integrands at higher loops guaranteed to be {\it big enough} for quantum field theories such as the Standard Model in any fixed number of dimensions. Amplitudes in such (renormalizable) theories can be represented in a basis with  $0$-gon power-counting, as this basis will include literally every Feynman diagram.

The much harder---and more subtle (and interesting)---problem is how to construct and organize bases of integrands with \emph{better}-than-$0$-gon power-counting such that the amplitudes of interesting quantum field theories can be represented. Take for example maximally supersymmetric Yang-Mills theory (sYM). Amplitudes in this theory are widely expected to be representable in terms of integrands with `box power-counting' in 4 dimensions; but until now, there has been no precise definition of such an integrand basis beyond the planar limit. 

In this section, we describe the basic ingredients required to describe integrand bases at three loops, and summarize what is found in the case of 3-gon power-counting in 4 dimensions. This particular case is interesting because this basis should span all maximally transcendental (another poorly defined notion) functions---as integrands with bubble power-counting are expected to universally have less than maximal transcendental weights.\footnote{Providing a precise definition of transcendental weight is complicated by the need for non-polylogarithmic integrands in general. See \emph{e.g.}~\cite{Bourjaily:2017bsb,Bourjaily:2018ycu,Bourjaily:2018yfy,Bourjaily:2019hmc,Vergu:2020uur}.} Thus, the basis we describe here should suffice to represent all scattering amplitudes in sYM beyond the planar limit. 

%================================================================================================================
\vspace{-0pt}\subsection{Three-Loop Integrand Bases: Basic Building Blocks}\label{subsec:three_loop_general_structure}\vspace{-0pt}
%================================================================================================================

%================================================================================================================
\vspace{-0pt}\subsubsection{Loop-Dependent Denominators: Ladders and Wheels}\label{subsubsec:three_loop_vacuum_graphs}\vspace{-0pt}
%================================================================================================================

At three loops, the Feynman propagator structures of all integrands can be classified as `wheel' or `ladder' type topologies. Specifically, let us define 
\eq{\fwbox{0pt}{\hspace{-15pt}\threeLoopWheelNamePrime{{\color{emphA}a_1}}{{\color{emphB}a_2}}{{\color{emphC}a_3}}{{\color{emphA}b_1}}{{\color{emphB}b_2}}{{\color{emphC}b_3}}\,\bigger{\Leftrightarrow}\hspace{2.5pt}\fig{-60.2pt}{0.75}{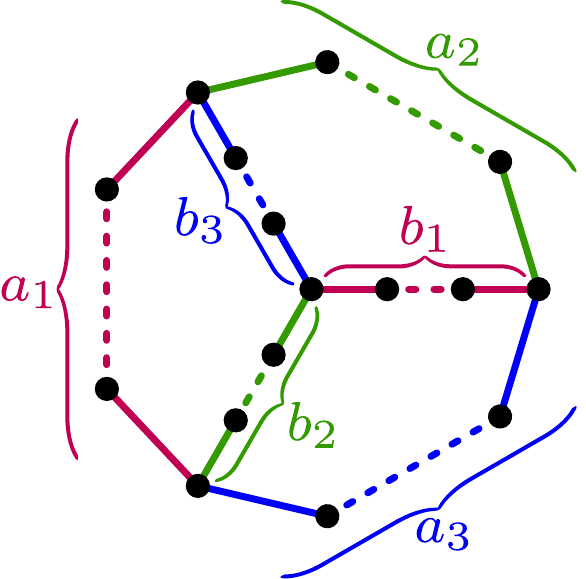}\hspace{-2pt},\hspace{4.5pt}\threeLoopLadderNamePrime{{\color{emphA}a_1}}{{\color{emphA}a_2}}{{\color{emphC}b_1}}{{\color{emphC}b_2}}{{\color{emphB}c_1}}{{\color{emphB}c_2}}\;\bigger{\Leftrightarrow}\fig{-36.25pt}{0.75}{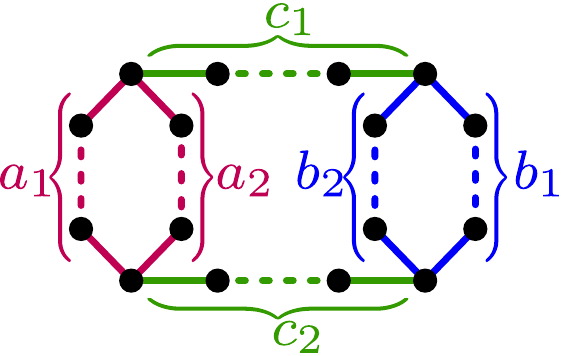}.}\label{three_loop_skeleton_graphs}}
The color-coding above is merely to help guide the eye toward notational meaning and conventions. To be clear, these two classes overlap for degenerate configurations. To disambiguate such cases, we conventionally require that none of the indices of a wheel integral vanish. For example, one can easily see that $\fwbox{40pt}{\threeLoopWheelName{{\color{emphA}a}}{{\color{emphB}b}}{{\color{emphC}c}}{{\color{emphA}d}}{{\color{emphB}e}}{{\bf 0}}}\,\simeq\fwbox{40pt}{\threeLoopLadderName{{\color{emphA}a}}{{\color{emphB}e}}{{\color{emphA}d}}{{\color{emphB}b}}{{\bf{\color{emphC}0}}}{{\color{emphC}c}}}$ is a graph isomorphism. Because of this, we choose to identify all degenerate wheels as instances of (degenerate) ladders. 

In addition to the overlap in name-space for the degenerate wheels and ladders, there are additional redundancies among the labels associated with standard graph isomorphisms. These are the analogues of the permutation-invariance among the indices $\{\cca{a},\ccb{b},\ccc{c}\}$ labeling the two-loop graphs $\twoLoopLadderNameText{a}{b}{c}$. When no indices vanish, the wheel integrands enjoy 24 symmetric relabelings, and the ladders enjoy 16 relabelings corresponding to the sizes of the automorphism groups of the graphs drawn in (\ref{three_loop_skeleton_graphs}), respectively. (There are more symmetries for certain degenerate configurations; for example: $\fwbox{40pt}{\threeLoopLadderName{{\color{emphA}a}}{{\color{emphA}b}}{{\color{emphC}c}}{{\color{emphC}d}}{{\bf{\color{emphB}0}}}{{\color{emphB}{\bf 0}}}}$ is permutation-invariant in its four non-zero labels.) As before, we could in principle solve the momentum conservation constraints and write all topologies in terms of three independent loop momenta, \emph{e.g.}
\eq{
    \tikzBox[0.7]{\wheelVacGraphWithLabels{1}{1}{1}{1}{1}{1}{2} \node[label={[label distance=1pt,rotate=-60]0:$\ccc{\ell_2{-}\ell_1}$}] at (-1.6,1.3) {};\node[label={[label distance=1pt,rotate=-300]0:$\ccb{\ell_1{-}\ell_3}$}] at (-2.1,-0.8) {};\node[label={[label distance=1pt,rotate=0]0:$\cca{\ell_3{-}\ell_2}$}] at (-1.4,-0.15) {};} \qquad\text{and}\hspace{55pt}\raisebox{26pt}{\tikzBox[0.7]{\ladderThreeVacGraph{1}{1}{1}{1}{1}{1}{2.2} \node[label={[label distance=1pt,rotate=0]0:$\cca{\ell_1}$}] at (-2.3,-0.8) {}; \node[label={[label distance=1pt,rotate=0]0:$\ccb{\ell_2}$}] at (-0.5,0.3) {};\node[label={[label distance=1pt,rotate=0]0:$\ccc{\ell_3}$}] at (1.4,-0.8) {};\node[label={[label distance=1pt,rotate=90]0:$\cca{\ell_1{-}\ell_2}$}] at (-0.5,-1.5) {};\node[label={[label distance=1pt,rotate=270]0:$\ccc{\ell_3{-}\ell_2}$}] at (0.75,-0.15) {};}}\hspace{20pt}.
}%
Often, we will only make use of graph theoretic notions of the relevant numerator spaces; however, when we actually compute the ranks of various numerator spaces, we do solve momentum conservation explicitly as indicated above. 

%================================================================================================================
\vspace{-0pt}\subsubsection{Loop-Dependent Numerators: Open Problems}\label{subsubsec:three_loop_tensor_structure}\vspace{-0pt}
%================================================================================================================

By dressing each of the propagators of each type (\ref{three_loop_skeleton_graphs}) with the corresponding space of generalized inverse-propagators, and specifying the dimension of spacetime, one may construct a complete basis of loop integrands sufficient to reproduce scattering amplitudes (to arbitrary multiplicity) in many theories. The total ranks of these spaces of numerators, however, grow very large; and we have not found a closed formula for them as we did at two loops (for dimensions less than five) in eq.~(\ref{eq:closed_form_rank_formula_2loop}). Analogous three-loop formulae would involve six indices and could be written
\eq{\begin{split}
    \mathfrak{w}_d^0(\cca{a_1},\ldots,\ccc{b_3}) &= \text{rank}\Big( 
    [\cca{\ell_1}]^{\cca{a_1}}[\ccb{\ell_2}]^{\ccb{a_2}}[\ccc{\ell_3}]^{\ccc{a_3}}
    [\cca{\ell_3{-}\ell_2}]^{\cca{b_1}}[\ccb{\ell_1{-}\ell_3}]^{\ccb{b_2}}[\ccc{\ell_2{-}\ell_1}]^{\ccc{b_3}}
    \Big)\,,\\
        \mathfrak{l}_d^0(\cca{a_1},\ldots,\ccb{c_2}) &= \text{rank}\Big( 
    [\cca{\ell_1}]^{\cca{a_1}}[\cca{\ell_1{-}\ell_2}]^{\cca{a_2}}[\ccb{\ell_2}]^{\ccb{c_1{+}c_2}}\, 
    [\ccc{\ell_3}]^{\ccc{b_1}}[\ccc{\ell_3{-}\ell_2}]^{\ccc{b_2}}
    \Big)\,.
\end{split}}
Again, it would be desirable to find a group-theoretic expression for the relevant numerator ranks similar to the simple one-loop expression in (\ref{pnum_ddim_rank}).

%================================================================================================================
\newpage\vspace{-0pt}\subsection{The Three-Loop Triangle Power-Counting Basis for Four Dimensions}\label{subsec:three_loop_triangle_power_counting}\vspace{0pt}
%================================================================================================================

Despite lacking a general rank count at three loops, however, it seems like a good idea to tackle this general problem in stages, starting with an integrand basis suitable for a theory with `good' power-counting, such as sYM. In four dimensions, the best-case would probably correspond to `box' power-counting $p\!=\!4$. But as with lower loops, there are good reasons to consider instead the space of integrands with next-to-optimal power-counting. In four dimensions, this corresponds to triangles. 

Why should we be interested in three loop integrands with triangle power-counting? After all, we expect that the best quantum field theories (in terms of ultraviolet behavior) should be expressible in terms of boxes. The answer is the same as at lower loops: insisting on integrands with box power-counting forces us to include topologies with more than $4L$ propagators, and such integrands generically satisfy relations that must be eliminated. In the best case scenario, these redundancies can be excluded by throwing out entire topological classes of integrands---as was (accidentally) the case for two loops in the planar limit. We suspect that such a strategy is doomed in general; but as with the concrete examples provided in refs.~\cite{Bourjaily:2019iqr,Bourjaily:2019gqu}, we suspect that nice integrand formulae exist for sYM beyond the planar limit even if we use a basis of integrands with next-to-optimal (namely, $p\!=\!3$) power-counting.\\

Recall that our definition of $p$-gon power-counting (in any number dimensions, and any loop-order) starts with a definition of \emph{scalar} integrands $\mathfrak{S}_p^L$. Recall that this consists of all vacuum graphs with \hyperlink{girth}{girth} $p$, such that all single-edge quotients have lower girth.

At three loops and $3$-gon power-counting, the set of scalars $\mathfrak{S}_3^3$ is given by 
\def\colorLabelsQ{1}
\vspace{-6pt}\eq{\fwbox{0pt}{\hspace{-25pt}\left\{\rule{0.0pt}{36pt}\right.\hspace{-7pt}\begin{array}{@{}c@{}}\fig{-20pt}{0.75}{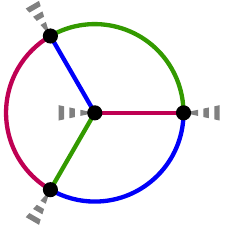}\\\threeLoopWheelName{1}{1}{1}{1}{1}{1}\end{array},
\begin{array}{@{}c@{}}\fig{-20pt}{0.75}{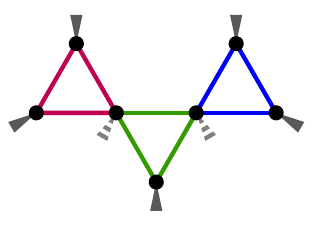}\\\fwbox{0pt}{\hspace{-12.pt}\threeLoopLadderName{3}{0}{0}{3}{1}{2}}\end{array},\!
\begin{array}{@{}c@{}}\fig{-20pt}{0.75}{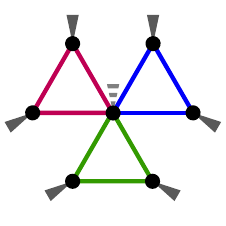}\\\fwbox{0pt}{\hspace{-12.pt}\threeLoopLadderName{3}{0}{0}{3}{0}{3}}\end{array},\!
\begin{array}{@{}c@{}}\fig{-20pt}{0.75}{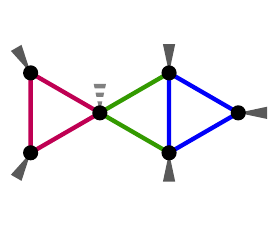}\\\hspace{-12.pt}\threeLoopLadderName{3}{0}{1}{2}{1}{1}\end{array},\!
\begin{array}{@{}c@{}}\fig{-20pt}{0.75}{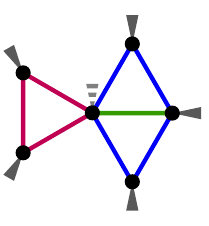}\\\hspace{-12.pt}\threeLoopLadderName{3}{0}{2}{2}{0}{1}\end{array},\!
\begin{array}{@{}c@{}}\fig{-20pt}{0.75}{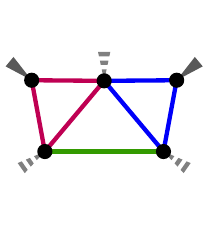}\\\hspace{-12.pt}\threeLoopLadderName{2}{1}{1}{2}{0}{1}\end{array},\!
\begin{array}{@{}c@{}}\fig{-20pt}{0.75}{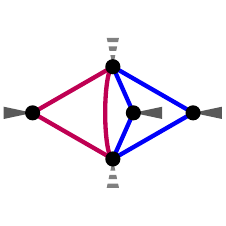}\\\hspace{-12.pt}\threeLoopLadderName{2}{1}{2}{2}{0}{0}\end{array}\hspace{-7pt}
\left.\rule{0.0pt}{36pt}\right\}\!\!.}
\label{3gon_scalar_graphs}
}
Notice that all but the first, sixth, and seventh of these are product-topologies. As always, there are multiple ways to label each of these graphs; the labeling we have chosen should be viewed as representative.

Having defined our basic scalar 3-gon power-counting topologies, we proceed to construct the numerator spaces for integrands with more propagators. The basic setup is almost identical to our more detailed two-loop discussion, which is why we are going to be relatively brief here. We need not dwell on the numerator decomposition of any product topologies, as their decomposition will follow trivially from our one- and two-loop discussions above. As before, we find that all (generic) integrands with more than $12\!=\!3\!\times\!4$ propagators are entirely decomposable; the ranks for these numerator spaces quoted below were obtained mostly from direct construction. 

Although the construction of $p$-gon power-counting numerator spaces follows directly from our discussion at two loops, it may be helpful to illustrate the non-triviality of this construction with a couple examples. Consider the ladder and wheel integrands, 
\vspace{-6pt}\eq{
    \threeLoopLadderName{{\color{emphA}3}}{{\color{emphA}1}}{{\color{emphC}2}}{{\color{emphC}2}}{{\color{emphB}1}}{{\color{emphB}2}} 
    \,= 
    %\fwboxL{53pt}{\hspace{-0pt}\fig{-10.5pt}{1}{ladder_312212_n1}}\,,
    \hspace{-14pt}\raisebox{-14.5pt}{\emphEdges
    \begin{tikzpicture}
        \ladderThreeGraph{3}{1}{2}{2}{1}{2}
    \end{tikzpicture}}\,,
    \quad\text{}\quad
    \threeLoopWheelName{{\color{emphA}3}}{{\color{emphB}2}}{{\color{emphC}1}}{{\color{emphA}2}}{{\color{emphB}1}}{{\color{emphC}1}}
    \,=
    %\fwboxL{45pt}{\fig{-19.5pt}{1}{wheel_321211_n1}}
    \hspace{-25pt}
    \raisebox{-24pt}{
    \begin{tikzpicture}
        \wheelGraph{3}{2}{1}{2}{1}{1}
    \end{tikzpicture}}\hspace{20pt}\,.\label{two_example_topologies}
\vspace{-6pt}}
For each of these topologies, the numerator space is defined as the product of translated inverse propagators for all sets of edges that, upon their collapse, would lead to an element of $\mathfrak{S}_3^3$ in (\ref{3gon_scalar_graphs}). From these spaces, the {\color{totalCount}total rank} may be computed (by brute force) in any number of dimensions, and the breakdown into {\color{topCount} top-level} degrees of freedom and contact-terms follows recursively by analogy with (\ref{contact_term_rank_decomposition}) at two loops. 

For the ladder example in (\ref{two_example_topologies}), we would find the total numerator vector-space to be given by 
\vspace{-7pt}\eq{
\underbrace{
\fwboxL{53pt}{\begin{array}{@{}c@{}r@{}}\hspace{-0pt}\fig{-10.5pt}{1}{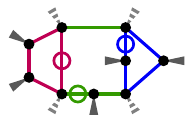}\hspace{0pt}&\\[-3.5pt]
\hspace{0pt}\threeLoopTensorNameL{0}{1}{0}{1}{0}{1}
&\fwboxR{0pt}{\text{{\footnotesize$\oplus$}}\hspace{-6pt}}\\[-2.5pt]\end{array}}
\fwboxL{53pt}{\begin{array}{@{}c@{}r@{}}\hspace{-0pt}\fig{-10.5pt}{1}{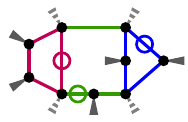}\hspace{0pt}&\\[-3.5pt]
\hspace{0pt}\threeLoopTensorNameL{0}{1}{1}{1}{0}{0}
&\fwboxR{0pt}{\text{{\footnotesize$\oplus$}}\hspace{-3.5pt}}\\[-2.5pt]\end{array}}
}_{\substack{\\[-3pt]\begin{turn}{90}$\prec$\end{turn}\\[-9pt]\fig{-20pt}{0.75}{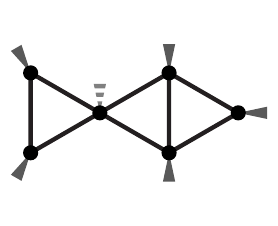}}}
\underbrace{
\fwboxL{53pt}{\begin{array}{@{}c@{}r@{}}\hspace{-0pt}\fig{-10.5pt}{1}{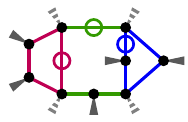}\hspace{0pt}&\\[-3.5pt]
\hspace{0pt}\threeLoopTensorNameL{0}{1}{0}{1}{0}{1}
&\fwboxR{0pt}{\text{{\footnotesize$\oplus$}}\hspace{-6pt}}\\[-2.5pt]\end{array}}
\fwboxL{53pt}{\begin{array}{@{}c@{}r@{}}\hspace{-0pt}\fig{-10.5pt}{1}{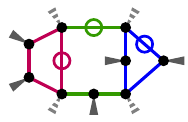}\hspace{0pt}&\\[-3.5pt]
\hspace{0pt}\threeLoopTensorNameL{0}{1}{1}{1}{0}{0}
&\fwboxR{0pt}{\text{{\footnotesize$\oplus$}}\hspace{-5pt}}\\[-2.5pt]\end{array}}
\fwboxL{53pt}{\begin{array}{@{}c@{}r@{}}\hspace{-0pt}\fig{-10.5pt}{1}{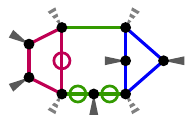}\hspace{0pt}&\\[-3.5pt]
\hspace{0pt}\threeLoopTensorNameL{0}{2}{0}{1}{0}{0}
&\fwboxR{0pt}{\text{{\footnotesize$\oplus$}}\hspace{-3.5pt}}\\[-2.5pt]\end{array}}
}_{\substack{\\[-3pt]\begin{turn}{90}$\prec$\end{turn}\\[-9pt]\fig{-20pt}{0.75}{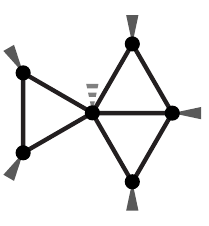}}}
\underbrace{
\fwboxL{53pt}{\begin{array}{@{}c@{}r@{}}\hspace{-0pt}\fig{-10.5pt}{1}{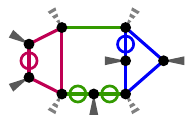}\hspace{0pt}&\\[-3.5pt]
\hspace{0pt}\threeLoopTensorNameL{1}{2}{0}{0}{0}{1}
&\fwboxR{0pt}{\text{{\footnotesize$\oplus$}}\hspace{-6pt}}\\[-2.5pt]\end{array}}
\fwboxL{53pt}{\begin{array}{@{}c@{}r@{}}\hspace{-0pt}\fig{-10.5pt}{1}{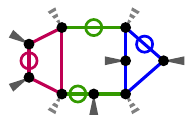}\hspace{0pt}&\\[-3.5pt]
\hspace{0pt}\threeLoopTensorNameL{1}{2}{1}{0}{0}{0}
&\fwboxR{0pt}{\text{{\footnotesize$\oplus$}}\hspace{-4pt}}\\[-2.5pt]\end{array}}
}_{\substack{\\[-3pt]\begin{turn}{90}$\prec$\end{turn}\\[-9pt]\fig{-20pt}{0.75}{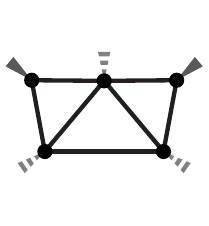}}}
\underbrace{
\fwboxL{53pt}{\begin{array}{@{}c@{}r@{}}\hspace{-0pt}\fig{-10.5pt}{1}{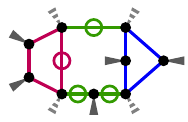}\hspace{0pt}&\\[-3.5pt]
\hspace{0pt}\threeLoopTensorNameL{1}{3}{0}{0}{0}{0}
&\fwboxR{0pt}{\text{{\footnotesize}}\hspace{-5pt}}\\[-2.5pt]\end{array}}
}_{\substack{\\[-3pt]\begin{turn}{90}$\prec$\end{turn}\\[-9pt]\fig{-20pt}{0.75}{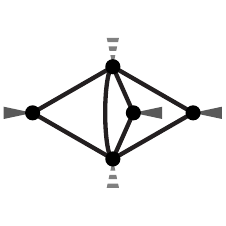}}}
\nonumber\vspace{-6pt}}
resulting in a loop-dependent numerator of
\vspace{-4pt}\eq{\mathfrak{N}_3\Big(\!\hspace{-1pt}\threeLoopLadderName{{\color{emphA}3}}{{\color{emphA}1}}{{\color{emphC}2}}{{\color{emphC}2}}{{\color{emphB}1}}{{\color{emphB}2}}\Big)\equivR\threeLoopTensorNameLb{0}{1}{0}{1}{0}{1}\!\oplus\!
\threeLoopTensorNameLb{0}{1}{1}{1}{0}{0}\!\oplus\!
\threeLoopTensorNameLb{0}{2}{0}{1}{0}{0}\!\oplus\!
\threeLoopTensorNameLb{1}{2}{0}{0}{0}{1}\!\oplus\!
\threeLoopTensorNameLb{1}{2}{1}{0}{0}{0}\!\oplus\!
\threeLoopTensorNameLb{1}{3}{0}{0}{0}{0}\,.
\label{eq:eg_num_span_3L_ladder}
\vspace{-4pt}}
This vector-space is the same in any number of dimensions, but its size and breakdown into contact-terms depends strongly on $d$. In four dimensions, it can readily be confirmed that
\vspace{-6pt}\begin{align}
\left|\hspace{-1pt}\mathfrak{N}_3\Big(\!\hspace{-1pt}\threeLoopLadderName{{\color{emphA}3}}{{\color{emphA}1}}{{\color{emphC}2}}{{\color{emphC}2}}{{\color{emphB}1}}{{\color{emphB}2}}\Big)\hspace{-1pt}\right|\hspace{-2pt}&=
\underset{d=4}{\text{rank}}\hspace{-0.5pt}\Big(\hspace{-1pt}
\threeLoopTensorNameLb{0}{1}{0}{1}{0}{1}\!\oplus\!
\threeLoopTensorNameLb{0}{1}{1}{1}{0}{0}\!\oplus\!
\threeLoopTensorNameLb{0}{2}{0}{1}{0}{0}\!\oplus\!
\threeLoopTensorNameLb{1}{2}{0}{0}{0}{1}\!\oplus\!
\threeLoopTensorNameLb{1}{2}{1}{0}{0}{0}\!\oplus\!
\threeLoopTensorNameLb{1}{3}{0}{0}{0}{0}
\Big)\nonumber\\
&={\color{totalCount}984}={\color{topCount}32}{\color{dim}\pl952}\,.\nonumber\\[-20pt]
\end{align}
Thus, even though the {\color{totalCount}total rank} of the numerator space is {\color{totalCount}984}, the number of degrees of freedom that are honestly associated to the $\threeLoopLadderName{{\color{emphA}3}}{{\color{emphA}1}}{{\color{emphC}2}}{{\color{emphC}2}}{{\color{emphB}1}}{{\color{emphB}2}}$ topology is {\color{topCount}32} and therefore relatively small. 

For the wheel example in (\ref{two_example_topologies}), we would find its numerator constructed according to the scalar contact-term topologies, 
\vspace{-6pt}$$\underbrace{
\fwboxL{45pt}{\begin{array}{@{}c@{}r@{}}\fig{-19.5pt}{1}{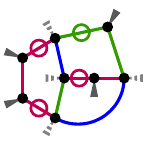}&\\[-9.5pt]
\hspace{0pt}\threeLoopTensorNameW{2}{1}{0}{0}{0}{1}
&\fwboxR{0pt}{\text{{\footnotesize$\oplus$}}\hspace{-5pt}}\\[-1.5pt]\end{array}}
}_{\substack{\\[-3pt]\begin{turn}{90}$\prec$\end{turn}\\[-6.5pt]\fig{-20pt}{0.675}{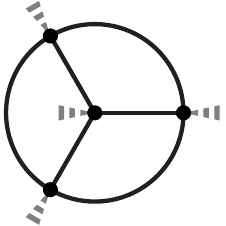}}}
\underbrace{
\fwboxL{42.5pt}{\begin{array}{@{}c@{}r@{}}\fig{-19.5pt}{1}{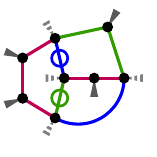}&\\[-9.5pt]
\hspace{-3pt}\threeLoopTensorNameW{0}{0}{0}{1}{1}{0}
&\fwboxR{0pt}{\text{{\footnotesize$\oplus$}}\hspace{-3pt}}\\[-1.5pt]\end{array}}
}_{\substack{\\[-3pt]\begin{turn}{90}$\prec$\end{turn}\\[-9pt]\fig{-20pt}{0.75}{scalar3_L302201_b}}}
\underbrace{
\fwboxL{45pt}{\begin{array}{@{}c@{}r@{}}\fig{-19.5pt}{1}{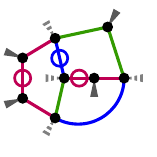}&\\[-9.5pt]
\hspace{-1pt}\threeLoopTensorNameW{1}{0}{0}{1}{0}{1}
&\fwboxR{0pt}{\text{{\footnotesize$\oplus$}}\hspace{-6pt}}\\[-1.5pt]\end{array}}
\fwboxL{45pt}{\begin{array}{@{}c@{}r@{}}\fig{-19.5pt}{1}{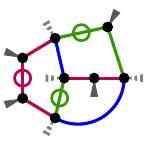}&\\[-9.5pt]
\threeLoopTensorNameW{1}{1}{0}{0}{1}{0}
&\fwboxR{0pt}{\text{{\footnotesize$\oplus$}}\hspace{-4pt}}\\[-1.5pt]\end{array}}
\fwboxL{45pt}{\begin{array}{@{}c@{}r@{}}\fig{-19.5pt}{1}{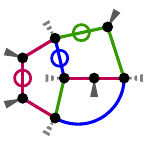}&\\[-9.5pt]
\threeLoopTensorNameW{1}{1}{0}{1}{0}{0}
&\fwboxR{0pt}{\text{{\footnotesize$\oplus$}}\hspace{-7pt}}\\[-1.5pt]\end{array}}
\fwboxL{45pt}{\begin{array}{@{}c@{}r@{}}\fig{-19.5pt}{1}{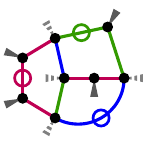}&\\[-9.5pt]
\threeLoopTensorNameW{1}{1}{1}{0}{0}{0}
&\fwboxR{0pt}{\text{{\footnotesize$\oplus$}}\hspace{-5pt}}\\[-1.5pt]\end{array}}
\fwboxL{45pt}{\begin{array}{@{}c@{}r@{}}\fig{-19.5pt}{1}{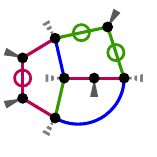}&\\[-9.5pt]
\threeLoopTensorNameW{1}{2}{0}{0}{0}{0}
&\fwboxR{0pt}{\text{{\footnotesize$\oplus$}}\hspace{-5pt}}\\[-1.5pt]\end{array}}
\fwboxL{45pt}{\begin{array}{@{}c@{}r@{}}\fig{-19.5pt}{1}{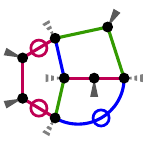}&\\[-9.5pt]
\threeLoopTensorNameW{2}{0}{1}{0}{0}{0}
&\fwboxR{0pt}{\text{{\footnotesize$\oplus$}}\hspace{-6pt}}\\[-1.5pt]\end{array}}
}_{\substack{\\[-3pt]\begin{turn}{90}$\prec$\end{turn}\\[-9pt]\fig{-20pt}{0.75}{scalar3_L211201_b}}}
\underbrace{
\fwboxL{45pt}{\begin{array}{@{}c@{}r@{}}\fig{-19.5pt}{1}{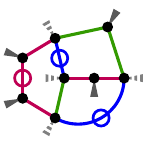}&\\[-9.5pt]
\threeLoopTensorNameW{1}{0}{1}{1}{0}{0}
&\\[-1.5pt]\end{array}}
}_{\substack{\\[-3pt]\begin{turn}{90}$\prec$\end{turn}\\[-9pt]\fig{-20pt}{0.75}{scalar3_L212200_b}}}
$$
resulting in a loop-dependent vector-space of numerators given by 
\eq{\begin{split}\mathfrak{N}_3\Big(\!\hspace{-1pt}\threeLoopWheelName{{\color{emphA}3}}{{\color{emphB}2}}{{\color{emphC}1}}{{\color{emphA}2}}{{\color{emphB}1}}{{\color{emphC}1}}\Big)\equivR&\phantom{\,\oplus\!}
\threeLoopTensorNameWb{2}{1}{0}{0}{0}{1}\oplus\!
\threeLoopTensorNameWb{0}{0}{0}{1}{1}{0}\oplus\!
\threeLoopTensorNameWb{1}{0}{0}{1}{0}{1}\oplus\!
\threeLoopTensorNameWb{1}{1}{0}{0}{1}{0}\\
&\!\!\oplus\!
\threeLoopTensorNameWb{1}{1}{0}{1}{0}{0}\oplus\!
\threeLoopTensorNameWb{1}{1}{1}{0}{0}{0}\oplus\!
\threeLoopTensorNameWb{1}{2}{0}{0}{0}{0}\oplus\!
\threeLoopTensorNameWb{2}{0}{1}{0}{0}{0}\oplus\!
\threeLoopTensorNameWb{1}{0}{1}{1}{0}{0}\,.
\end{split}
\label{eq:eg_num_span_3L_wheel}
}
As always, this vector-space for $3$-gon power-counting is the same for all spacetime dimensions. In $d\!=\!4$, 
\begin{align}
\left|\hspace{-1pt}\mathfrak{N}_3\Big(\!\hspace{-1pt}\threeLoopWheelName{{\color{emphA}3}}{{\color{emphB}2}}{{\color{emphC}1}}{{\color{emphA}2}}{{\color{emphB}1}}{{\color{emphC}1}}\Big)\hspace{-1pt}\right|\hspace{-2pt}&=
\underset{d=4}{\text{rank}}\hspace{-0.5pt}\Big(\hspace{-1pt}
\threeLoopTensorNameWb{2}{1}{0}{0}{0}{1}\oplus\!
\threeLoopTensorNameWb{0}{0}{0}{1}{1}{0}\oplus\!
\threeLoopTensorNameWb{1}{0}{0}{1}{0}{1}\oplus\!
\threeLoopTensorNameWb{1}{1}{0}{0}{1}{0}\nonumber\\
&\hspace{48pt}\!\!\oplus\!
\threeLoopTensorNameWb{1}{1}{0}{1}{0}{0}\oplus\!
\threeLoopTensorNameWb{1}{1}{1}{0}{0}{0}\oplus\!
\threeLoopTensorNameWb{1}{2}{0}{0}{0}{0}\oplus\!
\threeLoopTensorNameWb{2}{0}{1}{0}{0}{0}\oplus\!
\threeLoopTensorNameWb{1}{0}{1}{1}{0}{0}\Big)\nonumber\\
&={\color{totalCount}864}={\color{topCount}64}{\color{dim}\pl800}\,.\label{eg_num_span_3L_wheel_ranks}
\end{align}
As with the ladder example, we find that, although the {\color{totalCount}total} numerator space is quite large, the number of {\color{topCount}top-level} degrees of freedom is comparatively small.

For both of these examples---and all the other topologies relevant to triangle power-counting in four dimensions at three loops---these ranks were computed in two different ways. First, we computed the {\color{totalCount}total rank} by literally constructing the vector-spaces defined as in (\ref{eq:eg_num_span_3L_ladder}) and (\ref{eq:eg_num_span_3L_wheel}),  for example, and determining the {\color{black}rank} by brute force and determining the break-down into contact-terms according to the recursive definition analogous to (\ref{contact_term_rank_decomposition}) at two loops. Secondly, we determined the {\color{topCount}top-level} degrees of freedom of each topology by constructing the total vector-spaces and determining the rank spanned \emph{on the maximal cut of the graph}; and we used this data for all subtopologies of each graph to (recursively) infer the {\color{totalCount}total rank} of the space of numerators. That these two strategies produced the same rank counts gives us confidence in their correctness.

Following this procedure, we were able to construct numerator spaces and their breakdowns into {\color{topCount}top-level} degrees of freedom an contact-terms for all graphs with as many as $12 (\!=\!d\!\times\!L)$ propagators at three loops. We also verified that all graphs with more than $12$ propagators are entirely decomposable into contact-terms. The results of our analysis are summarized in Tables \ref{tab:3loop_res_1}, \ref{tab:3loop_res_2}, and \ref{tab:3loop_res_3} and also attached as \textsc{Mathematica} notebook in the ancillary files of this work's submission to the \texttt{arXiv}. Similarly to our two-loop discussion in section \ref{para:exceptional_cases} there can be exceptional cases of highly degenerate leg ranges where additional topologies are required (that we do not list explicitly here). 

As mentioned above, this basis of integrands should be sufficient to match the four-dimensional integrand of all-multiplicity scattering amplitudes in maximally supersymmetric ($\mathcal{N}\!=\!4$) sYM at three loops beyond the planar limit. As discussed in the following section, the brute-force construction of these vector-spaces is still quite far from providing us with a `nice' set of loop-integrand numerators suitable for the efficient representation of scattering amplitudes. Below, we describe the features desirable in a nice choice of basis elements---which goes well beyond the scope of our present analysis.

%========================================================================================
%=============     3-loop RANK TABLE 1                                 ==================
%========================================================================================
\newpage

\begin{table}[h!]\vspace{-0pt}$$\vspace{-0pt}\threeLoopWheelDataTablePrecompiled\vspace{-16pt}$$
\caption{
\label{tab:3loop_res_1}
Three-loop wheel integrand topologies consistent with triangle power-counting, and the decomposition of their numerators into {\color{topCount}top-level} and contact-term ranks.}\vspace{-30pt}\end{table}

%========================================================================================
%=============     3-loop RANK TABLE 2                                 ==================
%========================================================================================
\newpage

\begin{table}[h!]\vspace{-00pt}$$\vspace{-0pt}\threeLoopLadderDataTableOnePrecompiled\vspace{-16pt}$$
\caption{
\label{tab:3loop_res_2}
(1/2) Three-loop ladder integrand topologies consistent with triangle power-counting, and the decomposition of their numerators into {\color{topCount}top-level} and contact-term ranks.\vspace{-10pt}}\vspace{-40pt}\end{table}

%========================================================================================
%=============     3-loop RANK TABLE 3                                 ==================
%========================================================================================
\newpage

\begin{table}[h!]
\vspace{-0pt}
$$\vspace{-0pt}\threeLoopLadderDataTableTwoPrecompiled\vspace{-16pt}$$
\caption{
\label{tab:3loop_res_3}
(2/2) Three-loop ladder integrand topologies consistent with triangle power-counting, and the decomposition of their numerators into {\color{topCount}top-level} and contact-term ranks.\vspace{-10pt}}\vspace{-40pt}
\end{table}

%================================================================================================================
%    4. Discussion and Conclusions
%         
%================================================================================================================
\newpage\vspace{-6pt}\section{Open Problems and Future Directions}\label{sec:discussion_and_conclusions}\vspace{-6pt}
%================================================================================================================

%=========================================================================
\vspace{-0pt}\subsection{The Importance of Choosing Bases \emph{Wisely}}\label{subsec:choosing_bases_wisely}\vspace{-0pt}
%=========================================================================

The primary objective of this paper has been the enumeration and stratification of integrand bases from purely graph-theoretic considerations. Specifically, we have shown how one can determine (by direct construction) the \emph{number} of independent integrands for all relevant topologies necessary to express large classes of scattering amplitudes at one, two, and three loops in various spacetime dimensions. However, to construct a particular representation of an amplitude requires a \emph{choice of basis} and therefore one is required to select specific loop-momentum dependent numerators for each topology. Of course, we require those numerators to fill up the full numerator rank associated to a given integrand topology for a given power-counting and spacetime dimension. Implementing this procedure in an efficient (and elegant) way is an important open question with no obvious or unique answer. For the particular case of sYM at two loops in four spacetime dimensions, one arguably nice representation has been the subject of our related work \cite{Bourjaily:2019iqr,Bourjaily:2019gqu}. In order to give the reader another example of the substantial benefits gained from our basis partitioning described in this work, we consider some simple examples of an explicit basis construction. Furthermore, we comment on some statements (though not completely general) regarding `desirable' basis choices in this section. 

One strategy for basis construction is to follow the analysis done for three loops. Namely, construct arbitrary representatives of all basis elements, and use a computer algebra package (such as \textsc{Mathematica}) to construct the independent vector-spaces by brute force.  This method is not far from what was followed in traditional strategies of basis construction such as OPP \cite{Ossola:2006us,Mastrolia:2010nb}; it does furnish us with a valid and complete basis of loop integrands in which scattering amplitudes can be expressed, and these can be further refined according to certain criteria, see \emph{e.g.}~\cite{Ita:2015tya,Mastrolia:2016dhn}. 

The rank tables included in our present work provide the structure of the numerators consistent with a given power-counting and therefore allow the straightforward construction of such numerators. As an example, consider the integral topology \twoLoopLadderNameText{2}{2}{2} at two loops in $d\!=\!4$ spacetime dimensions, with triangle (\emph{i.e.}~$3$-gon) power-counting with five massless external legs:
\vspace{10pt}
\eq{\hspace{36pt}\tikzBox[0.7]{\twoLoopVacGraph{2}{2}{2}{2}\singleLegLabelled{(a1)}{180}{$p_1$};\singleLegLabelled{(v0)}{90}{$p_2$};\singleLegLabelled{(c1)}{0}{$\hspace{1pt}p_3$};\singleLegLabelled{(c2)}{-90}{$p_4$};\singleLegLabelled{(b2)}{180}{$p_5$};\arrow{(a1)}{(c2)}{143}{emphA};\arrow{(a1)}{(v0)}{37}{emphA};\arrow{(v0)}{(c1)}{-37}{emphC};\arrow{(c1)}{(c2)}{218}{emphC};\arrow{(c2)}{(b2)}{-90}{emphB};\arrow{(b2)}{(v0)}{-90}{emphB};\node[label={[label distance=\labelDist,rotate=0]0:$\textcolor{emphA}{\ell_1}$}] at ($(a1)!0.5!(c2)+(-0.6,-0.15)$) {};\node[label={[label distance=\labelDist,rotate=0]0:$\textcolor{emphC}{\ell_2}$}] at ($(v0)!0.5!(c1)+(-0.3,0.2)$) {};}\hspace{36pt}.
\vspace{15pt}\label{example_ladder_222}
}
For triangle power-counting, we would have a loop-momentum-dependent numerator given by \mbox{$[\cca{1}]\!\oplus\![\ccb{1\mi2}]\!\oplus\![\ccc{2}]$}. According to \mbox{Table \ref{tab:two_loop_counting_table}}, the dimension of this {\color{totalCount}total} numerator space is {\color{totalCount}12}, which can easily be seen to split into {\color{topCount}6} top-level (non-contact) numerators and 6 additional contact-term numerators---those that are proportional to one of the six inverse propagators of \twoLoopLadderNameText{2}{2}{2} leading to \twoLoopLadderNameText{2}{1}{2} topologies. Concretely, the space is spanned by numerators of the form
\eq{
\mathfrak{N}_3\!\big(\!\twoLoopLadderNameText{2}{2}{2}\big)= 
\underset{d=4}{\text{span}} \big\{[\cca{\ell_1}]\!\oplus\![\ccb{\ell_1{-}\ell_2}]\!\oplus\![\ccc{\ell_2}]\big\}.
}
As described earlier, there are many ways in which to choose representative bases of numerators, although the rank counting obviously does not depend on any particular choice. For the particular loop-momentum routing shown in (\ref{example_ladder_222}), one fairly arbitrary choice for the {\color{topCount}6} non-contact numerators would be\footnote{For $4$-gon power-counting, this integrand topology has only a single numerator and appears in the context of sYM with a special (loop-momentum independent) normalization \cite{Bern:2015ple}. It was later shown that this integrand satisfies a generalized `directional dual-conformal invariance' \cite{Bern:2018oao} which has further implications on the analytic structure of the result after integration \cite{Chicherin:2018wes}.}
\eq{
\widehat{\mathfrak{N}}_3\!\big(\!\twoLoopLadderNameText{2}{2}{2}\big)=\mathrm{span} \big\{1,\x{\cca{\ell_1}}{p_2},\,\x{\cca{\ell_1}}{p_4},\,\x{\ccc{\ell_2}}{p_2},\,\x{\ccc{\ell_2}}{p_4},\,\x{\ccb{\ell_1{-}\ell_2}}{p_2{+}p_4}
\big\}\,.\label{arbitrary_top_numerators_for_222}
}
(Notice that none of these generalized inverse propagators are propagators of the graph in (\ref{example_ladder_222}).) To determine the coefficients of these six integrands in the expansion of some amplitude under consideration requires the solution of a $6\times6$ linear system. If one is not careful in defining integrands that have \twoLoopLadderNameText{2}{2}{2} as subtopology, the linear system can become even larger since one has to take integrand topologies with more propagators into account. However, in the method of maximal cuts (see \emph{e.g.}~\cite{Bern:2007ct}) their coefficients are taken to be fixed by appropriate unitarity cuts so that one is still left with only {\color{topCount}6} unknowns. 

While this example is of course fairly trivial, analogous linear algebra problems quickly become prohibitive as the sizes of integrand bases grow. Therefore, it is clearly desirable to choose `nice' numerators---which partially pre-diagonalize the linear system we need to solve to represent amplitudes. These could be numerators designed to vanish at particular points in loop-momentum space, making the calculation of their coefficients (often) completely trivial. There is a considerable amount of literature on this subject---see \emph{e.g.}~\cite{Bourjaily:2015jna,Bourjaily:2017wjl,Bourjaily:2019iqr,Bourjaily:2019gqu}---although many open questions remain. In the context of sYM amplitudes, we studied some of these integrands extensively and constructed special numerators which diagonalized the system completely \cite{Bourjaily:2019iqr,Bourjaily:2019gqu} by demanding that the numerators vanished at certain special kinematic points, thus implementing the \emph{prescriptive} unitarity method of \cite{Bourjaily:2017wjl}. 

While we are not yet able to extend this approach to all integrand topologies for arbitrary power-counting, there are cases for which particular numerators help make various properties of amplitudes more manifest. For example, for the \twoLoopLadderNameText{2}{2}{2} topology, there exist certain \emph{chiral} numerators \cite{ArkaniHamed:2010gh} which provide a different (not necessarily better) basis choice for the {\color{topCount}6} non-contact degrees of freedom in $\widehat{\mathfrak{N}}_3\!\big(\!\twoLoopLadderNameText{2}{2}{2}\big)$:
\vspace{-2pt}\eq{
\widehat{\mathfrak{N}}_3\!\big(\!\twoLoopLadderNameText{2}{2}{2}\big)= \text{span}\big\{ \x{\cca{\ell_1}}{\cca{\ell_{1{\color{black},1}}^{*}}},\,\x{\cca{\ell_1}}{\cca{\ell_{1{\color{black},2}}^{*}}},\,\x{\ccc{\ell_2}}{\ccc{\ell_{2{\color{black},1}}^{*}}},\,\x{\ccc{\ell_2}}{\ccc{\ell_{2{\color{black},2}}^{*}}},\,\x{\ccb{\ell_1{-}\ell_2}}{\ccb{\ell_{12{\color{black},1}}^{*}}},\,\x{\ccb{\ell_1{-}\ell_2}}{\ccb{\ell_{12{\color{black},2}}^{*}}}\big\}\vspace{-2pt}
}
where $\ell^{*}_{x,i}$ can be chosen as the $i\!=\!1,2$ cut-solutions to the following cut equations
\vspace{-10pt}\begin{align}
\cca{\ell_{1{\color{black},i}}^{*}} \leftrightarrow\, &  
\big\{\ell^2_1 = (\ell_1+p_1)^2 = (\ell_1+p_1+p_3)^2=(\ell_1-p_5)^2 = 0\big\}\,, \nonumber\\
\ccc{\ell_{2{\color{black},i}}^{*}}\leftrightarrow\, &  
\big\{\ell^2_2 = (\ell_2+p_3)^2 = (\ell_2+p_3+p_5)^2=(\ell_2-p_1)^2 = 0\big\}\,,\label{chiral_numerators_for_222}\\
\ccb{\ell_{12{\color{black},i}}^{*}} \leftrightarrow\, &  
\big\{(\ell_1{-}\ell_2{-}p_3{-}p_4)^2 = (\ell_1{-}\ell_2{+}p_1{+}p_2)^2 = (\ell_1{-}\ell_2{+}p_2)^2=(\ell_1{-}\ell_2{-}p_4)^2 = 0\big\}\,. \nonumber \\[-20pt]\nonumber
\end{align}
In spinor-helicity notation, they may be written as
$\cca{\ell_{1{\color{black},1}}^{*}}\!=\!\frac{\ab{13}}{\ab{35}}\lam{5}\lamt{1}$, 
$\ccc{\ell_{2{\color{black},1}}^{*}}\!=\!\frac{\ab{35}}{\ab{51}}\lam{1}\lamt{3}$,
and $\ccb{\ell_{12{\color{black},1}}^{*}}\!=\! \frac{\ab{15}}{\ab{13}}\lam{3}\lamt{5}$. The $\ell^{*}_{x,i=2}$ solutions are obtained from these by replacing $\lam{i}\!\leftrightarrow\!\lamt{i}$ and $\ab{\,\cdot\,}\!\leftrightarrow\![\,\cdot\,]$. Of course, our choice of \emph{chiral} numerators is also somewhat arbitrary and different definitions of $\ell^{*}_{x,i}$ would have worked just as well.

As we will explain shortly, these chiral numerators are suitable to ameliorate certain IR singularities that can be present otherwise, and therefore represent better choices than the arbitrary choices in (\ref{arbitrary_top_numerators_for_222}). In the following subsection, we discuss some of the criteria by which good integrand bases could be chosen more wisely.

%================================================================================================================
\vspace{-0pt}\subsection{Choosing Bases According to Analytic Properties of Integrals}\label{subsec:analytic_properties}\vspace{-0pt}
%================================================================================================================

It would be desirable to establish a link between the choice of numerators to the analytic properties of the results of integration. In particular, the connections between IR divergences, transcendentality, and the behavior of integrands (and their numerators) on generalized unitarity cuts has been studied extensively from several perspectives \cite{ArkaniHamed:2010kv,ArkaniHamed:2012nw,Bourjaily:2012gy,Arkani-Hamed:2014bca,Bern:2014kca,Bern:2015ple,Bourjaily:2015jna,Bourjaily:2015jna,Bourjaily:2016mnp}, and we have learned some important lessons.  Moreover, there is mounting evidence that good integrands---those with wisely chosen numerators---are much easier to integrate directly (see \emph{e.g.}~ \cite{Bourjaily:2013mma,Bourjaily:2018aeq,Bourjaily:2019igt,Bourjaily:2019jrk,Bourjaily:2019vby}). 

For example, it is well understood that IR divergences arise from soft and collinear regions in loop-momentum space (see \emph{e.g.}~\cite{Sterman:1994ce}). Therefore, it is desirable to choose as many numerators as possible to vanish fast enough in precisely these regions so that the resulting integral is IR finite. In fact, a significant fraction of the chiral numerators introduced in \cite{ArkaniHamed:2010gh} satisfy this property, which will undoubtedly render them useful in determining analytic expressions for amplitudes. 

Another desirable aspect of loop integrands regards their differential structure. Indeed, many loop integrands can be chosen such that they may be expressed directly in $d\!\log$-form \cite{Arkani-Hamed:2014via,Arkani-Hamed:2014via,Bern:2014kca}; specifically, this is the case when an integrand only has logarithmic $dx/x$ poles throughout its cut structure. When this is the case, it is widely expected that---at least in most cases\footnote{Counterexamples exist that involve certain square roots in the arguments of the $d\!\log$ forms which can lead to more complicated functions \cite{Brown:2020rda}. We thank Claude Duhr for interesting discussions on this point.}---the result of loop integration will be a polylogarithmic function of maximal transcendentality.  On the other hand, if there are double (or higher) poles in the cut structure, then the post-integration result is expected to contain terms of lower transcendentality; see \emph{e.g.}~the discussion of the one-loop bubble integral in $d\!=\!4$ in \cite{Bern:2014kca}. 

In the remainder of this subsection, we elaborate on these general features and discuss their connection to the results of this work.

%================================================================================================================
\vspace{-0pt}\paragraph*{Manifesting Infrared Divergences and Infrared Finiteness}\label{subsubsec:ir_divergences}\vspace{-0pt}~\\[-12pt]
%================================================================================================================

\noindent 
The example given above for the integrand topology \twoLoopLadderNameText{2}{2}{2} illustrates an important point: we can often explicitly construct numerators for a given integral topology which eliminate IR poles---provided we are using a basis with sufficiently high power-counting. If the power-counting is limited, we can nonetheless use whatever freedom we have to attempt to eliminate at least a subset of IR singularities.

Let us return briefly to the example of the integrand topology \twoLoopLadderNameText{2}{2}{2} described above. For this example, there are potential collinear divergences in three regions: where $\cca{\ell_1}\!\sim\!p_1$, where $\ccc{\ell_2}\!\sim\!p_3$, or where $\ccb{\ell_1{-}\ell_2}{-}p_3{-}p_4\!\sim\!p_5$. It is easy to see that the numerators we chose in (\ref{chiral_numerators_for_222}) were each designed to eliminate one of the three IR divergent regions. However, with only $3$-gon power-counting, we would only be allowed to insert a single inverse propagator, making it impossible to make this integral IR finite in all regions. For example, the chiral numerator 
\eq{
 \x{\cca{\ell_1}}{\cca{\ell_{1{\color{black},1}}^{*}}} = \left(\cca{\ell_1}-\frac{\langle 13\rangle}{\langle 35\rangle}\lambda_5\widetilde{\lambda}_1 \right)^2\!\!\rightarrow 0 
\quad \mbox{as\; $\cca{\ell_1}\!\rightarrow\! \alpha \lambda_1\widetilde{\lambda}_1$}}
vanishes in the collinear region $\cca{\ell_1}\!\sim\!p_1$ but does not do anything to help the other two IR-divergent regions. Similar statements hold for the other five numerators given in (\ref{chiral_numerators_for_222}). An interesting open problem is to count, for a given integral topology, spacetime dimension and power-counting, the number of independent numerators that give rise to IR-finite integrals; and moreover, to partition the remaining (IR divergent) part of the basis into subspaces according to particular degrees of divergence after integration. For example, one could imagine constructing integrands that give rise to divergences of a particular degree $1/\epsilon^k$ in dimensional regularization. This partitioning was resolved in the planar sector at two loops by some of the authors \cite{Bourjaily:2015jna}, and each IR divergent integral was linked directly to a particular soft or collinear residue of a scattering amplitude's integrand. 

Why might such a decomposition be useful? Physically, the sorting of integrands according to their IR structure can help organize and recognize the expected intricate divergence structure in non-abelian gauge theory \cite{Gatheral:1983cz,Frenkel:1984pz,Catani:1998bh,Sterman:2002qn,Aybat:2006mz,Dixon:2008gr,Gardi:2009qi,Becher:2009cu,Almelid:2015jia} from the start. On the other hand, from an integration point of view, it has already been pointed out why one might prefer an IR finite basis of master integrals, see \emph{e.g.}~\cite{vonManteuffel:2014qoa}. Instead of shifting dimensions or doubling propagators, we would construct such a basis by designing appropriate loop-momentum-dependent numerators that eliminate all IR-divergences. Of course, actually achieving this goal must be left to future work.

%================================================================================================================
\newpage\vspace{-0pt}\paragraph*{Polylogarithmic Poles and Integral Transcendentality}\label{subsubsec:poles_transcendentality}\vspace{-0pt}~\\[-12pt]
%================================================================================================================

\noindent 
There is a close connection between the types of singularities of integrands and the degree of transcendental functions which result from integration. As mentioned above, integrands with only \emph{logarithmic singularities} which can always in principle be written in $d\log$ form, are expected to yield maximally transcendental functions, while the presence of multiple poles indicates a drop in transcendentality. 

Simple examples of integrals with only logarithmic singularities include the scalar box and triangle integrals at one loop in four dimensions:
\eq{\Bigg\{ \oneLoopGraph{3},\oneLoopGraph{4}\Bigg\} \,. \label{one_loop_tri_box}}
It is quite easy to see that taking any sequence of residues of the above integrands will always result in simple poles. This agrees with the well-known fact that both integrands integrate to functions of maximal transcendental weight---namely, ${\rm Li}_2$ and $\log^2$ functions. Moreover, there turns out to be a simple change of variables which makes this structure manifest. For the scalar box with propagators $\x{\ell}{Q_i}$, the integrand can be re-written as 
\vspace{4pt}\eq{\rule{0pt}{20pt}\hspace{-20pt}\fwbox{0pt}{d^4\ell\frac{(Q_1|Q_3)(Q_2|Q_4)}{(\ell| Q_1)(\ell|Q_2)(\ell|Q_3)(\ell|Q_4)} = \frac{1}{2} d\!\log\!\frac{\x{\ell}{Q_1}}{(\ell| Q_2)}d\!\log\!\frac{(\ell| Q_2)}{(\ell| Q_3)}d\!\log\!\frac{(\ell| Q_3)}{(\ell| Q_4)} d\!\log\!\frac{(\ell| \ell^{*}_1)}{(\ell| \ell^{*}_2)}\,,} \label{box_dlog_form}\vspace{2pt}
}
where $\ell^{*}_{1,2}$ are the two solutions to the cut equations $(\ell | Q_i) \!=\!0\,, \, i\!\in\!\{1,2,3,4\}$. Although checking whether or not the poles of any given integrand are logarithmic on all cuts is in principle straightforward (see \cite{Wasser:2018qvj,Henn:2020lye} for possible practical subtleties), finding an explicit (or compact) $d\!\log$-form for such an integrand remains something of an art, and no general procedure for obtaining this transformation is currently available (but see \cite{Wasser:2018qvj,Henn:2020lye} for a partial-fraction strategy---which, unfortunately, often yields rather unwieldy expressions for the $d\!\log$ forms).

If an integrand involves higher-order poles like ${dx}/{x^2}$ in its cut structure, it is empirically the case that the degree of transcendentality of the resulting integral after integration drops. In particular, this may mean it is a mixture of pieces with different transcendental weights. These higher-order poles can be located either in the UV region, corresponding to large loop momenta, $\ell\!\rightarrow\!\infty$, or in the IR region, corresponding to poles of higher degree in the denominator. The simplest example of the former type of pole is the scalar bubble integral at one loop in $d\!=\!4$. While the off-shell integrand does not have any poles of degree greater than one, when evaluated on the unitarity cut $\ell^2\!=\!(\ell\pl p_{1}\pl p_2)^2\!=\!0$, the integral becomes
\vspace{-6pt}\eq{
\begin{split}
&\raisebox{5pt}{\tikzBox[0.7]{\oneLoopGraphElement{2}{10}\legLabelled{(a1)}{180}{\normalsize{$p_{12}\hspace{8pt}$}};\legLabelled{(a0)}{0}{\normalsize{$\hspace{8pt}p_{34}$}};\arrow{($(a0)-(0,0.165)$)}{($(a1)-(0,0.165)$)}{180}{black};\arrow{($(a0)+(0,0.165)$)}{($(a1)+(0,0.165)$)}{0}{black};\node[label={[label distance=\labelDist,rotate=0]0:\footnotesize{$\ell$}}] at ($(a1)!0.5!(c2)+(-0.28,-0.05)$) {};}}\hspace{14pt}\Big\vert_{\text{cut}}= \int\frac{dz\,dw}{z}\,,
\hspace{10pt}\text{where\,\,}\ell\equivR\left(w\lambda_1+z\lambda_2\right)\left(\widetilde{\lambda}_1-\frac{w+1}{z}\widetilde{\lambda}_2\right)\,,
\label{eq:one_loop_bub_cut}
\end{split}
}
which has a double-pole at infinity when $w\!\rightarrow\!\infty$ (the double-pole is exposed by the usual inversion $w\!\to\!1/u$) which sends $\ell\!\rightarrow\!\infty$. As a result, we can expect a transcendental weight-drop in the post-integration result; this expectation is indeed realized in its expression from dimensional regularization \cite{Bern:1993kr,Bern:1995ix}.
\eq{
 \int\!\!\!d^{4-2\epsilon}\ell\;\Big[\oneLoopGraph{2}\Big]\propto \frac{1}{\epsilon} + \log\left(\frac{-s}{\mu^2}\right) + \tred{2} + \mathcal{O}(\epsilon)\,,
}
where $4\mi2\epsilon$ is the dimensionally-regulated spacetime dimension, $\mu^2$ is the usual renormalization scale, and $s\!\equivL\!(p_1\pl p_2)^2$. The UV divergence of the bubble integral around $d\!=\!4$ is encoded in the presence of the $1/\epsilon$ pole in the result. Often, there is a close link between higher poles at infinity of the integrand and UV divergences of the integrated answer based on simple power-counting arguments. However, in certain situations there are cases of higher poles at infinity that are not necessarily associated to the UV divergences; for a more detailed discussion of this subtle issue, see \emph{e.g.}~\cite{Bourjaily:2018omh,Edison:2019ovj}.

As mentioned above, the other common source of higher-degree poles can be found in the IR region, where loop momenta become soft and/or collinear. If we consider only integrals with simple propagators, these double-poles arise either from Jacobians generated when the integral is evaluated on cuts, or by the factorization of uncut propagators when evaluated on cuts. The simplest example where this occurs is for the topology \twoLoopLadderNameText{2}{2}{2} for four massless external particles (in four dimensions): 
\vspace{10pt}
\eq{
\fwboxR{0pt}{\mathcal{I}^{\ast}_{[\cca2,\ccb2,\ccc2]} = 
\qquad\hspace{11.5pt}} 
\tikzBox[0.7]{\twoLoopVacGraph{2}{2}{2}{2}\singleLegLabelled{(a1)}{180}{$p_1$};\singleLegLabelled{(v0)}{90}{$p_2$};\singleLegLabelled{(c1)}{0}{$\hspace{1pt}p_4$};\singleLegLabelled{(b2)}{180}{$p_3$};\arrow{(a1)}{(c2)}{143}{emphA};\arrow{(a1)}{(v0)}{37}{emphA};\arrow{(v0)}{(c1)}{143}{emphC};\arrow{(c1)}{(c2)}{38}{emphC};\arrow{(c2)}{(b2)}{-90}{emphB};\arrow{(b2)}{(v0)}{-90}{emphB};\node[label={[label distance=\labelDist,rotate=0]0:$\textcolor{emphA}{\ell_1}$}] at ($(a1)!0.5!(c2)+(-0.6,-0.15)$) {};\node[label={[label distance=\labelDist,rotate=0]0:$\textcolor{emphC}{\ell_2}$}] at ($(c1)!0.5!(c2)+(-0.25,-0.15)$) {};}\label{IR_double_pole_example}
\vspace{4pt}}
We first cut $\cca{\ell_1}^2\!=\!0$ which sets $\cca{\ell_1}\!=\!\lambda_{\cca{\ell_1}}\widetilde{\lambda}_{\cca{\ell_1}}$ on-shell. On the support of this cut, the propagator involving $p_1$ factorizes $(\cca{\ell_1}\pl p_1)^2= \ab{\cca{\ell_1} 1} [\cca{\ell_1} 1]$, and sending both factors to zero sets $\cca{\ell_1}\!=\!\alpha \lam{1}\lamt{1}$---\emph{i.e.} $\cca{\ell_1}$ becomes proportional to $p_1$. Of course, this cut has merely localized $\cca{\ell_1}$ to the collinear region of $p_1$. Similarly, on the other side of the diagram by cutting $\ccc{\ell_2}^2$ and both factors in $(\ccc{\ell_2}\pl p_4)^2$ we set $\ccc{\ell_2}\!=\!\beta \lambda_4\widetilde{\lambda}_4$. On this cut, the two central propagators become
\eq{
(\ccb{\ell_1+\ell_2})^2 = \alpha\beta s_{14}\,, \qquad (\ccb{\ell_1+\ell_2}-p_3)^2 = \alpha\beta s_{14} - \alpha s_{13} - \beta s_{34}\,.\label{residue_for_IR_double_pole_example}}
If we consider this integrand topology with scalar numerator $s_{14}^2$, the residue on this cut evaluates to
\eq{
\underset{(\ref{residue_for_IR_double_pole_example})}{\rm Res}\Big[ 
\mathcal{I}^{\ast}_{[\cca2,\ccb2,\ccc2]}
\Big] =\int\!\!\!d\alpha\,d\beta \frac{s_{14}^2}{\alpha\beta (\alpha\beta s_{14} + \alpha s_{13} + \beta s_{34})}\xrightarrow{\beta=0}\int\frac{d\alpha}{\alpha^2},}
which exposes the double-pole at $\alpha\!=\!0$. According to general wisdom, this double-pole should be reflected in the structure of the integrated result. Indeed,  performing the integral yields lower transcendental terms as can be seen by evaluating this integral in dimensional regularization and expanding in $\epsilon$, \cite{Tausk:1999vh}.

If we work with box power-counting, the above scalar numerator is the only possibility (up to rescaling by external kinematic dependent factors that do not change the analysis). Therefore, this double-pole of the integrand and the associated transcendentality drop is unavoidable. However, for triangle (and higher) power-counting, we can choose a numerator for (\ref{IR_double_pole_example}) which cancels this pole---for example, $(\ccc{\ell_2}\pl p_1)^2$---and write down an integrand which is of uniform transcendentality. However, this procedure requires a bit of care, as loop-dependent numerators can accidentally introduce additional double-poles at infinity (which would in turn re-introduce a drop in transcendentality).

Establishing a hierarchy for integrand basis elements in accordance with their highest degree poles anywhere in their cut structure serves as an integrand-level proxy for the transcendental weight of the resulting integrated answers and is an important open problem which we leave for future work. Besides higher degree poles, there is at least one further essential part of the story we have neglected: namely, the presence of new types of singularities which are not simply poles of some rational function.

%================================================================================================================
\vspace{-0pt}\paragraph*{Non-Polylogarithmic Singularities: Elliptics and Beyond}\label{subsubsec:elliptics}\vspace{-0pt}~\\[-12pt]
%================================================================================================================

\noindent 
It is now a well-known fact that---at sufficiently high multiplicity and/or loop-order---almost all scattering amplitudes in almost all quantum field theories involve non-polylogarithmic structures. The appearance of such structures can be avoided at low multiplicity and loop-order, but these additional structures eventually appear necessary (even if only in local integrand representations). Recently, the analysis of simple examples of such non-polylogarithmic pieces have attracted considerable interest in the high energy physics community, for both practical reasons and formal motivations alike (see, \emph{e.g.}~\cite{Bourjaily:2017bsb,Bourjaily:2018ycu,Bourjaily:2018yfy,Bourjaily:2019hmc,Vergu:2020uur}).

From an integrand perspective, there are many different types of singularities of Feynman integrals beyond single and multiple poles, and the complete list is not known in general.  Among the best known examples in four dimensions is the two-loop double-box integral with a scalar numerator (which corresponds to \twoLoopLadderNameText{3}{1}{3} in the notation of this paper), which is known to be elliptic \cite{CaronHuot:2012ab,Paulos:2012nu,Nandan:2013ip,Bourjaily:2017bsb}:
\eq{
\begin{split}
\fwboxR{0pt}{\mathcal{I}_{[\cca3,\ccb1,\ccc3]} = \hspace{5pt}}
\tikzBox[0.7]{
\twoLoopVacGraph{3}{1}{3}{1.25}\leg{(a1)}{145};\leg{(a2)}{-145};\leg{(c1)}{35};\leg{(c2)}{-35};\leg{(v0)}{90};\leg{(v1)}{-90};}
\label{fig:elliptic_dbl_box}
\end{split}}
Evaluating the integrand on one of the two solutions to the maximal (hepta-)cut when all propagators in the graph are put on-shell yields an integral over one remaining parameter,
\eq{
{\rm Res}\,\,{\cal I}_{[\cca3,\ccb1,\ccc3]} = \int \frac{dx}{\sqrt{x^4+\alpha x^3 + \beta x^2 + \gamma x +\delta}}\,,
\label{eq:elliptic_max_cut}
}
where $x$ parameterizes the last unfixed degree of freedom in $\cca{\ell_1}$, $\ccc{\ell_2}$ and the coefficients $\alpha,\ldots,\delta$ only depend on external kinematics. There are no poles in this expression, including the pole at $x\!\rightarrow\!\infty$. The nature of the singularity obtained here is aptly called \emph{elliptic} as the integration over $x\!\in\!\mathbb{R}$ would give an elliptic function. Indeed, evaluating the integral (\ref{fig:elliptic_dbl_box}), one finds that the result is not a polylogarithm, but rather an elliptic function \cite{Bourjaily:2017bsb}. 

In general, there is an obvious connection between post-integration results and the types of singularities which appear in the associated integrand. If the only singularities are logarithmic, the integral is expected to evaluate to a sum of (generalized) polylogarithms. However, we do not currently have sufficient knowledge about the space of functions for Feynman integrals with non-logarithmic singularities in their unitarity cut structure. Trying to make progress on this very difficult question is an active area of research in both physics and mathematics.

%================================================================================================================
\vspace{-0pt}\paragraph*{Transcendental Filtration and Integrand Stratification}\label{subsubsec:filtration}\vspace{-0pt}~\\[-12pt]
%================================================================================================================

\noindent 
An important question from the integrand perspective is whether or not we can establish a hierarchy of numerators for a given integrand topology based on the type of singularities which appear in their cuts. Such a classification would imply that we can take any integrand topology and (for a given power-counting) divide the numerator basis into groups classified by the presence of single poles (logarithmic singularities), higher-degree poles, elliptic singularities, etc. 

As an example, let us take the \twoLoopLadderNameText{4}{1}{3} topology, the so-called penta-box integral:
\eq{
\begin{split} 
\tikzBox[0.7]{
\twoLoopGraph{4}{1}{3}{1.25}}
\label{fig:penta_box}
\end{split}}
With box power-counting, the \twoLoopLadderNameText{4}{1}{3} topology is allowed to carry numerators drawn from the basis structure $[\cca{\ell_A}]$. According to our basis counting summarized in \mbox{Table \ref{tab:two_loop_counting_table}}, in four spacetime dimensions these numerators consist of {\color{topCount}2} top-level and 4 contact-term degrees of freedom (each of which cancels one propagator of the left loop leading to \twoLoopLadderNameText{3}{1}{3} topologies). Specifically, we can span the {\color{totalCount}full-rank} basis by two (familiar) chiral numerators and four double box contact-terms,
\eq{ [\cca{\ell_A}] = \text{span}\big\{ {\color{topCount}(\ell_A | \ell^{*}_{A{\color{black},1}})},\, {\color{topCount}(\ell_A | \ell^{*}_{A{\color{black},2}})},\, {\color{dim}(\ell_A | Q_1){\color{black},}\, (\ell_A | Q_2){\color{black},}\, (\ell_A | Q_3){\color{black},}\, (\ell_A | Q_4) } \big\} \,,
}
where the special external kinematic points $\cca{\ell^{*}_{A{\color{black},i}}}$ are again taken as the two solutions to the cut equations $(\cca{\ell_A} | Q_1)\!=\!(\cca{\ell_A} | Q_2)\!=\!(\cca{\ell_A} | Q_3)\!=\!(\cca{\ell_A} | Q_4)\!=\!0$. In the generic case of massive external momenta, all double-box integrals are of the form (\ref{fig:elliptic_dbl_box}) and therefore have elliptic maximal cuts (\ref{eq:elliptic_max_cut}). \emph{A priori}, the two top-level integrands have support on four different elliptic cuts associated to the various pinchings of propagators to yield \twoLoopLadderNameText{3}{1}{3} topologies. On one of the heptacuts, the numerators take the schematic form,
\eq{
{\rm Res}\Big[{\cal I}_{(\ref{fig:penta_box})}\Big] = \int\!\!\!dx \frac{N(x)}{(x-x_1)(x-x_2)\sqrt{x^4+\alpha x^3 + \beta x^2 + \gamma x +\delta}},}
where $x_1$ and $x_2$ are the two positions of leading singularities associated with cutting the last propagator, and the numerator $N(x)$ is a quadratic function of $x$. It is easy to see that no choice of numerator $N(x)$ can remove the elliptic singularity from the heptacut, and it remains an important open question whether or not there is \emph{any} choice of numerator for \twoLoopLadderNameText{4}{1}{3} for which the final result is polylogarithmic (without elliptic contributions). While we cannot remove the elliptic \emph{cut} completely, it might be possible that there is some special choice of numerator which removes all elliptic effects from the integrated result. 

We should also note that for special cases when some of the corners in \twoLoopLadderNameText{4}{1}{3} or \twoLoopLadderNameText{3}{1}{3} are massless (or vanishing) momenta, all singularities are logarithmic and the integrated result does indeed evaluate to generalized polylogarithms \mbox{\cite{Drummond:2010cz,Papadopoulos:2015jft,Gehrmann:2018yef}}.

%================================================================================================================
\vspace{-0pt}\subsection{An Alternative Proposal: Classification by Poles at Infinity}\label{subsec:why_reducible_integrands}\vspace{-0pt}
%================================================================================================================

Our discussion of integrand bases has been most detailed for two and three loops. While a similar analysis can be extended to higher loops, our definition of $p$-gon power-counting suffers from some unfortunate features. This is easiest to see in the case of planar integrands, where a more familiar (and powerful) definition of \emph{planar}-power-counting exists. Recall that a definition of power-counting for plane integrands can be given by demanding that every loop momentum variable (encoded by the dual of the plane graph) scales like a $p$-gon at infinity. 

Starting at four loops, there exist planar integrands which we would classify as `scalar' $p$-gons, but which would admit non-trivial loop-dependent numerators while still satisfying \emph{planar} $p$-gon power-counting. The simplest examples of these are graphs with a $(p\pl1)$-gon surrounded by $p$-gons: 
\eq{\tikzBox[0.7]{\oneLoopGraphElement{3}{0}\draw[int] (a1) to[bend left=90] (a2);\draw[int] (a2) to[bend left=90] (a3);\draw[int] (a3) to[bend left=90] (a1);}\,,\hspace{10pt} 
\tikzBox[0.7]{\oneLoopGraphElement{4}{2}\coordinate (e1) at ($(a1)+(-60:0.666*\edgeLen)$);\draw[int] (a1) -- (e1) -- (a4);\coordinate (e2) at ($(a1)+(150:0.666*\edgeLen)$);\draw[int] (a1) -- (e2) -- (a2);\coordinate (e3) at ($(a2)+(60:0.666*\edgeLen)$);\draw[int] (a2) -- (e3) -- (a3);\coordinate (e4) at ($(a3)+(-30:0.666*\edgeLen)$);\draw[int] (a3) -- (e4) -- (a4);\leg{(e1)}{-90};\leg{(e2)}{180};\leg{(e3)}{90};\leg{(e4)}{0};}\hspace{8pt}\,, \hspace{10pt} 
\tikzBox[0.7]{\oneLoopGraphSkeleton{5}{0}\def\edgeLen{0.5*\figScale}\draw[int](a1)--(a2)--(a3)--(a4)--(a5)--(a1);\coordinate (e1) at ($(a3)+(90:\edgeLen*0.7)$);\coordinate (e2) at ($(a4)+(90-72:\edgeLen*0.7)$);\coordinate (e3) at ($(a5)+(-54:\edgeLen*0.7)$);\coordinate (e4) at ($(a1)+(-54-72:\edgeLen*0.7)$);\coordinate (e5) at ($(a2)+(-54-72*2:\edgeLen*0.7)$);\draw[int](a3)--(e1);\draw[int](a4)--(e2);\draw[int](a5)--(e3);\draw[int](a1)--(e4);\draw[int](a2)--(e5);\draw[int](e1)--(e2)--(e3)--(e4)--(e5)--(e1);\leg{(e1)}{90};\leg{(e2)}{90-72};\leg{(e3)}{90-72*2};\leg{(e4)}{90-72*3};\leg{(e5)}{90-72*4};}\hspace{5pt},\hspace{15pt}
\tikzBox[0.7]{\oneLoopGraphElement{5}{0}\coordinate (e1) at ($(a1)+(-90:\edgeLen*0.6)$);\coordinate (e2) at ($(e1)+(0:\edgeLen*0.6)$);\draw [int] (a1) -- (e1) -- (e2) -- (a5);\coordinate (e3) at ($(a1)+(-162:\edgeLen*0.6)$);\coordinate (e4) at ($(a2)+(-162:\edgeLen*0.6)$);\draw [int] (a1) -- (e3) -- (e4) -- (a2);\coordinate (e5) at ($(a2)+(126:\edgeLen*0.6)$);\coordinate (e6) at ($(a3)+(126:\edgeLen*0.6)$);\draw [int] (a2) -- (e5) -- (e6) -- (a3);\coordinate (e7) at ($(a3)+(54:\edgeLen*0.6)$);\coordinate (e8) at ($(a4)+(54:\edgeLen*0.6)$);\draw [int] (a3) -- (e7) -- (e8) -- (a4);\coordinate (e9) at ($(a4)+(-18:\edgeLen*0.6)$);\coordinate (e10) at ($(a5)+(-18:\edgeLen*0.6)$);\draw [int] (a4) -- (e9) -- (e10) -- (a5);\leg{(e1)}{-135};\leg{(e2)}{-45};\leg{(e3)}{-127};\leg{(e4)}{-217};\leg{(e5)}{171};\leg{(e6)}{81};\leg{(e7)}{99};\leg{(e8)}{9};\leg{(e9)}{27};\leg{(e10)}{-63};}\hspace{10pt} ,\ldots\label{fig:4L_pc_fail}}
All of these integrands meet our definition of being `scalar' $p$-gon integrands, as all daughter topologies have girth strictly less than $p$. Nevertheless, as planar integrands, we may use the preferential loop-momentum routing of the dual graph to see that each of these may admit a non-trivial numerator while remaining `$p$-gons' according to \emph{planar}-power-counting. The third and fourth examples above are particularly poignant as they would be rendered dual-conformal with such a numerator, while if given the numerator `$1$', they would not be. 

As these examples illustrate, \emph{our definition} of power-counting---while applicable to all graphs---is more restrictive than what is traditionally used in the planar limit. This is particularly problematic because it is clear, for example, that our definition of $p\!=\!4$ power-counting fails to include dual-conformal integrands at sufficiently high loop order (namely, 5). It is well-known that amplitudes in planar sYM may be expressed in terms of dual-conformal integrands; moreover, a complete basis of such integrands requires non-zero coefficients for 4-particle amplitudes, as seen in \mbox{\cite{Bourjaily:2011hi,Bourjaily:2015bpz,Bourjaily:2016evz}}. But our definition of `box' power-counting fails to include all such integrands---and, in fact, therefore, all amplitudes in planar sYM at sufficiently high loop orders. Thus, our stratification, while being well defined for all graphs, does not have the property that amplitudes in sYM can be expressed in terms of integrands with box power-counting beyond 4 loops or triangle power-counting beyond 7 loops. 

To be clear: it would not be hard to alter our definition of the box-power-counting basis to include dual-conformal integrands in the planar limit. Our rule for generating numerators according to graph inclusions would permit us to define a basis that `scales like' \emph{any} set of specified graphs \emph{with any pre-chosen, loop-dependent numerators}. The real problem is that we do not know any clear rule for adding such numerators into the space we use to recursively define a basis of integrands. This remains an important open problem that must be left to future work. Specifically: \emph{Is there a non-planar, graph-theoretic definition of a basis of integrands with `4-gon power-counting' such that all amplitudes in sYM are inside this basis?}

More generally, it remains an open problem to define any well-behaved integrand basis of integrands guaranteed to include the amplitudes of `nice' quantum field theories to all loop-orders. In the planar limit, we have the basis of integrands dictated by dual-conformal invariance---which is in fact stronger than mere planar-box-power-counting; and there is a great deal of evidence that amplitudes in planar sYM can be represented in such a basis. We know of no non-planar analogue of box-power-counting (let alone dual-conformal invariance) that should suffice for amplitudes in sYM beyond the planar limit to all orders of perturbation theory. \\

One promising strategy---at least, formally---would be to define power-counting in terms of a hierarchy of poles `at infinity'. The real challenge here is to make this definition useful without specific reference to loop momentum routing or to a brute force survey of potential singularities of integrands. Nevertheless, the poles at infinite loop-momentum have been extensively studied in many papers \cite{Arkani-Hamed:2014via,Bern:2014kca,Bern:2015ple,Bourjaily:2018omh}, mainly in the context of particular theories. For example, it was conjectured and later verified \cite{Bourjaily:2018omh} up to three loops that sYM integrands are free of \emph{any} poles at infinity to all orders of perturbation theory, suggesting something like a non-planar analogue of dual-conformal symmetry. (In contrast, amplitudes in ${\cal N}\!=\!8$ supergravity have higher-degree poles at infinity that grow with multiplicity and loop-order.)

To make this more precise, we may organize loop integrands by the maximal degree of singularity encountered as the momentum flowing through \emph{any} edge goes to infinity via any sequence of residues. This maximal degree is formally well-defined, if hard to detect in practice. Specifically, this requires that one perform all possible cuts, and list the degrees of all singularities which send $\ell\!\rightarrow\!\infty$:
\begin{equation}
\underset{\!\text{cuts}\!}{\mathrm{max}}\Big\{
    {\rm Cut}\big[{\cal I}\big] \xrightarrow[\ell\rightarrow\infty]{} {\cal O}(\ell^s)
\Big\}\,.
\end{equation}
The maximal degree $s$ over the set of all cuts for the momentum $\ell$ flowing through any edge is then defined to be the degree of a given integrand's pole(s) at infinity. Obviously, from a practical point of view, this is not a very constructive approach, as it requires that many checks be made to decide $s$ for a given integrand. Despite these practical limitations, it certainly gives us a unique answer. This definition extends to all loops and provides a hierarchy of loop integrands based on this degree. 

We may illustrate this at one loop in $d\!=\!4$. In fact, our definition above overlaps with the standard definition of power-counting with the following identifications:
\begin{center}
\begin{tabular}{| l|l|}
\hline
  \fwbox{90pt}{\text{\textbf{power-counting}}}  &\fwbox{115pt}{\text{\textbf{degree of pole}}} \\
 \hline\hline 
 $p=4$ (box)        & $s=-2$ (no pole)\\ 
 \hline
 $p=3$  (triangle)   & $s=-1$ (single pole) \\   
 \hline
 $p=2$  (bubble)     & $s=0$ (double-pole)\\ 
 \hline
 $p=1$  (tadpole)    & $s=1$ (triple pole)\\ 
 \hline
 $p=0$  (constant)   & $s=2$ (quadruple pole)\\ 
 \hline
\end{tabular}
\end{center}
As an example, in eq.~(\ref{eq:one_loop_bub_cut}) we computed a double-cut of the scalar bubble integral and identified the double-pole at infinity for $w\!\rightarrow\!\infty$. One may readily verify that a double-pole at infinity is indeed the worst possible singularity on any cut of the bubble integral. The analogous procedure now extends to higher loops without any conceptional problem: the (maximal) degree of poles at infinity is always well defined as the supremum of degrees we encounter over the space of all sequences of cuts.

At two loops, for the scalar planar double-box \twoLoopLadderNameText{3}{1}{3} and the non-planar double-box \twoLoopLadderNameText{3}{2}{2}, the degree at infinity is $s\!=\!-2$---\emph{i.e.}, no pole at infinity---which is in agreement with the definition of the power-counting we have used throughout this paper. However, there is a difference for certain \twoLoopLadderNameText{2}{1}{2} topologies, where our graph-theoretic power-counting definition allows for one degree of freedom (the scalar numerator) within triangle power-counting $p\!=\!3$. From the pole at infinity perspective, we see that a special degeneration of this integrand in fact has not just a single pole, but a double-pole at infinity:
\vspace{-8pt}\eq{\begin{split}
\fwboxR{0pt}{\mathcal{I}^{\ast\ast}_{[\cca2,\ccb1,\ccc2]} =\hspace{35pt}} \tikzBox[0.7]{\twoLoopVacGraph{2}{1}{2}{2}\legLabelled{(a1)}{180}{$p_{12}\hspace{5pt}$};\legLabelled{(c1)}{0}{$\hspace{5pt}p_{34}$};\arrow{(a1)}{(c2)}{143}{emphA};\arrow{(a1)}{(v0)}{37}{emphA};\arrow{(v0)}{(c1)}{143}{emphC};\arrow{(c1)}{(c2)}{38}{emphC};\arrow{(c2)}{(v0)}{-90}{emphB};\node[label={[label distance=\labelDist,rotate=0]0:$\textcolor{emphA}{\ell_1}$}] at ($(a1)!0.5!(c2)+(-0.6,-0.15)$) {};\node[label={[label distance=\labelDist,rotate=0]0:$\textcolor{emphC}{\ell_2}$}] at ($(c1)!0.5!(c2)+(-0.25,-0.15)$) {};}
\end{split}
\label{eq:G212_bdy_dbl_pole_infty}
}
An intuitive way to understand this fact for this particular diagram is to realize that we generate a bubble when we remove the central propagator. This can be exposed by first cutting all propagators
\vspace{-2pt}\eq{
\cca{\ell_1}^2=(\ccb{\ell_1+\ell_2})^2=(\cca{\ell_1}{+}p_1{+}p_2)^2=
\ccc{\ell_2}^2=(\ccc{\ell_2}{+}p_3{+}p_4)^2=0\,,\vspace{-2pt}
}
and further localizing $\cca{\ell_1}$ on the composite leading singularity, which double-cuts $(\ccb{\ell_1\pl\ell_2})^2\!=\!0$ and sets $\cca{\ell_1}\!=\!-\ccc{\ell_2}$. The residue on this cut has Jacobian $1/p_{12}^2$, and the resulting expression is just a one-loop cut bubble integral in $\ccc{\ell_2}$,
\vspace{-4pt}\eq{
\text{Cut}\left[{\cal I}^{\ast\ast}_{[\cca2,\ccb1,\ccc2]}\right] \simeq 
\text{Cut}\left[\int\!\!\frac{d^4\ccc{\ell_2}}
{\ccc{\ell_2}^2(\ccc{\ell_2}{+}p_3{+}p_4)^2}\right]
\vspace{-4pt}}
which has a double-pole at infinity, as discussed in (\ref{eq:one_loop_bub_cut}). Note that while it is necessary to choose some parameterization for this test and express the degrees of freedom in loop momenta in a particular way (in addition to an arbitrariness of how the loop momenta are to be routed) for performing cuts, the degree of poles at infinity is unambiguous and does not depend on the choices made. 

We suspect that tracking the degree of poles at infinity will be the best---and morally (more) correct way---to define the power-counting at arbitrary loop-order. Moreover, we conjecture that amplitudes in sYM may be represented by a basis that meets \emph{this} definition of `box power-counting' to all orders. \\

Another natural question is whether or not it is desirable to expand a four-dimensional integrand with overall box power-counting---however this basis may be defined---in terms of a basis of integrands with the same property. We have argued in previous papers where we explicitly constructed all multiplicity two-loop supersymmetric Yang-Mills amplitudes \cite{Bourjaily:2019iqr,Bourjaily:2019gqu} that the answer is: not necessarily. For example, we have found that amplitudes in $\mathcal{N}\!=\!8$ supergravity require arbitrarily bad power-counting bases (the degree of which grows with multiplicity) starting at two loop \emph{despite} the fact that amplitudes in this theory are known to have bounded behavior in the ultraviolet.  

The reader may wonder why we have not used the proposed `poles at infinity' criterion for our two- and three-loop bases described in sections \ref{sec:two_loops} and \ref{sec:three_loops}, but rather resorted to a more purely graph-theoretic definition. The answer was already alluded to above: the pole at infinity check is not a constructive approach and it can be computationally prohibitive. We do not \emph{a priori} know the degree of poles at infinity of a given integrand, and to check this we would have to perform all possible cuts, which is extremely laborious. Moreover, the degree can differ between the general case and boundary cases. For \twoLoopLadderNameText{2}{1}{2}, for example, the general case with external momenta flowing into the middle vertices has a single pole at infinity $s\!=\!-1$; in contrast, the boundary case with three-point vertices in the middle has a double-pole at infinity ($s\!=\!0$) as discussed below (\ref{eq:G212_bdy_dbl_pole_infty}). This makes it harder to do the bookkeeping of various integrand topologies. While this all is just a technical inconvenience, it makes it hard to implement our conjecture systematically at present. Minimally, our graph-theoretic definition of power-counting should suffice to construct box-power-counting bases big enough for sYM through at least four loops---with $3$-gon power-counting through at least seven loops. 

In the future, we would like to use the results, technology, and the counting of basis elements from the current work and transfer this knowledge to a new framework which uses poles at infinity as the primary condition. This almost certainly requires that we build new tools to systematically detect the degree of the poles without performing all possible cuts.

%================================================================================================================
\vspace{-0pt}\subsection{Building Amplitudes: Beyond the Bases of Integrands}\label{subsec:beyond_basis}\vspace{-0pt}
%================================================================================================================

As was discussed above, integrands with more than $d\!\times\!L$ propagators can always be reduced to simpler topologies for an integrand basis with $p\!<\!d$ power-counting. However, there are certain situations when integrands with more propagators are convenient and useful for the representation of amplitudes. For example, they may be more finite in the ultraviolet/infrared than their decomposition into daughters would suggest.

The best example is perhaps the two-loop MHV integrand in planar sYM, which was written in \cite{ArkaniHamed:2010kv,ArkaniHamed:2010gh,Bourjaily:2015jna}:
\eq{\fwboxR{0pt}{\mathcal{A}_{\text{MHV, planar}}^{\text{2-loops}}=}\sum_{\substack{1\leq a<b<c\\b<c<d<a+n}}\hspace{10pt}\tikzBox[0.7]{\twoLoopVacGraph{4}{1}{4}{2}
\singleLegLabelled{(a3)}{-90-30}{$a$};\singleLegLabelled{(a1)}{90+30}{$b$};\optLeg{(a2)}{180};
\singleLegLabelled{(c1)}{90-30}{$c$};\singleLegLabelled{(c3)}{-90+30}{$d$};\optLeg{(c2)}{0};\optLeg{(v0)}{90};\optLeg{(v1)}{-90};
\node at ($(a2)+(0:0.55)$) []{\footnotesize{$\mathfrak{n}(\!\cca{\ell_A}\!)$}};\node at ($(c2)+(180:0.55)$) []{\footnotesize{$\mathfrak{n}(\!\ccc{\ell_C}\!)$}};
}\label{eq:2L_mhv_amp_chi_dbl_pent}
}
where $\mathfrak{n}(\ell_i)$ represents a particular choice of loop-dependent numerators for the graph. In this work's terminology, this formula involves particular choices for integrands of the \twoLoopLadderNameText{4}{1}{4} topology. In particular, the numerators are chosen to match all cuts in field theory of a \emph{different} topology---namely, \twoLoopLadderNameText{4}{0}{4}, a.k.a.~the kissing-boxes depicted in eq.~(\ref{eq:kissing_boxes_os_diags}). In contrast to the integrand \twoLoopLadderNameText{4}{0}{4}, which would have a scalar numerator in box power-counting, the double pentagon integrand \twoLoopLadderNameText{4}{1}{4} can be assigned numerators to match all four kissing-box cuts separately. However, it comes at the cost of introducing spurious cuts, which cancel in the sum over terms in (\ref{eq:2L_mhv_amp_chi_dbl_pent}). The representation (\ref{eq:2L_mhv_amp_chi_dbl_pent}) is a miraculously simple expression, and it is natural to ask about its origin and if we can use the underlying principle in different cases too. The answer is a bit complicated. 

It is suggestive that each double-pentagon matches one kissing-box leading-singularity of the amplitude, and indeed the numerator counting works along these lines: provided that we \emph{exclude} from our basis the \twoLoopLadderNameText{4}{0}{4},\,\twoLoopLadderNameText{5}{0}{4}, and \twoLoopLadderNameText{5}{0}{5} integrands, we would find {\color{topCount}4} (as opposed to {\color{topCount}3}) independent {\color{topCount}top-level} numerators for box power-counting \emphEdges
\begin{align}
    \begin{split}
        \twoLoopLadderNameText{4}{1}{4} \,=\;
        \tikzBox[0.7]{ \twoLoopGraph{4}{1}{4}{1}}
        \;\Leftrightarrow\,
        \mathfrak{N}_4\!\big(\!\twoLoopLadderNameText{4}{1}{4}\big)= \cca{[\ell_A]}\ccc{[\ell_C]}\,,
        \qquad
        \underset{d=4}{\text{rank}}\big[\mathfrak{N}_4\!\big(\!\twoLoopLadderNameText{4}{1}{4}\big)\big] = \tensorDecomp{36}{4}{32}
    \end{split}
\end{align}
matching exactly the four kissing-box leading singularities,\resetGraphDefaults
\eq{
\label{eq:kissing_boxes_os_diags}
\hspace{10pt}\tikzBox[0.7]{\twoLoopVacGraph{4}{0}{4}{1}\singleLegLabelled{(a2)}{90}{};\singleLegLabelled{(a4)}{-90}{};\singleLegLabelled{(c2)}{-90}{};\singleLegLabelled{(c4)}{90}{};\optLeg{(v0)}{90};\leg{(a3)}{180};\leg{(c3)}{0};\node at (v0) [fullmhv] {};\node at (a2) [fullmhvBar] {};\node at (a3) [fullmhv] {};\node at (a4) [fullmhvBar] {};\node at (c2) [fullmhvBar] {};\node at (c3) [fullmhv] {};\node at (c4) [fullmhvBar] {};}
\hspace{25pt}
\tikzBox[0.7]{\twoLoopVacGraph{4}{0}{4}{1}\singleLegLabelled{(a2)}{90}{};\singleLegLabelled{(a4)}{-90}{};\singleLegLabelled{(c2)}{-90}{};\singleLegLabelled{(c4)}{90}{};\optLeg{(v0)}{90};\leg{(a3)}{180};\leg{(c3)}{0};\node at (v0) [fullmhv] {};\node at (a2) [fullmhv] {};\node at (a3) [fullmhv] {};\node at (a4) [fullmhv] {};\node at (c2) [fullmhvBar] {};\node at (c3) [fullmhv] {};\node at (c4) [fullmhvBar] {};}
\hspace{25pt}
\tikzBox[0.7]{\twoLoopVacGraph{4}{0}{4}{1}\singleLegLabelled{(a2)}{90}{};\singleLegLabelled{(a4)}{-90}{};\singleLegLabelled{(c2)}{-90}{};\singleLegLabelled{(c4)}{90}{};\optLeg{(v0)}{90};\leg{(a3)}{180};\leg{(c3)}{0};\node at (v0) [fullmhv] {};\node at (a2) [fullmhvBar] {};\node at (a3) [fullmhv] {};\node at (a4) [fullmhvBar] {};\node at (c2) [fullmhv] {};\node at (c3) [fullmhv] {};\node at (c4) [fullmhv] {};}
\hspace{25pt}
\tikzBox[0.7]{\twoLoopVacGraph{4}{0}{4}{1}\singleLegLabelled{(a2)}{90}{};\singleLegLabelled{(a4)}{-90}{};\singleLegLabelled{(c2)}{-90}{};\singleLegLabelled{(c4)}{90}{};\optLeg{(v0)}{90};\leg{(a3)}{180};\leg{(c3)}{0};\node at (v0) [fullmhv] {};\node at (a2) [fullmhv] {};\node at (a3) [fullmhv] {};\node at (a4) [fullmhv] {};\node at (c2) [fullmhv] {};\node at (c3) [fullmhv] {};\node at (c4) [fullmhv] {};}
\hspace{10pt}.}
\emphEdges
The special numerator that is used in (\ref{eq:2L_mhv_amp_chi_dbl_pent}) was chosen to \emph{only} have support on the MHV-compatible solution to these cuts.

So far, this all looks reasonably natural but there is an important subtlety. Apart from the kissing-box leading singularity there are many more leading singularities of the amplitude which also must be matched. Moreover, by using these `chiral' double-pentagon integrands, we introduce a number of unphysical singularities (term-by-term) which only cancelled in the sum. None of that was used in \cite{ArkaniHamed:2010kv,ArkaniHamed:2010gh} to find the formula (\ref{eq:2L_mhv_amp_chi_dbl_pent}), but it is \emph{post-facto}-guaranteed by global residue theorems (GRTs) which connect various singularities of the amplitude together in a consistent framework.\\

The difficulty is that it is very hard to write down (or guess) expressions such as (\ref{eq:2L_mhv_amp_chi_dbl_pent}) and verify that they give the correct amplitude exactly because not all singularities are matched correctly by construction. This is in stark contrast to the generalized unitarity approach \cite{Bern:1994zx,Bern:1994cg,Britto:2004nc} or its prescriptive refinement \cite{Bourjaily:2017wjl}, where we do match each physical cut separately and we can be confident that the resulting expression is correct.  However, the prescriptive approach here is most straightforward to use within boosted $p\!<\!d$-gon power-counting. For $d\!=\!4$, we would find that the nine-propagator graphs such as \twoLoopLadderNameText{4}{1}{4} would be fully decomposable; but we would be able to meet all the eight-propagator cuts distinctly and directly. With boosted power-counting, it is therefore easy to match one leading singularity at a time while integrands such as \twoLoopLadderNameText{4}{1}{4} (required for box power-counting in $d\!=\!4$) would contribute to many leading singularities in violation of the prescriptive philosophy. Therefore, we would be hard pressed to recover an expression like (\ref{eq:2L_mhv_amp_chi_dbl_pent}) from the prescriptive approach. In \cite{Bourjaily:2017wjl}, we showed how to use prescriptive unitarity to reconstruct general two and three-loop $n$-point amplitudes in planar sYM preserving the box power-counting in the planar limit, but these constructions required very judicious choices (and some magic) to work. Today, we do not know how such tricks and `magic' can be (or even could be) generalized to higher loop-orders.

We know that if we go beyond the planar limit, for which we generated integrand bases in this paper, the prescriptive approach to constructing amplitudes works without any problems (other than increasing complexity); but the search for \emph{simple} representations of amplitudes with lower power-counting become increasingly difficult to find because of the presence of color factors. In \cite{Bourjaily:2019gqu}, we used the prescriptive approach to determine the integrand for the full (non-planar) two-loop $n$-point MHV amplitude in sYM by carefully constructing the basis of integrands with triangle power-counting (while making some convenient choices) and matching all of the relevant leading singularities,
\eq{\fwboxR{0pt}{\mathcal{A}_{\text{MHV, non-planar}}^{\text{2-loops}}=\;\;}\hspace{-8pt}\sum_{\substack{\text{inequivalent}\\\text{leading singularities $\mathfrak{f}$}\\[-5pt]~}}\hspace{-10pt}\mathfrak{f}\,\times\,\mathcal{I}_{\mathfrak{f}}\;.\label{two_loop_mhv_amplitudes}}
As discussed above, this representation (as it involves integrands with $3$-gon power-counting) involves poles at infinity in virtually every integrand basis element, while it is known (at two loops) that the final amplitude is free of these poles. The absence of poles at infinity for non-planar amplitudes suggested an extension of dual-conformal symmetry beyond the planar sector in sYM and was partially explored in \cite{Bern:2014kca,Bern:2015ple}. From the perspective of our work here, the lack of poles at infinity in the final amplitude raises the question whether or not we can write the amplitude using integrands with box power-counting only to make the behavior at infinity manifest term-by-term. This would necessarily introduce integrals with 9 or more propagators such as \twoLoopLadderNameText{3}{3}{3}, \twoLoopLadderNameText{4}{2}{3}, or \twoLoopLadderNameText{4}{2}{4}:
\begin{align}
    \begin{split}
    \begin{tikzpicture}
        \twoLoopGraph{3}{3}{3}{1}
     \end{tikzpicture}
    \qquad
     \begin{tikzpicture}
        \twoLoopGraph{4}{2}{3}{1}
     \end{tikzpicture}
    \qquad
    \begin{tikzpicture}
        \twoLoopGraph{4}{2}{4}{1}
     \end{tikzpicture}         
    \end{split}
\end{align}

Such integrands would have support on many leading singularities at the same time---and could only be chosen to match particular subsets. Thus, they must conspire to match everything at once in some spectacular way to give the correct expression for a scattering amplitude. This would have to follow from GRTs (including color information) that link various color-dressed on-shell functions together into a bigger interrelated network. Using these relations practically requires one to enumerate all the independent GRTs, which is hard to do systematically.

Na\"ively, one can just use the traditional version of generalized unitarity with some definition of box power-counting, write an ansatz for numerators and color factors which multiply a given topology and check all cuts. Apart from being a rather involved exercise for high multiplicity (let alone general $n$-point), the resulting solution would suffer from a major problem: the coefficients would generically be some complicated sums of products of kinematics and color factors (with different graph structures), where the latter do not necessarily correspond to any particular color-dressed on-shell function. The color factors would appear as essentially \emph{ad hoc} expressions---appearing merely to satisfy the constraints of generalized unitarity. This makes perfect sense because the color factors are tied to the leading singularity pictures (with 8 or fewer propagators) while our integrands can have 9 or more propagators, and can not be translated to color factors in this way. On the hand, the prescriptive unitarity approach treatment of color factors is very straightforward: each integrand is multiplied by a single color-dressed on-shell function directly and the color factor can be read off from the graph directly; for further details and powerful illustrations, see \cite{Bourjaily:2019iqr,Bourjaily:2019gqu}. 

Of course, there can in principle exist a fundamentally new approach that would allow us to use a minimal set of building block integrals, such as \twoLoopLadderNameText{3}{3}{3}, \twoLoopLadderNameText{4}{2}{3}, or even \twoLoopLadderNameText{4}{2}{4} with some magic numerators. Similar to the planar amplitude written in terms of the chiral pentagon expansion in eq.~(\ref{eq:2L_mhv_amp_chi_dbl_pent}), each integrand would match many leading singularities at the same time. It would be great to find some way to constructively find such numerators based on some underlying principles. We know that such an approach must combine both color and kinematics in a non-trivial way, and at the same time must exploit the richness of the GRTs. We would love to pursue this path in the future.

%================================================================================================================
\vspace{-0pt}\subsection{Concluding Remarks}\label{subsec:concluding_remarks}\vspace{-0pt}
%================================================================================================================

In this work, we have discussed a systematic approach to the construction of loop-integrand bases for general, non-planar scattering amplitudes of general quantum field theories. We provided a graph-theoretic definition of `power-counting' and used it as constructive tool to organize and stratify integrand bases according to their UV behavior. We illustrated our approach by explicitly enumerating all diagram topologies and their associated numerator degrees of freedom for various power-counting at one, two, and three loops. These results provide the number of independent basis elements required to express scattering amplitudes in different quantum field theories. The same graph-theoretic implementation is in principle applicable to any loop order where the size of the integrand basis grows. 

We also discussed the limitations and open problems in our approach. In particular, we compared our choices of power-counting with a different, and in our opinion ultimately better, framework of poles at infinity. The further exploration along these lines at higher loops is the main open question for the future.

%================================================================================================================
%    Acknowledgements 
%================================================================================================================
\vspace{\fill}\vspace{-4pt}
\section*{Acknowledgements}%
\vspace{-4pt}
\noindent The authors gratefully acknowledge fruitful conversations with Bo Feng, Franz Herzog, Andrew McLeod, Ben Page, J.J.~Stankowicz, and Ellis Yuan. This work was performed in part at the Aspen Center for Physics, which is supported by National Science Foundation grant PHY-1607611, and the Harvard Center of Mathematical Sciences and Applications. 
This project has been supported by an ERC Starting Grant \mbox{(No.\ 757978)}, a grant from the Villum Fonden \mbox{(No.\ 15369)}, and by a grant from the Simons Foundation (341344, LA) (JLB).
The research of J.T.\ is supported in part by U.S. Department of Energy grant DE-SC0009999 and by the funds of University of California. E.H.\ is supported by the U.S. Department of Energy under contract DE-AC02-76SF00515.

\appendix
%================================================================================================================
%    Appendix A:  
%================================================================================================================
\newpage\vspace{-6pt}\section[\mbox{Graph Theory for Integrand Basis Building}]{\mbox{\hspace{0pt}Graph Theory for Integrand Basis Building}}\label{appendix:graph_theory}%
\vspace{-0pt}

In this appendix, we introduce a few formal definitions that should be useful to better understand the results described in this work. Beyond clarifying terms and their usage, we hope that this list of definitions and concepts are useful for the sake of precision.

In what follows, we describe the main ingredients (mostly graph-theoretic) required in our work. Definitions are organized loosely by narrative and logical flow.\\[-28pt]

\defn{internalEdges}{internal edges \textnormal{(`propagators')}}{edges $e_{\text{int}}\in\Gamma$ of a graph $\Gamma$ (of edge-connectivity at least 2) which connect pairs of vertices of valency at least two. They are indicated diagrammatically as lines with constant, medium stroke and represent standard Feynman propagators of a scalar field theory. Most often, internal edges are colored black, unless subsets of edges are being highlighted to emphasize some additional structure. 
}
\defn{externalEdges}{external edges \textnormal{(`legs')}}{edges $e_{\text{ext}}\in\Gamma$ of a graph $\Gamma$ which connect at least one monovalent vertex.\footnote{These `external' monovalent vertices are never drawn in our figures.} To make these edges visually distinct, and to avoid confusion with internal edges defined above, external edges are drawn as wedges.\\ \indent There are two types of external edges, drawn differently as wedges:\vspace{-4pt}
\begin{itemize}
\item[{\color{hred}\rule[2.5pt]{10pt}{1pt}}]\subdefn{necessaryEdges}{necessary \textnormal{external edges}}{drawn as `\fig{0pt}{1}{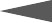}'---those external edges connected to a vertex of valency exactly three. These legs are called `necessary' because they allow us to differentiate the momentum flowing through pairs of internal edges separated by the 3-valent vertex.~\\[-8pt]

These edges are always drawn in our figures as solid wedges, (with slightly gray coloring) to distinguish them from internal edges.~\\[-8pt]
 
Physically, \emph{necessary} edges denote any \emph{non-empty} subset of external particles. As these subsets can be of arbitrary (but never empty) size, we consider graphs 
\eq{\fig{-12pt}{1}{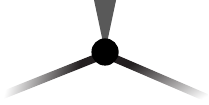}\bigger{\simeq}\fig{-12pt}{1}{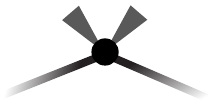}\bigger{\simeq}\fig{-12pt}{1}{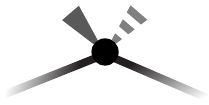}\,\vspace{5pt}}
}
\item[{\color{hred}\rule[2.5pt]{10pt}{1pt}}]\subdefn{optionalEdges}{optional \textnormal{external edges}}{drawn as `\fig{0pt}{1}{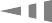}'---those external edges connected to (internal) vertices of valency at least four. These are called `optional' because they need not carry any momentum to allow us to differentiate the momentum flowing through the multiples of internal edges to which they connect. ~\\

These edges are always drawn in our figures as `sliced' greyscale wedges to emphasize their optionality.~\\[-8pt]

Physically, \emph{optional} edges should be understood of representing any subset of external particles' momenta---\emph{including empty subsets of external particles.} 
}
\end{itemize}\vspace{-24pt}
}
\defn{skeletonGraph}{skeleton graph $\widetilde{\Gamma}$}{for a given graph $\Gamma$, we define its skeleton $\widetilde{\Gamma}$ to be the subgraph obtained by deleting all external edges and vertices. For example, 
\eq{\fig{-17pt}{0.33}{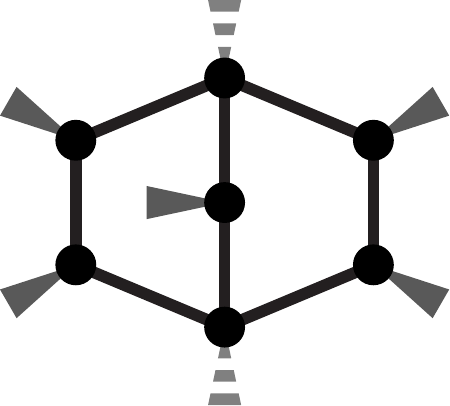}\,\,\equivL\Gamma\,\,\bigger{\mapsto}\,\,\widetilde{\Gamma}\equivL\fig{-17pt}{0.33}{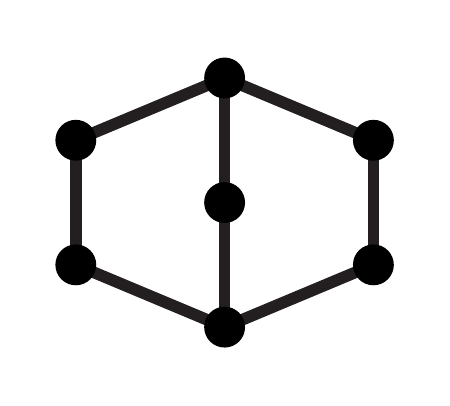}\,.}
This concept is useful in the discussion of the parent-daughter covering relations in which we are interested as external edges never carry internal loop momenta and therefore never play any role in numerator decompositions. 
}
\defn{loopNumber}{\textnormal{(number of)}~loops}{the first Betti number of the graph, unless otherwise specified. The qualification in the previous sentence is to clarify that---very occasionally---we will have reason to discuss loop integrands that involve no propagators for some or all of the internal loop momenta; for example, we may speak of the `$L$ loop integrand' $d^L\vec{\ell}\times 1$. In such cases, there are no loop-dependent propagators to draw; but we hope that the `loop order' of such integrands is always clear from the surrounding context.}
\defn{simpleCycle}{\textnormal{(simple)}~cycle}{a connected subgraph of a given graph with loop number 1. Throughout this work, the term `cycle'  should always be understand to mean (what is more formally described as) a \emph{simple} cycle.}
\defn{girth}{girth}{the length of the shortest cycle of a given graph.}
\defn{cycleBasis}{cycle basis}{for a graph with $L$ loops, a choice of $L$ simple cycles such that every internal edge is an element of at least one cycle.}
\defn{routing}{\textnormal{(loop momentum)} routing}{a choice of cycle basis for a given graph. We usually take these cycles to be oriented (even though the internal edges do not have any intrinsic orientation). A routing for a graph is equivalent to a choice of loop-momentum variables up to translation (associated with internal degrees of freedom not fixed by momentum conservation at every vertex in the graph).

One point that is worth clarifying is that our definition of loop-momentum \emph{routing} is slightly broader than what is more common in the physics literature. Frequently physicists discuss (especially non-planar, multiloop) integrands' loop momentum dependence by choosing a subset of $L$ (oriented) edges of the graph whose graph-complement is a $1$-forest (a tree graph). Such a choice of edges is always possible, and is understood as dictating that `the momentum flowing through edge $i$ of the graph to be loop momentum $\ell_i$'. This convention is considerably more restrictive than what we mean by routing here: not only does the more familiar prescription eliminate the all translational invariance of each loop momentum, but it also prevents us from choosing routings such as
\eq{\fig{-38pt}{1.5}{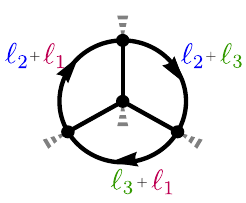}\;\text{with cycle basis}\quad\left\{\rule{0pt}{30pt}\right.\fwbox{50pt}{\fig{-36pt}{1.25}{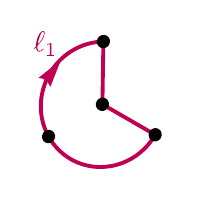}},\fwbox{50pt}{\fig{-36pt}{1.25}{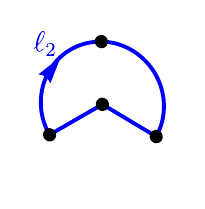}},\fwbox{50pt}{\fig{-36pt}{1.25}{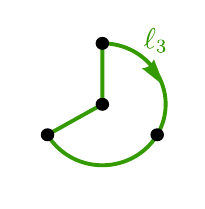}}\left.\rule{0pt}{30pt}\right\}\,.}
}

\subsection*{\mbox{\hspace{0pt}Covering Relations: `Parents' and `Daughters' (a.k.a. `Contact-Terms')}}\label{appendix:graph_theory:covering}

Graphs may be partially ordered according to quotients generated by {\it internal} {\bf edge contractions}. Thus, we say that $\Gamma_d\!\prec\!\Gamma_p$ if the \subdefn{daughter}{{\bf daughter}} $\Gamma_d$ can be obtained by contracting some number of (exclusively) internal edges of the \subdefn{parent}{{\bf parent}} graph $\Gamma_p$. To be perhaps overly pedantic, we always use the freedom of (\ref{eq:graph_equivalence_ext_props}) to delete excess external edges. Alternatively, we could speak exclusively of skeleton graphs.

This partial ordering on the set of Feynman graphs provides us with covering relations which prove extremely useful. In particular, we use the symbol $\partial$ to denote the set of daughters of a given parent obtained by a single edge-contraction. A skeleton graph $\Gamma$ with $n_e$ internal edges will have $n_e$ daughters---that is, $|\partial(\Gamma)|\!=\!n_e$. 

In the colloquial vernacular of physicists, daughter graphs are often called {\bf contact-terms} of their parents. This is because a sufficiently general space of loop-dependent numerators can include terms proportional to the inverse of any propagator (which are represented by the internal edges of the graphs). Thus, the vector-space of rational differential forms on the space of loop momenta with denominators fixed as Feynman propagators associated with a particular graph include all daughter integrals unless bounds are specified for loop-dependent factors in the numerators of these integrands.

%================================================================================================================
%    References (& /Document)
%================================================================================================================
\newpage

%\bibliographystyle{physics}
%\bibliography{amplitude_refs}
%\end{document}
\providecommand{\href}[2]{#2}\begingroup\raggedright\endgroup

\end{document}